\documentclass[aps,pra,reprint,groupedaddress,amsmath,amssymb,%
eqsecnum,longbibliography]{revtex4-1}
\pdfoutput=1

\usepackage{hyperref}

\newtheorem{theorem}{Theorem}

\newcommand{\rmd}{\mathrm{d}}

\newcommand{\rmb}{\mathrm{b}}
\newcommand{\rmf}{\mathrm{f}}

\DeclareMathOperator{\im}{im}

\DeclareMathOperator{\tr}{tr}

\DeclareMathOperator{\rank}{rank}
\DeclareMathOperator{\real}{Re}
\DeclareMathOperator{\imag}{Im}

\newcommand{\ket}[1]{\ensuremath{\lvert #1 \rangle}}
\newcommand{\bra}[1]{\ensuremath{\langle #1 \rvert}}
\newcommand{\pket}[1]{\ensuremath{\lvert #1 )}}

\newcommand{\inprod}[2]{\ensuremath{\langle #1 \vert #2 \rangle}}
\newcommand{\outprod}[2]{\ensuremath{\lvert #1 \rangle \langle #2 \rvert}}
\newcommand{\matelm}[3]{\ensuremath{\langle #1 \lvert #2 \rvert #3 \rangle}}
\newcommand{\pinprod}[2]{\ensuremath{( #1 \vert #2 )}}

\newcommand{\pmatelm}[3]{\ensuremath{( #1 \lvert #2 \rvert #3 )}}

\newcommand{\abs}[1]{\lvert #1 \rvert}
\newcommand{\norm}[1]{\lVert #1 \rVert}
\newcommand{\expect}[1]{\langle #1 \rangle}

\newcommand{\babs}[1]{\biggl\lvert #1 \biggr\rvert}

\newcommand{\newinprod}[2]{\ensuremath{( #1 \vert #2 )}}

\newcommand{\dzarea}{\ensuremath{\rmd^{2b} z}}

\begin{document}

\title{Subsystems and time in quantum mechanics}

\author{Bradley A. Foreman}
\affiliation{Department of Physics, The Hong Kong University of Science and 
Technology, Clear Water Bay, Kowloon, Hong Kong, China}


\begin{abstract}
This paper investigates the relationship between subsystems and time in a closed
nonrelativistic system of interacting bosons and fermions. It is possible to
write any state vector in such a system as an unentangled tensor product of
subsystem vectors, and to do so in infinitely many ways. This requires the
superposition of different numbers of particles, but the theory can describe in
full the equivalence relation that leads to a particle-number superselection
rule in conventionally defined subsystems. Time is defined as a functional of
subsystem changes, thus eliminating the need for any reference to an external
time variable. The dynamics of the unentangled subsystem decomposition is
derived from a variational principle of dynamical stability, which requires the
decomposition to change as little as possible in any given infinitesimal time
interval, subject to the constraint that the state of the total system satisfy
the Schr\"odinger equation. The resulting subsystem dynamics is deterministic.
This determinism is regarded as a conceptual tool that observers can use to make
inferences about the outside world, not as a law of nature. The experiences of
each observer define some properties of that observer's subsystem during an
infinitesimal interval of time (i.e., the present moment); everything else must
be inferred from this information. The overall structure of the theory has some
features in common with quantum Bayesianism, the Everett interpretation, and
dynamical reduction models, but it differs significantly from all of these. The
theory of information described here is largely qualitative, as the most
important equations have not yet been solved. The quantitative level of
agreement between theory and experiment thus remains an open question.
\end{abstract}


\maketitle

\tableofcontents

\section{Introduction}

\begin{quote}
\textit{It seems to me that for a systematic foundation of quantum mechanics one
needs to begin with the composition and decomposition of quantum systems.} \\
\phantom{x} \hfill --- W. Pauli \cite{[] [{, p.\ 3.}] AtmanspacherFuchs2014,
Pauli1985, [{Another translation of Ref.\ \cite{Pauli1985} is provided by
Howard: ``Quite independently of \emph{Einstein}, it appears to me that, in
providing a systematic foundation for quantum mechanics, one should \emph{start}
more from the composition and separation of systems than has until now (with
Dirac, e.g.) been the case. --- This is indeed --- as Einstein has
\emph{correctly} felt --- a very fundamental point in quantum mechanics, which
has, moreover, a direct connection with your reflections about the \emph{cut}
and the possibility of its being shifted to an arbitrary place.'' See }] []
Howard1997b}
\end{quote}

In physics, as in everyday life, we must divide the world conceptually into
subsystems before we can say anything about it. We cannot comprehend the
undivided world; we can only talk about how its parts differ from or change
relative to each other. In quantum mechanics the role of subsystem
decompositions is, if anything, even more important. As Wheeler has said, ``we
are first able to play the game when with chalk we have drawn a line across the
empty courtyard'' \cite{Wheeler1981}. If we decline to draw such a line, ``then
physics has vanished, and only a mathematical scheme remains'' \cite{[{}] [{,
p.\ 58.}] Heisenberg1930}.

The ``mathematical scheme'' that Heisenberg refers to is the Schr\"odinger
equation for the time evolution of a closed system. If the world is not divided
into subsystems, time may pass on a sheet of paper, but nothing can be said to
\emph{happen}, as there is no point of contact between theory and experiment.
Indeed it is not even clear what the time variable in the Schr\"odinger
equation \emph{means} if one cannot talk about relative changes between
different parts of the system.

This paper explores the relationship between subsystems and time in
nonrelativistic quantum mechanics. It springs from an examination of four
interrelated questions: (1)~how to construct subsystem decompositions without
entanglement; (2)~how to define time in a closed system; (3)~how to define a
dynamics of interacting subsystems without entanglement; and (4)~how to extract
information from these subsystems.

Cursory answers to these four questions are as follows: (1)~Unentangled
subsystem decompositions can be constructed in Fock space, using superpositions
of different numbers of particles. (2)~Time can be defined as a functional of
subsystem changes. (3)~An entanglement-free dynamics 
can be derived by maximizing the stability of the subsystem decomposition.
(4)~An observer obtains information from only one subsystem, and only during
the present moment of time. All else must be inferred.

A brief introduction to each of these four subject areas is given below. It must
be emphasized at the outset that this paper is exploratory in nature; the theory
of information presented here has not yet reached a stage of development that
would permit a direct comparison with experiment.

\subsection{Definition of subsystems}

\label{sec:intro_define_subsystems}

In quantum mechanics, subsystems are traditionally defined using a tensor
product of vector spaces \cite{CohTan1977}. Subsystems defined in this way
inevitably become entangled by interactions \cite{Schrodinger1935b}. Although
entanglement is now commonly regarded as a resource \cite{Horodecki2009}, its
consequences still leave many people uneasy with the foundations of quantum
mechanics \cite{[] [{, p.\ 102.}] Weinberg2015, Leggett2005, [] [{, p.\ 185.}]
Isham1995, Laloe2012, Bell2004}.

But is this definition necessary? Another approach has been
advocated by Primas and Amann \cite{Primas1983, Primas1987, Primas1990a,
*Primas1990b, Primas1991, Primas1993, Primas1994a, Primas1994b, Primas2000,
Amann1988, Amann1991a, Amann1991b, *Amann1991c, Amann1993, Amann1995,
AmannPrimas1997b, AmannAtmanspacher1999}. It is based on the concept of a
\emph{quantum object}, which Primas defines as ``an open quantum system which in
spite of its interaction with the environment is not [entangled] with the
environment'' \cite{Primas1987}. Objects may change their properties but they
are required to ``keep their identity in the course of time'' \cite{Primas1987}.
The concept of an object is similar to that of a quasiparticle \cite{Nozieres1997,
Kaxiras2003, Jain2007}. However, quasiparticles are designed to minimize
interactions, whereas objects are designed to minimize entanglement.

Algebraic ``dressing'' techniques have been developed for the construction of
quantum objects in some simple models, but only in the context of certain
approximations and asymptotic limits \cite{Primas1983, Primas1987, Primas1990a,
*Primas1990b, Primas1991, Primas1993, Primas1994a, Primas1994b, Primas2000,
Amann1988, Amann1991a, Amann1991b, Amann1991c, Amann1993, Amann1995,
AmannPrimas1997b, AmannAtmanspacher1999}. One approximation considered essential
by Primas is neglecting the Pauli exclusion principle at the level of the
interactions between subsystems. Thus, electrons in different subsystems are
treated as distinguishable. However, this contradicts a basic principle of
elementary-particle physics.

Is it possible to construct quantum objects without neglecting the Pauli
exclusion principle? The answer is yes, if we define a subsystem decomposition
as a tensor product of \emph{vectors} rather than a tensor product of vector
\emph{spaces}. Of course, the answer also depends on how we define a system.

The system studied in this paper is a closed nonrelativistic system of
interacting fermions and bosons. The dimensions of the underlying
single-particle Hilbert spaces are assumed to be finite; the dimension of the
resulting fermion Fock space is then finite too, but that of the boson Fock
space is infinite. For example, one might consider a finite volume in coordinate
space and cut off wavelengths shorter than some given value. The use of such a
cutoff is congruent with the current understanding of the standard model of
elementary-particle physics as an effective field theory \cite{WeinbergVol1}.
Bell has called this type of system a ``serious part of quantum mechanics'' that
covers a ``substantial fragment of physics'' \cite{Bell1990}, thus
making it (in his opinion) a worthy object of study in the field of quantum
foundations \cite{Bell1987, Bell1990}.

It is shown in Secs.\ \ref{sec:tensor_prod_indistinguishable} and
\ref{sec:mathematics} that any state vector $\ket{\psi}$ in such a system can be
written \emph{exactly} as an unentangled tensor product of an arbitrary number
of subsystem vectors $\ket{u_k}$. Furthermore, for any given number of
subsystems, this can be done in infinitely many ways. Despite the lack of
entanglement, knowledge of the whole ($\ket{\psi}$) does not imply knowledge of
the parts ($\ket{u_k}$), because the subsystem vectors are not confined to
subspaces. On the contrary, $\ket{u_k}$ occupies the same Fock space as
$\ket{\psi}$.

\subsection{Definition of time}

\label{sec:intro_define_time}

Such a decomposition is only meaningful if it persists over time. A crucial
question that must be addressed is how to \emph{define} time in a closed system
without making reference to an external time variable \cite{BarbourBertotti1982,
PageWootters1983, Wootters1984, UnruhWald1989, Pegg1991, Isham1992, Page1994,
Barbour1994a, Barbour1994b, GambiniPortoPullin2004a, GambiniPortoPullin2004b,
GambiniPortoPullin2009, Poulin2006, MilburnPoulin2006, AlbrechtIglesias2008,
Arce2012, Moreva2014}. The given subsystem decomposition is very useful in this
regard. It allows time to be defined as a \emph{functional} that organizes
information about subsystem changes. 

Section \ref{sec:kinematics} begins the preparatory work for this definition by
developing ways to quantify differences between subsystem decompositions. The
main focus is on geometric concepts related to the Fubini--Study metric in
Hilbert space. Section \ref{sec:time_functional} then defines time as a
functional of two infinitesimally different (but otherwise arbitrary) subsystem
decompositions. This functional is defined so as to maximize the amount of
change that can be expressed in the form of Schr\"odinger dynamics. It is a
functional of the entire subsystem decomposition; this eliminates 
the need to single out any particular subsystem as a clock.

\subsection{Subsystem dynamics}

\label{sec:intro_dynamics}

The time functional can be used to formulate a dynamics of subsystems by means
of a variational principle of \emph{dynamical stability}. This principle
requires the subsystem decomposition to change as little as possible in any
given infinitesimal time interval, subject to the constraint that the total
state vector $\ket{\psi}$ of the closed system satisfy the Schr\"odinger
equation. However, the interacting subsystems derived from this principle do
\emph{not} satisfy the Schr\"odinger equation.

The concept of \emph{subsystem} dynamics does not even exist in the usual
tensor-product-of-vector-spaces formulation of subsystems. The tensor-product
decomposition is just given arbitrarily, with no connection (in principle)
between different times (although it is often taken to be time independent). The
subsystem dynamics developed here (in Sec.\ \ref{sec:dynamical_stability})
therefore has no parallel in orthodox quantum mechanics.

Dynamically stable subsystems are \emph{quantum objects} in the sense defined
above. Does this mean that their observable properties can be regarded as
elements of a ``free-standing reality'' \cite{FuchsPeres2000, Fuchs2003}?
Deriving such a description was in fact a large part of the original motivation
for this study. However, the answer turns out to be an emphatic \emph{no}.
Dynamically stable subsystems still have an unavoidable element of subjectivity.

There are two fundamental reasons for this. One is that the subjective choice of
a \emph{number} of subsystems is essential to the dynamics. The results depend
explicitly on this number, and it has to be put in by hand. Its value cannot be
derived from the principle of dynamical stability itself.

The second reason is that the resulting dynamics is \emph{deterministic}. This
investigation began with a vague expectation that the dynamical stability
problem might not have unique solutions, thereby necessitating the introduction
of ``objective'' probabilities at a fundamental level. This could be regarded as
a new type of decoherence mechanism that is inherent in the dynamics of quantum
objects.

However, this expectation proved to be false. The variation problem has a
unique solution, so the resulting subsystem dynamics is deterministic.
This means that the principle of dynamical stability cannot explain the 
lack of determinism exemplified by the ``quantum jumps'' of 
orthodox quantum mechanics \cite{[] [{, p.\ 36.}] Dirac1958, vonNeumann1955}.

If the principle of dynamical stability cannot be regarded as the foundation for
a law of nature, what is it good for? It is essentially just a tool for
observers to use. They can use it to infer something about the properties of
subsystems in the past or the future from whatever information they have about
those subsystems now. There is no guarantee that these inferences will agree
with their experiences, because the subsystem dynamics is not viewed as a law
governing the behavior of anything ``real.'' It is instead viewed as an
instantiation of Wheeler's aphorism that ``the only law is the law that there is
no law'' \cite{Wheeler1973}. That is, nature is not governed by laws; laws are
just useful conceptual tools.

How are these results affected by the inclusion of superselection rules
\cite{WickWightmanWigner1952, HegerfeldtKrausWigner1968, WickWightmanWigner1970,
Wightman1995}? One must be careful in answering this question, because the
standard rules are derived from the tensor-product-of-subspaces definition of
subsystems, which is not relevant in the present context. (The standard rules
are also highly controversial even in the proper context
\cite{AharonovSusskind1967a, AharonovSusskind1967b, AharonovRohrlich2005,
Mirman1969, Mirman1970, Mirman1979, Lubkin1970, Zeh1970, Zurek1982,
GiuliniKieferZeh1995, Giulini2003a, Giulini2009a, WeinbergVol1,
DowlingBartlettRudolphSpekkens2006, BartlettRudolphSpekkens2007, Earman2008}.)
One must therefore go back to the underlying \emph{cause} of a superselection
rule (i.e., the lack of an external reference frame
\cite{BartlettRudolphSpekkens2007}) and reexamine its consequences for the
definition of subsystems used here.

In the absence of an external reference frame, certain subsystem decompositions
become observationally indistinguishable from one another. One can account for
this indistinguishability by introducing \emph{equivalence classes}
\cite{Jauch1964} of subsystem decompositions. For the reference frame that gives
rise to a particle-number superselection rule, the subsystem dynamics with or
without such equivalence classes is qualitatively the same, as shown in Sec.\
\ref{sec:superselection}. That is, the dynamics remains fully deterministic.
However, the quantitative dynamics can be quite different in these two cases.

An intriguing consequence of the Pauli exclusion principle is that the subsystem
dynamics depends on the \emph{order} in which the subsystem states $\ket{u_k}$
are multiplied. Ignoring this ordering would give rise to an apparent
decoherence effect. However, when the ordering of subsystems is accounted for,
the subsystem dynamics remains deterministic, as shown in Sec.\
\ref{sec:subsystem_permutations}.

\subsection{Information about subsystems}

\label{sec:intro_information}

Section \ref{sec:information_theory} addresses the question of precisely how
inferences are to be drawn from information about the properties of subsystems.
Because the subsystems are not entangled, we can use the methods of
Bayesian inference familiar from \emph{classical} probability theory
\cite{Jaynes1989, Jaynes2003, BernardoSmith2000, Appleby2005a, Appleby2005b,
Jeffrey2004}. This automatically ensures compliance with Bohr's injunction that
``however far the phenomena transcend the scope of classical physical
explanation, the account of all evidence must be expressed in classical terms''
\cite{Bohr1949, BohrVol2}.

These subsystems are nonclassical, but due to their lack of entanglement, their
properties can be regarded as ``beables'' rather than ``observables'' (in the
sense of Bell \cite{Bell2004}). That is, the subsystems can be considered to
\emph{have} certain properties, independently of whether they are ``measured.''
However, it should be emphasized that these properties are in general only
\emph{inferred} to exist, and that they are \emph{contextual} in the sense that
they depend on the subjective choice of a number of subsystems. As noted above,
they cannot be regarded as elements of a free-standing reality.

How are we to reconcile the determinism of the subsystem dynamics with the
``quantum jumps'' of textbook quantum mechanics? This can be done by imposing
limitations on the information an observer has access to. An observer is assumed
to experience directly only those beables associated with \emph{one} subsystem,
and only for an infinitesimal interval of time. This interval is called the
\emph{present moment} of time. The properties of all other subsystems, as well
as those of the observer's own subsystem in the past and future of the present
moment, must be inferred from this information.

These limitations are not due to an observer's ignorance, in the literal sense
of ignoring possibly accessible information. The observer simply has no access
to anything other than her own experiences. These experiences are not predicted
by anything in the theory. The degree to which an observer's inferences of an
outside world match her experiences in successive present moments determines how
useful her image of this outside world is. The degree of mismatch can be
regarded as an effective ``quantum jump.''

The conceptual structure of this theory of information draws heavily upon the
ideas of quantum Bayesianism (or QBism) developed by Caves, Fuchs, Schack, and
Mermin \cite{CavesFuchsSchack2002, CavesFuchsSchack2007, Fuchs2003, Fuchs2010,
Mermin2012a, *Mermin2012b, FuchsSchack2013, FuchsMerminSchack2014, Mermin2014b,
Mermin2014a, *Mermin2014c, Mermin2014d, Mermin2016}. The main difference is that
an observer is treated not as a black box but as a subsystem like any other.
This allows multiple observers to be treated as equals, thus providing a
framework for meaningful descriptions of intersubjective agreement.
``Objective'' properties can then be defined as those features of the common
worldview that survive when more observers are added to the framework.

The theory of information is developed only to the level of a bare skeleton,
however, because I have not solved the resulting system of equations. The
discussion of this part of the theory is therefore mostly qualitative, and no
proof can be offered that it leads to predictions in agreement with experiment.
The challenge of adding flesh to this skeleton nevertheless opens up many
promising avenues for future research.

\subsection{A guide for the reader}

As an aid to navigation, this subsection indicates which parts of the paper are
most essential on a first reading. Overall, the most difficult parts are those
dealing with technical details of boson vector spaces; these are largely
relegated to a series of appendices. On a first reading, one should focus
attention primarily on the properties of systems containing only fermions. 

The beginning of Sec.\ \ref{sec:tensor_prod_indistinguishable} shows that,
contrary to statements in the literature, systems of indistinguishable particles
do have a tensor product structure. Readers who are willing to take this for
granted can go straight to Eqs.\ (\ref{eq:uU0})--(\ref{eq:uvwUVW}), referring
back to earlier material as needed for definitions of notation. The paragraph
below Eq.\ (\ref{eq:U_fermi}) establishes when a fermion creator is invertible.
Equation (\ref{eq:U_bf}) shows that for systems containing bosons, the complex
numbers in the fermion matrix (\ref{eq:U_fermi}) are replaced by commuting
operators.

Section \ref{sec:mathematics} explains why invertibility of subsystems
is important and develops some mathematical tools for inversion.
Readers who are willing to take invertibility for granted can go straight
to the exponential representation of Eqs.\ (\ref{eq:U_exp_X}) and 
(\ref{eq:delta_eA}).  The definition of the word ``quasiclassical''
and its relation to classical probability theory in Sec.\ \ref{sec:quasiclassical}
are also worth noting.

Section \ref{sec:kinematics} deals with various kinematical aspects of the
definition of subsystems. The definition of observable quantities in Secs.\
\ref{sec:observables} and \ref{sec:relational_properties} should be included in
a first reading, but Secs.\ \ref{sec:permutations} and
\ref{sec:subsystem_differences} can be glossed over and referred back to when
needed. The geometric properties defined in Sec.\ \ref{sec:subsystem_geometry}
are essential as they are used throughout the rest of the paper.

Section \ref{sec:time_functional} develops various connections between
subsystems and time. Those willing to accept that interacting subsystems never
satisfy the Schr\"odinger equation can skip Sec.\
\ref{sec:interacting_not_Schroedinger}. The remaining subsections all contain
essential material. The key ideas of Sec.\ \ref{sec:time_functional_defn} are
the conceptual foundation for the time functional in Eqs.\ (\ref{eq:Schr_ideal})
and (\ref{eq:lambda_defn}) and the resulting functional in Eq.\
(\ref{eq:Delta_t}). The central result of Sec.\
\ref{sec:properties_time_functional} is the inequality (\ref{eq:time_energy}).

The topic of Sec.\ \ref{sec:dynamical_stability} is subsystem dynamics. The
concept of dynamical stability is quantified in Eq.\ (\ref{eq:chi_definition})
of Sec.\ \ref{sec:dynamical_stability_functional}. The simple case of a
time-independent total system state $\ket{\psi}$ is considered first in Sec.\
\ref{sec:time_independent_psi}. The solution for the unique maximum of the
dynamical stability functional is given in Eq.\ (\ref{eq:Dx_soln}). The
extension to time-dependent $\ket{\psi}$ in Secs.\ \ref{sec:time_dependent_psi},
\ref{sec:real_matrix}, and \ref{sec:solve_dynamical_stability} can be glossed
over on a first reading. The main result, shown in Eq.\ (\ref{eq:Delta_x_real}),
is qualitatively the same as the previous solution (\ref{eq:Dx_soln}), apart
from a necessary change of notation. Section \ref{sec:model_calculations} can be
skipped on a first reading, but the results of Sec.\ \ref{sec:arbitrary_number}
are essential for understanding why the subsystem decompositions used here
cannot be considered objective.

Section \ref{sec:superselection} discusses a system lacking a phase reference,
which would lead to a particle-number superselection rule in textbook quantum
mechanics. In such a system, the time functional (\ref{eq:Delta_t}) can be
replaced with the renormalized functional (\ref{eq:Delta_t_K}). After a similar
renormalization of other variables, the overall solution for the dynamically
stable subsystem change has the same form as before [i.e., Eq.\
(\ref{eq:Delta_x_real})]. This is an important result, but the details of the
derivation are not needed in any later sections of the paper. The same is true
for Sec.\ \ref{sec:subsystem_permutations}, which deals with the effect of
subsystem permutations on subsystem dynamics.

Section \ref{sec:information_theory} describes how information is extracted from
the preceding theory. It contains many key concepts but only one equation,
so it can be read in its entirety the first time around. The same is true for
the conclusions in Sec.\ \ref{sec:conclusions}.

\section{Tensor products and indistinguishable particles}

\label{sec:tensor_prod_indistinguishable}


The purpose of this section is to define clearly what is meant by the tensor
product of many-particle quantum states. Only a small part of the material
presented here is entirely new, but establishing a clear notation at the outset
helps to simplify calculations in subsequent sections of the paper.

The most convenient tensor product for systems of many indistinguishable
particles has an additional algebraic structure that accounts for symmetry or
antisymmetry with respect to particle permutations. The precise form of this
algebra is defined uniquely by the geometry of Hilbert space, in the form of a
cluster decomposition property for the inner product in Fock space. The
resulting tensor algebra is the same as that of particle creation operators in
Fock space.  Special care is needed to ensure closure of the algebra 
for systems containing bosons.

\subsection{Definition of a tensor product}

\label{sec:tensor}

Given two vector spaces $\mathcal{V}_1$ and $\mathcal{V}_2$, the tensor product
of vectors $\ket{u} \in \mathcal{V}_1$ and $\ket{v} \in \mathcal{V}_2$ is
written as $\ket{u} \otimes \ket{v}$. The tensor product is bilinear but not
commutative \cite{Szekeres2004}. That is, it must be distributive over vector
addition:
\begin{subequations} \label{eq:distributive}
\begin{align}
\ket{u} \otimes (\ket{v} + \ket{w}) & = \ket{u} \otimes \ket{v} + \ket{u}
\otimes \ket{w} , \\ (\ket{u} + \ket{z}) \otimes \ket{v} & = \ket{u} \otimes
\ket{v} + \ket{z} \otimes \ket{v} ,
\end{align}
\end{subequations}
as well as bilinear with respect to scalar multiplication:
\begin{equation}
\alpha (\ket{u} \otimes \ket{v}) = (\alpha \ket{u}) \otimes \ket{v} = \ket{u}
\otimes (\alpha \ket{v}) . \label{eq:bilinear}
\end{equation}
The set of linear combinations of such tensor products defines the vector space
$\mathcal{V}_1 \otimes \mathcal{V}_2$. The tensor product of three or more
vectors is associative:
\begin{equation}
(\ket{u} \otimes \ket{v}) \otimes \ket{w} = \ket{u} \otimes (\ket{v} \otimes
\ket{w}) , \label{eq:associative}
\end{equation}
thus defining the vector space $\mathcal{V}_1 \otimes \mathcal{V}_2 \otimes
\mathcal{V}_3$ uniquely.

\subsection{Indistinguishable particles}

\label{sec:indistinguishable}

The vector space for a system of $n$ identical particles is
\begin{equation}
\mathcal{H}^{n} = \underbrace{\mathcal{H} \otimes \mathcal{H} \otimes \cdots
\otimes \mathcal{H}}_{n \text{ times}} ,
\end{equation}
in which $\mathcal{H}$ is the Hilbert space of a single particle. The Fock space
$\mathcal{F} (\mathcal{H})$ for a system with an indefinite number of identical
particles is defined as the direct sum \cite{NegeleOrland1998,
ReedSimonVol1, Geroch1985}
\begin{equation}
\mathcal{F} (\mathcal{H}) = \bigoplus_{n=0}^{\infty} \mathcal{H}^{n} =
\mathcal{H}^{0} \oplus \mathcal{H} \oplus \mathcal{H}^{2} \oplus
\cdots ,
\end{equation}
where the zero-particle space $\mathcal{H}^{0}$ consists of all scalar multiples
of the vacuum state $\ket{0}$ (the null vector is written as $0$). In
the tensor algebra of Fock space, $\ket{0}$ is the multiplicative identity:
\begin{equation}
\ket{0} \otimes \ket{\psi} = \ket{\psi} \otimes \ket{0} = \ket{\psi} \qquad
\forall \ket{\psi} \in \mathcal{F} (\mathcal{H})  .  \label{eq:mult_ident}
\end{equation}
By construction, $\mathcal{F} (\mathcal{H})$ is therefore closed under tensor
multiplication \footnote{Here we are considering tensor multiplication only at
the formal level. See Appendix \ref{app:rigged} for a discussion of the
subtleties that arise when normalization is considered.}:
\begin{equation}
\mathcal{F} (\mathcal{H}) \otimes \mathcal{F} (\mathcal{H}) = \mathcal{F}
(\mathcal{H}) .
\end{equation}

According to the symmetrization postulate \cite{Messiah1962,
MessiahGreenberg1964}, 
the only physically meaningful states in $\mathcal{F} (\mathcal{H})$ are those
satisfying the symmetry condition
\begin{equation}
S \ket{\psi} = \ket{\psi} , \label{eq:symmetry}
\end{equation}
where $S$ is a projector for states of appropriate symmetry (i.e., totally
symmetric for bosons and totally antisymmetric for fermions).  It is defined by
\begin{equation}
S = \sum_{n=0}^{\infty} \Pi_n S_n \Pi_n , \qquad S_n = \frac{1}{n!} \sum_{\sigma} \varepsilon (\sigma) \sigma , \label{eq:symmetrizer}
\end{equation}
in which $\Pi_n$ is the projector for the $n$-particle subspace $\mathcal{H}^n$
and $S_n$ is the projector for the symmetric or antisymmetric states of
$\mathcal{H}^n$ (with $S_0 = S_1 = 1$). The sum over $\sigma$ covers the $n!$
permutation operators in $\mathcal{H}^n$. For fermions, $\varepsilon (\sigma)$
is the sign of the permutation $\sigma$:
\begin{equation}
\varepsilon (\sigma) = 
\begin{cases}
+1 & \text{if } \sigma \text{ is even} , \\
-1 & \text{if } \sigma \text{ is odd} . \\
\end{cases}
\end{equation}
For bosons, $\varepsilon (\sigma) = 1$. The subspace of vectors in $\mathcal{F}
(\mathcal{H})$ that satisfy the symmetry constraint (\ref{eq:symmetry}) is
denoted $\mathcal{F}_{s} (\mathcal{H})$, while the corresponding subspace of
$\mathcal{H}^{n}$ is denoted $S(\mathcal{H}^{n})$.

\subsection{The \texorpdfstring{$\psi$}{psi} product}

\label{sec:psi_product}

It is convenient at this point to introduce another tensor product that
automatically accounts for all symmetry requirements. This binary operator is
defined by
\begin{equation}
\ket{u^{(p)}} \odot \ket{v^{(q)}} = c(p, q) S (\ket{u^{(p)}} \otimes
\ket{v^{(q)}}) , \label{eq:psi_product}
\end{equation}
in which $\ket{u^{(p)}} \in S(\mathcal{H}^{p})$, $\ket{v^{(q)}} \in
S(\mathcal{H}^{q})$, and $c(p, q)$ is a numerical coefficient to be defined
below. The linearity of $S$ and bilinearity of $\ket{u} \otimes \ket{v}$ then
give a unique extension of $\ket{u} \odot \ket{v}$ to the case of general
$\ket{u}, \ket{v} \in \mathcal{F}_{s} (\mathcal{H})$.

For fermions, $\ket{u} \odot \ket{v}$ is the same as the exterior or wedge
product $\ket{u} \wedge \ket{v}$ \cite{Szekeres2004, Misner1973,
LoomisSternberg1990, Abraham1988, KostrikinManin1989, Hassani2013}, while for
bosons, $\ket{u} \odot \ket{v}$ is known as the symmetric product
\cite{Szekeres2004, Abraham1988, KostrikinManin1989, Hassani2013}. For brevity,
the name ``$\psi$ product'' is used here as an umbrella term covering both cases
in the context of many-particle quantum states of generic symmetry. Note that
the $\psi$ product is commutative only for bosons, since \cite{Szekeres2004,
LoomisSternberg1990, Abraham1988, KostrikinManin1989,
Hassani2013}
\begin{equation}
\ket{u^{(p)}} \odot \ket{v^{(q)}} = \zeta^{pq} \, \ket{v^{(q)}}
\odot \ket{u^{(p)}} , \label{eq:signed_product}
\end{equation}
where $\zeta = +1$ for bosons and $\zeta = -1$ for fermions.

The coefficient $c(p, q)$ is partially defined by requiring the $\psi$
product to be associative. As shown in Appendix \ref{app:associative}, this
requires $c(p, q)$ to have the form \cite{Abraham1988}
\begin{equation}
c(p, q) = \frac{f(p + q)}{f(p) f(q)} , \label{eq:cf}
\end{equation}
in which $f(1) = 1$ but $f(n)$ is otherwise arbitrary. Most authors choose
either $f(n) = 1$ \cite{Szekeres2004, KostrikinManin1989} or $f(n) = n!$
\cite{Misner1973, LoomisSternberg1990, Abraham1988, Hassani2013}.

However, for applications in quantum mechanics, it is more convenient to choose
$f(n) = \sqrt{n!}$ \footnote{See pp.\ 603 and 622--623 of Messiah
\cite{Messiah1962}.}, due to the following cluster decomposition theorem. The
theorem refers to a case in which $\mathcal{H} = \mathcal{H}_1 \oplus
\mathcal{H}_2$, where the subspaces $\mathcal{H}_1$ and $\mathcal{H}_2$ are
orthogonal. The Fock space therefore factors as $\mathcal{F}_{s} (\mathcal{H}) =
\mathcal{F}_{s} (\mathcal{H}_1) \odot \mathcal{F}_{s} (\mathcal{H}_2)$.

\begin{theorem}[Cluster decomposition] \label{thm:cluster}
Let $\ket{s t} = \ket{s} \odot \ket{t}$ and $\ket{u v} = \ket{u} \odot \ket{v}$
be vectors in $\mathcal{F}_{s} (\mathcal{H})$, where $\ket{s}, \ket{u} \in
\mathcal{F}_{s} (\mathcal{H}_1)$ and $\ket{t}, \ket{v} \in \mathcal{F}_{s}
(\mathcal{H}_2)$. Then the inner product $\inprod{s t}{u v}$ factors as
\begin{equation}
\inprod{s t}{u v} = \inprod{s}{u} \inprod{t}{v} \label{eq:cluster_decomp}
\end{equation}
for all such vectors if and only if $\abs{f(n)} = \sqrt{n!}$.
\end{theorem}
The factorization (\ref{eq:cluster_decomp}) is called a cluster decomposition
\cite{Peres1995} because the subspaces $\mathcal{H}_1$ and $\mathcal{H}_2$
typically refer to different regions in coordinate space. Theorem
\ref{thm:cluster} is proved in Appendix \ref{app:cluster_decomposition}.

The phase of $f(n)$ is not determined by this theorem. However, choosing $f(n)$
to be real and positive means that the coefficient of $\ket{u^{(p)}} \otimes
\ket{v^{(q)}}$ in $\ket{u^{(p)}} \odot \ket{v^{(q)}}$ is also real and positive.

The $\psi$ product is required here to have the cluster decomposition
property and to satisfy this phase convention. The coefficient in equation
(\ref{eq:psi_product}) is thus given uniquely by 
\begin{equation}
c(p, q) = \sqrt{\frac{(p+q)!}{p! q!}} . \label{eq:cpqf}
\end{equation}

It is sometimes said that ``a tensor-product structure is not present'' in
systems of indistinguishable particles \cite{Zanardi2002}. This is misleading,
however, because the $\psi$ product has all of the essential properties of a
tensor product (see Sec.\ \ref{sec:tensor}). Only the physically meaningless
tensor-product structure of a system of \emph{distinguishable} particles is
lacking.

\subsection{Subsystem creators}

\label{sec:creation}

The algebra of the $\psi$ product defined above is the same as the familiar
algebra of particle creation operators in Fock space. To see this, one can start
by defining a general $n$-particle $\psi$ product
\begin{equation}
\ket{\alpha_1 \alpha_2 \cdots \alpha_n} = \ket{\alpha_1} \odot \ket{\alpha_2}
\odot \cdots \odot \ket{\alpha_n} , \label{eq:ket_general}
\end{equation}
in which $\ket{\alpha_k} \in \mathcal{H}$, and the set $\{ \ket{\alpha_k} \}$ is
not assumed to be linearly independent or normalized. This vector is related to
the corresponding unsymmetrized $n$-particle tensor product by
\begin{equation}
\ket{\alpha_1 \alpha_2 \cdots \alpha_n} = \sqrt{n!} \, S (\ket{\alpha_1} \otimes
\ket{\alpha_2} \otimes \cdots \otimes \ket{\alpha_n}) , \label{eq:ket_relation}
\end{equation}
which follows from the definition of $\ket{u} \odot \ket{v}$ in Eqs.\
(\ref{eq:psi_product}) and (\ref{eq:cpqf}). The product of two such states has
the composition property (due to associativity)
\begin{equation}
\ket{\alpha_1 \cdots \alpha_p} \odot \ket{\beta_1 \cdots \beta_q} =
\ket{\alpha_1 \cdots \alpha_p \beta_1 \cdots \beta_q} .
\end{equation}

Consider now the case in which the set $\{ \ket{\alpha_k} \}$ is orthonormal,
with repetition of the same state permitted. Let $n_{\alpha}$ be the number of
times a particular state $\ket{\alpha} \in \mathcal{H}$ occurs in equation
(\ref{eq:ket_general}), with $\sum_{\alpha} n_{\alpha} = n$. For fermions, $\ket{\alpha} \odot \ket{\alpha} = 0$, so
values of $n_{\alpha} > 1$ merely give rise to the null vector. Excluding such
cases, the normalization of the vector (\ref{eq:ket_general}) is given by
\cite{NegeleOrland1998}
\begin{equation}
\inprod{\alpha_1 \cdots \alpha_n}{\alpha_1 \cdots \alpha_n} = \prod_{\alpha}
n_{\alpha} ! \; . \label{eq:normalization}
\end{equation}
That is, nonzero fermion states are normalized to unity, but this is true for
bosons only if no single-particle state is repeated. Unit vectors are useful in
some contexts (e.g., Appendix \ref{app:rigged}), but here the normalization
(\ref{eq:normalization}) is more convenient.

The particle creation operator $a_{\lambda}^{\dagger}$ can now be defined
as \cite{NegeleOrland1998}
\begin{equation}
a_{\lambda}^{\dagger} \ket{\lambda_1 \cdots \lambda_n} = \ket{\lambda} \odot
\ket{\lambda_1 \cdots \lambda_n} = \ket{\lambda \lambda_1 \cdots \lambda_n} ,
\label{eq:creation_operator}
\end{equation}
which maps $S(\mathcal{H}^{n})$ into $S(\mathcal{H}^{n+1})$. The vectors $\{
\ket{\lambda_{i}} \}$ need not be orthonormal, although the operator commutation
relations are simpler if they are \cite{NegeleOrland1998}. A product of creation
operators can thus be used to generate any simple product state from the vacuum:
\begin{equation}
\ket{\lambda_1 \cdots \lambda_n} = a_{\lambda_1}^{\dagger} \cdots
a_{\lambda_n}^{\dagger} \ket{0} . \label{eq:create_vacuum}
\end{equation}
The absence of numerical factors in this equation is due to the fact that no
normalization convention is imposed on $\ket{\lambda_1 \cdots \lambda_n}$
\cite{NegeleOrland1998}.

Any vector $\ket{u} \in \mathcal{F}_{s} (\mathcal{H})$ can therefore be
generated from the vacuum by
\begin{equation}
\ket{u} = U \ket{0} , \label{eq:uU0}
\end{equation}
in which $U$ is a linear combination of products of creation operators,
including (in general) a scalar term for the creation of no particles. For
convenience, the operator $U$ is called the \emph{creator} of the state
$\ket{u}$. The lengthier phrase ``creation operator'' is reserved for the
creator $a_{\lambda}^{\dagger}$ of a single-particle state $\ket{\lambda}$;
thus, the set of creation operators is a proper subset of the creators.

Given another such state $\ket{v} = V \ket{0}$, the $\psi$ product of $\ket{u}$
and $\ket{v}$ is
\begin{subequations} \label{eq:uvUV}
\begin{align}
\ket{u} \odot \ket{v} & = (U \ket{0}) \odot \ket{v} \\ & = U (\ket{0} \odot
\ket{v}) \\ & = U \ket{v} \\ & = UV \ket{0} .
\end{align}
\end{subequations}
The algebra of the vectors $\ket{u}$ and $\ket{v}$ is therefore the same as the
algebra of the creators $U$ and $V$. This can be seen even more clearly when a
redundant vacuum state is appended to $\ket{v}$:
\begin{equation}
\ket{u} \odot \ket{v} \odot \ket{0} = UV \ket{0} .
\end{equation}
This result can be extended to any number of $\psi$ products; for
example, if $\ket{w} = W \ket{0}$, then
\begin{equation}
\ket{u} \odot \ket{v} \odot \ket{w} \odot \ket{0} = UVW \ket{0} .
\label{eq:uvwUVW}
\end{equation}

Of course, it is only meaningful to write such equations if the vector defined
by this product is normalizable. No difficulty arises for fermions, because the
dimension of the fermion Fock space is finite; fermion creation operators are
therefore bounded. However, boson creators are unbounded (with respect
to the topology defined by the usual Hilbert space norm); the $\psi$ product of
two normalized boson vectors could therefore be unnormalizable. One must take
care to choose a boson vector space that is closed under the $\psi$ product.

The construction of such a vector space is described in Appendix
\ref{app:rigged}. The result, denoted $\mathcal{F}_{\psi} (\mathcal{H}_{\rmb})$,
is a dense subspace of $\mathcal{F}_{s} (\mathcal{H}_{\rmb})$, where
$\mathcal{H}_{\rmb}$ is the Hilbert space of a single boson. It should be noted
that $\mathcal{F}_{\psi} (\mathcal{H}_{\rmb})$ is a Fr\'echet space
\cite{ReedSimonVol1, Horvath2012} rather than a Hilbert space. This distinction
can be ignored for many purposes, because the inner product from
$\mathcal{F}_{s} (\mathcal{H}_{\rmb})$ remains well defined in the subspace
$\mathcal{F}_{\psi} (\mathcal{H}_{\rmb})$; the only change is that the
Fr\'echet-space topology is not defined by this inner product.

\subsection{Matrix notation for fermions}

\label{sec:fermion_matrix}

In a fermion system, the Fock space has a finite dimension $2^{d}$, in which
$d$ is the dimension of the single-fermion Hilbert space $\mathcal{H}_{\rmf}$.
It is therefore convenient to introduce matrix representations for the (always
bounded) fermion creators $U$ and $V$.

Let $\mathcal{H}_{\rmf}$ be spanned by an orthonormal basis $\{ \ket{e_{k}} \}$,
in which $k \in \{ 0, 1, 2, \ldots, d-1 \}$. The Fock space $\mathcal{F}_{s}
(\mathcal{H}_{\rmf})$ is then spanned by the basis $\{ \ket{f_i} \}$, where the
integer $i \in \{ 0, 1, 2, \ldots, 2^{d} - 1 \}$ has the binary representation
\begin{equation}
i = \sum_{k=0}^{d-1} i_{k} 2^{k} \qquad (i_k \in \{ 0, 1 \}) ,
\label{eq:binary_notation}
\end{equation}
in which $i_k$ is the $k$th binary digit of $i$. If $i_k = 1$, the state
$\ket{e_k}$ is occupied in $\ket{f_i}$; otherwise, it is unoccupied. Thus, for
example, when $d = 4$, the basis vector $\ket{f_5}$ can be written in various
ways as
\begin{equation}
\ket{f_5} = \ket{0101} = \ket{e_2} \odot \ket{e_0} .
\end{equation}
In this notation, $\ket{f_0}$ is just the vacuum state $\ket{0}$.

As a simple example, consider the case $d = 2$, for which a general vector
$\ket{u} \in \mathcal{F}_{s} (\mathcal{H}_{\rmf})$ can be written as
\begin{equation}
\ket{u} = \sum_{i=0}^{3} c_{i} \ket{f_i} , \qquad c_{i} \equiv \inprod{f_i}{u} .
\label{eq:u_Fermi}
\end{equation}
The $\psi$ products of $\ket{u}$ with the
basis vectors of $\mathcal{F}_{s} (\mathcal{H}_{\rmf})$ are then given by
$\ket{u} \odot \ket{f_0} = \ket{u}$, $\ket{u} \odot \ket{f_{1}} = c_{0}
\ket{f_{1}} + c_{2} \ket{f_{3}}$, $\ket{u} \odot \ket{f_{2}} = c_{0}
\ket{f_{2}} - c_{1} \ket{f_{3}}$, and $\ket{u} \odot \ket{f_{3}} = c_{0}
\ket{f_{3}}$. The matrix representing $U$ in this basis is therefore
\begin{equation}
U = 
\begin{pmatrix}
c_{0} & 0        & 0        & 0 \\
c_{1} & c_{0}  & 0       & 0 \\
c_{2} & 0        & c_{0} & 0 \\
c_{3} & c_{2} & -c_{1} & c_{0}
\end{pmatrix} . \label{eq:U_fermi}
\end{equation}

In general, $U$ has the form of a lower triangular matrix whenever the basis $\{
\ket{f_{i}} \}$ is arranged in order of nondecreasing particle number $\abs{i}
\equiv \sum_{k=0}^{d-1} i_k$. (Such an arrangement is called graded
lexicographic ordering \cite{Cox2007}.) Hence, $U$ is invertible if and only if
$c_0 \ne 0$, since the determinant of a triangular matrix is just the product of
its diagonal elements.

\subsection{Different types of particles}

\label{sec:different_types}

Up to this point it has tacitly been assumed that the system under consideration
contains only one type of particle. To combine different types of either bosons
or fermions, one can simply take a direct sum of the single-particle Hilbert
spaces before constructing the Fock space \cite{[] [{, pp.\ 166--167 and
227--237.}] Sudbery1986}. All boson (fermion) creation operators then commute
(anticommute) with each other. Such a description is possible because the
operators for different fermions can be chosen arbitrarily to either commute or
anticommute \cite{[] [{, pp.\ 250 and 475.}] LandauLifshitz1977, [] [{, p.\
94.}] LandauLifshitz1980}. (These two choices are related by a Jordan--Wigner
transformation \cite{JordanWigner1928, LiebSchultzMattis1961,
SchultzMattisLieb1964, Araki1961, [] [{, Sec.\ 4-4.}] StreaterWightman1989}.)

In this mode of description, which is commonly used in the treatment of isospin
\cite{Messiah1962, Sudbery1986, LandauLifshitz1977}, there are only two types of
particles: bosons and fermions. Different types of bosons or different types of
fermions are treated as formally indistinguishable; the distinction is
maintained only at the level of quantum numbers within the single-particle
Hilbert spaces $\mathcal{H}_{\rmb}$ and $\mathcal{H}_{\rmf}$ \footnote{In an
alternative mode of description, different types of bosons and fermions are
distinguished using an unsymmetrized tensor product (see Appendix
\ref{app:different}). Formulas such as Eq.\ (\ref{eq:u_not_Schmidt})
then become much more cumbersome, but the physics is the same. In particular,
the vector space $\mathcal{E}$ of Eq.\ (\ref{eq:total_E}) is not affected by
this change of convention; only the labeling and phase of the basis vectors is
altered \cite{Sudbery1986}.}.

Bosons and fermions are combined together into one system by means of an
ordinary (unsymmetrized) tensor product. A general vector in the tensor-product
space
\begin{equation}
\mathcal{E} = \mathcal{F}_{\psi} (\mathcal{H}_{\rmb}) \otimes \mathcal{F}_{s}
(\mathcal{H}_{\rmf}) \label{eq:total_E}
\end{equation}
is thus of the form
\begin{equation}
\ket{u} = \sum_{j=0}^{\infty} \sum_{i=0}^{2^{d}-1} c_{ji} \ket{b_j} \otimes
\ket{f_i} , \label{eq:u_expansion}
\end{equation}
in which $\{ \ket{b_j} \}$ and $\{ \ket{f_i} \}$ are fixed orthonormal bases for
the boson and fermion Fock spaces (see Sec.\ \ref{sec:fermion_matrix} for the
definition of $\{ \ket{f_i} \}$). This can also be written as
\begin{equation}
\ket{u} = \sum_{i=0}^{2^{d}-1} \ket{u_i} \otimes \ket{f_i} ,
\label{eq:u_not_Schmidt}
\end{equation}
in which $\ket{u_i} = \sum_{j} c_{ji} \ket{b_j}$. This is not a Schmidt
decomposition, since the basis $\{ \ket{f_i} \}$ is independent of $\ket{u}$;
the set $\{ \ket{u_i} \}$ is therefore generally not orthonormal.

In the matrix notation of Sec.\ \ref{sec:fermion_matrix}, $\ket{u}$ can be
written as a $2^{d}$-component ``spinor,'' given here explicitly (in decimal and
binary) for the case $d = 2$:
\begin{equation}
\ket{u} =
\begin{pmatrix}
\ket{u_{0}} \\ \ket{u_{1}} \\ \ket{u_{2}} \\ \ket{u_{3}}
\end{pmatrix} =
\begin{pmatrix}
\ket{u_{00}} \\ \ket{u_{01}} \\ \ket{u_{10}} \\ \ket{u_{11}}
\end{pmatrix} . \label{eq:u_spinor}
\end{equation}
The boson-fermion creator $U$ then has the form
\begin{equation}
U = 
\begin{pmatrix}
U_{0} & 0         & 0        & 0 \\
U_{1} & U_{0}  & 0        & 0 \\
U_{2} & 0        & U_{0} & 0 \\
U_{3} & U_{2}  & -U_{1} & U_{0}
\end{pmatrix} , \label{eq:U_bf}
\end{equation}
in which the constants $c_i$ of Eq.\ (\ref{eq:U_fermi}) are replaced by
boson creators $U_i$ (defined by $\ket{u_i} = U_i
\ket{0}_{\rmb}$).

As shown in Appendix \ref{app:different}, the definition of the $\psi$ product
$\ket{u} \odot \ket{v}$ can easily be extended to the vector space
$\mathcal{E}$. All of the main conclusions of Secs.\ \ref{sec:psi_product} and
\ref{sec:creation}, including the cluster decomposition property and the
algebraic equivalence shown in Eq.\ (\ref{eq:uvwUVW}), remain valid in
$\mathcal{E}$.

\section{Basic tools for subsystem analysis}

\label{sec:mathematics}


Let us turn now to an investigation of the conditions under which a general ket
$\ket{\psi} \in \mathcal{E}$ can be written as a $\psi$ product of the
form $\ket{\psi} = \ket{u} \odot \ket{v}$.  This equation is equivalent to 
the operator product
\begin{equation}
\Psi = U V , \label{eq:Psi_U_V}
\end{equation}
in which $\Psi$, $U$, and $V$ are the creators of the states $\ket{\psi}$,
$\ket{u}$, and $\ket{v}$, respectively.

\subsection{Invertible subsystem creators}

\label{sec:invertible}

Such an equation can clearly be written for any \emph{invertible} subsystem
creator $U$, since one can then take
\begin{equation}
V = U^{-1} \Psi \label{eq:V_Uinv_Psi}
\end{equation}
for arbitrary $\Psi$. Likewise, if $V$ is invertible, one can always let $U =
\Psi V^{-1}$. This approach can easily be extended to more general products
such as $\Psi = UVW$, since (for example) if $U$ and $W$ are invertible one
can always let $V = U^{-1} \Psi W^{-1}$.

Under what conditions is a creator invertible? This question has already been
answered for fermion systems; in Sec.\ \ref{sec:fermion_matrix}, it was noted
that a fermion creator $U$ is invertible if and only if $c_0 \ne 0$, where $c_0
= \inprod{f_0}{u} = \inprod{0}{u}$ is the vacuum component of $\ket{u}$.
Invertibility is thus a weak constraint: the set of invertible fermion creators
is uncountable and forms a smooth manifold.

A similar condition can be derived for systems containing both bosons and
fermions. The main result is expressed here in the form of a theorem (proved in
Appendix \ref{app:invertibility_theorem}), using the notation of Sec.\
\ref{sec:different_types}.
\begin{theorem}[Invertibility] \label{thm:invertible}
For a boson-fermion creator $U$, the following statements are equivalent:
(a)~The linear map $U : \mathcal{E} \to \mathcal{E}$ is invertible. (b)~The
associated boson creator $U_0 : \mathcal{F}_{\psi} \to \mathcal{F}_{\psi}$ is
invertible. (c)~The corresponding boson state $\ket{u_0} = U_{0} \ket{0}_{\rmb}$
is a coherent state.
\end{theorem}
The coherent states mentioned in part (c) of the theorem are just the familiar
Glauber states \cite{Glauber1963c, KlauderSkagerstam1985, BlaizotRipka1986,
NegeleOrland1998, Perelomov1986, ZhangFengGilmore1990, Gazeau2009}, as defined
in Eq.\ (\ref{eq:coherent}) of Appendix \ref{app:rigged}.

The constraint imposed by Theorem \ref{thm:invertible} remains very weak. For
$U$ to be invertible, the boson state $\ket{u_0}$ associated with the fermion
vacuum $\ket{f_0} = \ket{0}_{\rmf}$ must be a coherent state, but all of the
other kets $\ket{u_i}$ in Eq.\ (\ref{eq:u_not_Schmidt}) (i.e., those with $i \ne
0$) are completely arbitrary.

\subsection{Why invertibility is important}

\label{sec:invertible_important}

Equation (\ref{eq:V_Uinv_Psi}) is not the most general solution of Eq.\
(\ref{eq:Psi_U_V}). The most general solution has the form
\cite{CampbellMeyer2009, BenIsrael2003, StewartSun1990, Nashed1976}
\begin{equation}
V = U^{-} \Psi + (1 - U^{-} U) Y , \label{eq:gen_inv}
\end{equation}
in which $U^{-}$ is a generalized inverse of $U$ \cite{CampbellMeyer2009,
BenIsrael2003, StewartSun1990, Nashed1976} and the creator $Y$ is arbitrary. For
finite-dimensional matrices, such a solution exists if and only if $\Psi \in \im
U$, where $\im U$ denotes the image (or range) of $U$. When $U$ is invertible,
Eq.\ (\ref{eq:gen_inv}) reduces to Eq.\ (\ref{eq:V_Uinv_Psi}), but invertibility
of $U$ is not necessary for the existence of the solution (\ref{eq:gen_inv}).

Nevertheless, all subsequent analysis in this paper is based on the study of
\emph{invertible} subsystem creators. This is done for two reasons. The first is
simply the pragmatic reason that the algebra of the subsystem creators is much
simpler when invertibility is assumed at the outset.

The second reason is more technical but also more compelling. For simplicity,
consider the Moore--Penrose generalized inverse $A^{-}$ of a finite-dimensional
square matrix $A$ \cite{CampbellMeyer2009, BenIsrael2003, StewartSun1990}. In
this case, $(A + \delta A)^{-}$ is well known to be continuous at $\delta A = 0$
if and only if the rank of $(A + \delta A)$ is the same as the rank of $A$ for
all perturbations $\delta A$ in some finite neighborhood of $\delta A = 0$
\cite{CampbellMeyer2009, BenIsrael2003, StewartSun1990}. But if $A$ does not
have full rank, there exist matrices $A + \delta A$ with $\rank (A + \delta A) >
\rank A$ for arbitrarily small $\norm{\delta A} > 0$. Hence, $(A + \delta
A)^{-}$ is continuous at $\delta A = 0$ if and only if $A$ has full rank---i.e.,
if and only if $A$ is invertible.

Continuity of the subsystem (\ref{eq:gen_inv}) is essential because the
subsystem dynamics in Sec.\ \ref{sec:dynamical_stability} is derived from a
variational principle. In order for subsystems to be stationary states of the
dynamical stability functional, they must first be continuous with respect to
small variations. Subsystems derived from generalized inverses are consequently
not used in this paper.

A simple corollary of Theorem \ref{thm:invertible} is that all invertible
subsystems must have a vacuum component. This implies that all nontrivial (i.e.,
not purely vacuum) invertible subsystems must have an indefinite number of
particles. A pragmatic reason for allowing this type of subsystem was pointed
out by Bell \cite{Bell1976}:
\begin{quote}
The real world is made of electrons and protons and so on, and as a result the
boundaries of natural objects are fuzzy, and some particles in the boundary can
only doubtfully be assigned to either object or environment. I think that
fundamental physical theory should be so formulated that such artificial
divisions are manifestly inessential.
\end{quote}
The superposition of different numbers of particles is therefore essential to
the definition of the subsystems considered here. This does not imply that the
physical limitations leading to particle-number superselection rules must be
ignored in this theory. A detailed discussion of this topic is, however,
postponed until Sec.\ \ref{sec:superselection}.

\subsection{Functions of subsystem creators}

Before writing down any formulas for the inverse of a creator, it is helpful to
start by establishing some general properties of creators and functions of
creators. Creators are actually easier to work with than generic operators. The
reason for this is that any creator $A$ can be decomposed into even and odd
parts:
\begin{equation}
A = A_{+} + A_{-} ,
\end{equation}
in which $A_{+}$ ($A_{-}$) comprises all terms with an even (odd) number of
fermion creation operators. From Eq.\ (\ref{eq:signed_product}) we see that an
even creator commutes with any creator, whereas two odd creators anticommute:
\begin{equation}
A_{+} B_{\pm} = B_{\pm} A_{+} , \qquad 
A_{-} B_{-} = - B_{-} A_{-} . \label{eq:creator_commute_anti}
\end{equation}
The commutator $[ A, B ] \equiv AB - BA$ of two creators is therefore given by
\begin{equation}
[ A, B ] = 2 A_{-} B_{-} , \label{eq:creator_commutator}
\end{equation}
which commutes with any other creator $C$, since $A_{-} B_{-}$ is even. Equation
(\ref{eq:creator_commute_anti}) also implies that all odd creators are
nilpotent:
\begin{equation}
A_{-}^2 = 0 . \label{eq:A_minus_nilpotent}
\end{equation}
The binomial expansion of any integer power of a creator thus contains
only two terms:
\begin{equation}
A^{n} = (A_{+} + A_{-})^{n} = A_{+}^{n} + n A_{+}^{n-1} A_{-} .
\label{eq:An_binomial}
\end{equation}
All functions $f(A)$ defined by a power series can be expanded likewise
as
\begin{equation}
f(A) = f(A_{+}) + A_{-} f' (A_{+}) ,
\end{equation}
in which $f'(x) = \rmd f / \rmd x$.

\subsection{Inversion formula}

\label{sec:inversion_formula}

Let us now return to the case of an invertible creator $U$. Given that $U_0$ is
invertible, we can write $U$ as
\begin{equation}
U = U_0 (1 + Z) = (1 + Z) U_0 , \label{eq:Z}
\end{equation}
in which $Z \equiv U_0^{-1} U - 1 = U U_0^{-1} - 1$. Calculating $U_0^{-1}$ is
trivial, because the operator $U_0$ for a coherent state is an exponential
function [see Eq.\ (\ref{eq:coherent})]. In the matrix notation of Eq.\
(\ref{eq:U_bf}), $Z$ has the same lower-triangular form as $U$:
\begin{equation}
Z = 
\begin{pmatrix}
Z_{0}  & 0         & 0        & 0 \\
Z_{1}  & Z_{0}  & 0        & 0 \\
Z_{2}  & 0         & Z_{0} & 0 \\
Z_{3} & Z_{2}  & -Z_{1} & Z_{0}
\end{pmatrix}
\qquad (\text{for } d = 2) . \label{eq:Z_bf}
\end{equation}
The key difference is that $Z_0 = 0$, by definition
of $Z$. The operator $Z$ is therefore nilpotent---i.e., $Z^{k} = 0$ for some
finite integer $k$. Since the matrix size is $2^d \times 2^d$, we see
immediately that $k \le 2^{d}$.

However, we can obtain a stronger bound on $k$ by noting from Eq.\
(\ref{eq:A_minus_nilpotent}) that the square of any \emph{monomial} function of
the fermion creation operators is also zero. Since the maximum degree of any
such monomial is $d$, inspection of Eq.\ (\ref{eq:An_binomial}) shows that
\begin{equation}
Z^k = 0 \qquad \forall k > \lceil (d/2) \rceil ,
\label{eq:Z_nilpotent}
\end{equation}
in which $\lceil x \rceil$ is the ceiling function. One can then use the
geometric series to obtain the inversion formula
\begin{equation}
U^{-1} = U_0^{-1} \sum_{n=0}^{\lceil (d/2) \rceil} (-1)^{n} Z^{n} .
\label{eq:U_inverse_series}
\end{equation}

\subsection{The exponential representation}

In a similar fashion, we can use Eq.\ (\ref{eq:Z}) to calculate the
logarithm of $U$:
\begin{equation}
\ln U = \ln U_0 + \ln (1 + Z) ,
\end{equation}
in which $\ln (1 + Z)$ is given by the power series
\begin{equation}
\ln (1 + Z) = \sum_{n=1}^{\lceil (d/2) \rceil} \frac{(-1)^{n+1}}{n} Z^n .
\end{equation}
Every invertible creator can therefore be represented as an 
exponential function:
\begin{align}
U & = \exp X = e^{X_0} e^{(X - X_0)} \nonumber \\ 
& = e^{X_0} \sum_{n=0}^{\lceil (d/2) \rceil} 
\frac{(X - X_0)^n}{n!} , \label{eq:U_exp_X}
\end{align}
in which $X = \ln U$. The intermediate steps in Eq.\ (\ref{eq:U_exp_X}) used the
facts that $X_0 = \ln U_0$ is even and that $X - X_0$ is nilpotent.

The exponential representation (\ref{eq:U_exp_X}) plays a crucial role in the
remainder of this paper. For both computation and analysis, it is preferable to
take the creator $X$ as fundamental and \emph{define} the subsystem $U$ as $U =
\exp X$. This guarantees that regardless of any changes in $X$, $U$ always
remains invertible, its inverse being given simply by $U^{-1} = \exp (-X)$. In
this approach, the inversion formula (\ref{eq:U_inverse_series}) becomes
redundant.

The Campbell--Baker--Hausdorff formula for two creators $A$ and $B$ can now be
used to show that
\begin{align}
e^{A} e^{B} & = e^{A + B + [A, B] / 2} = 
e^{A + B + A_{-} B_{-}} \nonumber \\
& = e^{A + B} e^{A_{-} B_{-}} = e^{A + B} (1 + A_{-} B_{-}) ,
\end{align}
in which Eq.\ (\ref{eq:creator_commutator}) and $(A_{-} B_{-})^2 = 0$ were used.
Adding $e^{A} e^{B}$ and $e^{B} e^{A}$ then yields the very useful formula
\begin{equation}
e^{A+B} = \{ e^{A} , e^{B} \} , \label{eq:expAplusB}
\end{equation}
in which $\{ A, B \} \equiv \frac12 (AB + BA)$ denotes the symmetrized product.
As a special case of Eq.\ (\ref{eq:expAplusB}), note that if $A$ is varied by
$\delta A$, the corresponding first-order variation in $e^{A}$ is given by
\begin{align}
\delta (e^{A}) & \equiv e^{A + \delta A} - e^{A} \nonumber \\ 
& = \{ e^{\delta A} - 1 , e^{A} \} \nonumber \\
& = \{ \delta A , e^{A} \} , \label{eq:delta_eA}
\end{align}
in which terms of second and higher order in $\delta A$ were discarded in the
final step. This yields a simple expression for the derivative of $e^{A}$ with
respect to a parameter $s$:
\begin{equation}
\frac{\partial}{\partial s} (e^{A}) = 
\left\{ \frac{\partial A}{\partial s} , e^{A} \right\}  .
\end{equation}
This result is much simpler than the corresponding formulas for a
general operator $A$ \cite{Wilcox1967}. Several other identities for the
symmetrized product of creators are collected together in Appendix
\ref{app:creator_identities}.

\subsection{Quasiclassical subsystems}

\label{sec:quasiclassical}

The representation of quantum states by exponential functions has a long
history, dating back at least to the WKB approximation of 1926. The modern
theory of generalized coherent states \cite{KlauderSkagerstam1985,
Perelomov1986, ZhangFengGilmore1990, Gazeau2009} also relies heavily on
exponential representations. In the latter approach, the coherent states are
tied to a particular Lie group (namely, the dynamical group associated with the
Hamiltonian of the total system), and the exponential functions used to
construct the coherent states are \emph{unitary} operators involving both
creation and annihilation operators. Such unitary operators have the advantage
of convenient normalization properties, but they are not useful in the present
context because their algebra is not isomorphic to the algebra of the $\psi$
product.

The exponential representation (\ref{eq:U_exp_X}) can be viewed as a further
generalization of the coherent-state concept, in that the creator $X$ is not
tied to any Lie group. Indeed, $X$ is almost entirely arbitrary, the only
restrictions being the algebraic closure constraint of Appendix \ref{app:rigged}
(which requires all $\psi$ products of subsystems to be normalizable) and the
invertibility constraint requiring $U_0 = \exp X_0$ to be a Glauber coherent
state (\ref{eq:coherent}) (which implies that $X_0$ is a \emph{linear} function
of the boson creation operators). But the latter constraint is not independent
of the former, because the algebraic closure condition of Appendix
\ref{app:rigged} was used in Appendix \ref{app:invertibility_theorem} to derive
the linearity of $X_0$.

Subsystems permitting an exponential representation $U = \exp X$ will be
referred to as \emph{quasiclassical} subsystems in this paper. However, since
the word ``quasiclassical'' has various other connotations (including the WKB
approximation and the Glauber coherent states \cite{CohTan1977}), it is
important to be clear about the sense in which this label is being used here.
In this paper, ``quasiclassical'' is just a synonym for ``invertible.''

This sense of the word quasiclassical does not imply the use of any
approximation. As emphasized in Sec.\ \ref{sec:invertible}, the product $\Psi =
UV$ provides an exact representation for arbitrary states $\Psi$. As discussed
in Sec.\ \ref{sec:invertible_important}, invertibility of $U$ is the minimal
restriction needed to ensure continuity of $V$ when $U$ is varied and $\Psi$ is
held constant. Note that when $U$ is taken to be quasiclassical, $V$ is
quasiclassical if and only if $\Psi$ is quasiclassical---but $\Psi$ need not be
quasiclassical.

The fact that all coherent states are invertible is one possible reason for
designating the latter as quasiclassical. A more significant reason is that the
existence of an unentangled product $\ket{\psi} = \ket{u} \odot \ket{v}$ permits
the use of \emph{classical} probability theory. Since different subsystem
decompositions $\ket{u} \odot \ket{v}$ are mutually exclusive alternatives for
the representation of $\ket{\psi}$, one's state of knowledge or belief about the
suitability of various decompositions can be described using ordinary Bayesian
probability theory. The implications of this idea are developed further in the
next section.

\section{Kinematics of subsystems without subspaces}

\label{sec:kinematics}


This section explores several topics that are independent of and
logically prior to any concept of subsystem dynamics, thus falling
into the category of kinematics.  These include the definition of
observable quantities, the description of differences between two
subsystem decompositions, and subsystem geometry.

For generality, let $\ket{\psi}$ be decomposed into a product of $m$
subsystems, where $m \ge 2$:
\begin{equation}
\ket{\psi} = \ket{u_1} \odot \ket{u_2} \odot \cdots \odot \ket{u_{m}} .
\label{eq:psi_u_decomp}
\end{equation}
Here and below, the subscript $k$ in $\ket{u_k}$ is just a label for the $m$
different subsystems ($k \in \{ 1, 2, \ldots, m \}$). The subscripts introduced
previously in Eq.\ (\ref{eq:u_not_Schmidt}) are henceforth retired as they have
no further use.

\subsection{Observables and beables}

\label{sec:observables}

Given such a subsystem decomposition, how are we to define the 
observable quantities of the theory?

A fundamental hypothesis of the present paper is that \emph{all} observables are
calculated from the subsystem vectors $\ket{u_k}$. In other words, no observable
quantity is calculated directly from the total system vector $\ket{\psi}$. This
is not to say that $\ket{\psi}$ is meaningless; the subsystem dynamics depends
on $\ket{\psi}$. However, $\ket{\psi}$ does not appear anywhere in the 
definition of observables.

Apart from this change, the mathematical apparatus used to define observables is
nearly the same as that in ordinary many-particle quantum mechanics
\cite{NegeleOrland1998, BlaizotRipka1986, Merzbacher1998}. That is, observables
are represented by hermitian operators $A$ that are totally symmetric with
respect to permutations of identical particles. For each such operator, one can
calculate a number
\begin{equation}
\expect{A}_k \equiv \frac{\matelm{u_k}{A}{u_k}}{\inprod{u_k}{u_k}}
\label{eq:mean_value}
\end{equation}
for each subsystem $\ket{u_k}$. In ordinary quantum mechanics, only the
eigenvalues of $A$ are observable; the numbers $\expect{A}_k$ are therefore
interpreted as mean values.

Here, however, the numbers $\expect{A}_k$ are taken to be directly observable.
This means that observables are defined using what Bell has called the ``density
of stuff'' interpretation \cite{Bell1990}, first introduced by Schr\"odinger
\cite{[] [{; English translation in Ref.\ \cite{Schrodinger1982}.}]
Schrodinger1926c, [] [{, pp.\ 41--44.}] Schrodinger1982}. A similar mass-density
interpretation has been used in the dynamical reduction theory of Ghirardi
\emph{et al}.\ \cite{GhirardiPearleRimini1990, GhirardiGrassiBenatti1995,
BassiGhirardi2003}, but here $\expect{A}_k$ can describe properties other than
mass density.

According to Bell \cite{Bell2004}, the numbers $\expect{A}_k$ can then be
classified as ``beables'' rather than observables. That is, subsystem
$\ket{u_k}$ is taken to \emph{possess} the property $\expect{A}_k$ independently
of any measurement (the concept of ``measurement'' having no place in the
fundamental postulates of the theory).

Such an interpretation of $\expect{A}_k$ is possible for the reason already
explained in Sec.\ \ref{sec:quasiclassical}---namely, that different subsystem
decompositions can be described using classical probability theory.
Without entanglement, there is no need for the properties
of the subsystems to be described as \emph{potential} rather
than \emph{actual}.

Of course, it remains to be seen whether this interpretation of $\expect{A}_k$
can give rise to experimental predictions similar to those obtained from
ordinary quantum mechanics. This is a difficult problem for which only
qualitative results are obtained in this paper.  Further discussion of this topic
is presented in Sec.\ \ref{sec:information_theory}.

\subsection{Relational properties}

\label{sec:relational_properties}

One aspect of observables that deserves special attention is that not all
quantities $\expect{A}_k$ are necessarily observable. For example, since the
total system is taken to be closed (i.e., not interacting with anything else),
quantities such as absolute position or orientation in space have no meaning
within the theory. One can of course calculate numbers for these quantities
within a given model, but such numbers are meaningless due to the lack of any
external reference frame.

The only meaningful properties are therefore relational properties. For example,
although the absolute position of the center of mass of a given subsystem is
meaningless, it is meaningful to talk about the relative distance between the
centers of mass of two subsystems. Another meaningful quantity would be the mass
density of a subsystem relative to the position of its own center of mass.

This type of restriction has received much attention lately in regard to the
connection between reference frames (or their lack) and superselection rules
\cite{BartlettRudolphSpekkens2007}. The consequences of such restrictions within
the present theory will be investigated in detail (for a specific example) in
Sec.\ \ref{sec:superselection}. Until then, however, it will be assumed that
there are no restrictions (in principle) on the observable quantities
$\expect{A}_k$.

\subsection{Subsystem permutations}

\label{sec:permutations}

Let us now return to the subsystem decomposition
(\ref{eq:psi_u_decomp}), expressed in terms of creators:
\begin{equation}
\Psi = U_{1} U_{2} \cdots U_{m} .
\end{equation}
If the only observables are the numbers (\ref{eq:mean_value}), 
no observable quantity depends on the \emph{order} in which
the subsystems are multiplied.  Any permutation
\begin{equation}
\Psi_{\pi} = U_{\pi (1)} U_{\pi (2)} \cdots U_{\pi (m)}
\label{eq:psi_pi}
\end{equation}
yields the same observables, where $\pi$ denotes one of the $m!$
permutations of the integers $(1, 2, \ldots, m)$. The value of the product
does, however, depend on this order, as indicated by the subscript on
$\Psi_{\pi}$.

According to the results of Sec.\ \ref{sec:quasiclassical}, all but one of the
subsystems $U_k$ are taken to be quasiclassical. For definiteness, let $U_1$ be
the one subsystem that need not be quasiclassical. In the permutation
(\ref{eq:psi_pi}), $U_1$ is at position $k = k_v$, where
\begin{equation}
\pi (k_v) = 1 , \qquad k_v = \pi^{-1} (1) .
\end{equation}
We can then solve Eq.\ (\ref{eq:psi_pi}) for $U_1$, obtaining
\begin{equation}
V \equiv U_1 = U_{\pi (k_v - 1)}^{-1} \cdots U_{\pi (1)}^{-1}
\Psi_{\pi} U_{\pi (m)}^{-1} \cdots U_{\pi (k_v + 1)}^{-1} .
\end{equation}
As noted here, the symbol $V$ or $\ket{v} = V \ket{0}$ will often be used to
refer to this non-quasiclassical subsystem. 

\subsection{Subsystem differences}

\label{sec:subsystem_differences}

The next topic is how to describe small differences between two subsystem
decompositions. Consider two sets of subsystems, $\{ U_k \}$ and $\{ U_k' \}$,
whose products are $\Psi$ and $\Psi'$, respectively. The quasiclassical
subsystems can be given an exponential representation
(\ref{eq:U_exp_X}), for which the difference $\Delta X_k \equiv X_k' - X_k$ is
assumed to be small.  The corresponding difference $\Delta
U_k \equiv U_k' - U_k$ is then given (for $k \ne 1$) by Eq.\ (\ref{eq:delta_eA}):
\begin{align}
\Delta U_k & = \exp (X_k + \Delta X_k) - \exp X_k \nonumber
\\ &= \{ \Delta X_k, U_k \} , \label{eq:Delta_Uk}
\end{align}
in which terms beyond the first order in $\Delta X_k$ have been neglected.
Likewise, the first-order difference between $(U_k')^{-1}$ and $U_k^{-1}$ 
is (for $k \ne 1$)
\begin{align}
\Delta U_k^{-1} & = \exp (-X_k - \Delta X_k) - \exp (-X_k ) \nonumber
\\ &= - \{ \Delta X_k, U_k^{-1} \} .
\end{align}
The difference $\Delta V = V' - V = U_1' - U_1$ is taken to be determined by the
values of $\{ \Delta U_k^{-1} \}$ and $\Delta \Psi = \Psi'- \Psi$. To define
$\Delta V$, it is helpful to introduce linear functionals $V_{X}$ and
$\tilde{V}_k [Y]$ such that when $X = \Psi$ and $Y = U_{k}^{-1}$ we have
\begin{equation}
V_{\Psi} = V = \tilde{V}_k [U_{k}^{-1}]  \qquad (k \ne 1) .
\end{equation}
To first order in small quantities, $\Delta V$ is then given by
\begin{equation}
\Delta V = V_{\Delta \Psi} + \sum_{k=2}^{m} \tilde{V}_k [\Delta U_k^{-1}] .
\label{eq:Delta_V}
\end{equation}
Note that $\Delta V$ depends (implicitly) on the choice of permutation $\pi$ in
Eq.\ (\ref{eq:psi_pi}).

For practical calculations, the exponents $\ket{x_k} = X_k \ket{0}$ are
usually expanded in some orthonormal basis $\{ \ket{e_{ki}} \}$:
\begin{equation}
\ket{x_k} = \sum_{i} c_{ki} \ket{e_{ki}} \qquad 
(k \ne 1) , \label{eq:xk_cki}
\end{equation}
in which $c_{ki} = \inprod{e_{ki}}{x_k}$. The corresponding change $\ket{\Delta
x_k}$ is thus
\begin{equation}
\ket{\Delta x_k} = \sum_{i} \Delta c_{ki} \ket{e_{ki}} , \quad 
\Delta c_{ki} = \inprod{e_{ki}}{\Delta x_k} . \label{eq:Dxk}
\end{equation}
Combining this expression with Eq.\ (\ref{eq:Delta_Uk}) then
gives
\begin{equation}
\ket{\Delta u_k} = \sum_{i} \Delta c_{ki} \ket{f_{ki}} ,
\label{eq:Delta_uk}
\end{equation}
in which the creator of $\ket{f_{ki}}$ is defined to be
\begin{equation}
f_{ki} = \{ e_{ki}, U_k \} . \label{eq:fki}
\end{equation}
For a given value of $k$, the set $\{ \ket{f_{ki}} \}$ is generally not
orthonormal, but it is linearly independent if and only if the set $\{
\ket{e_{ki}} \}$ is. [This can be shown easily using Eq.\
(\ref{eq:switch_basis}).] From Eq.\ (\ref{eq:Delta_uk}), we now see that
$\ket{\Delta u_k}$ and $\ket{\Delta x_k}$ are related by
\begin{equation}
\ket{\Delta u_k} = \biggl( \sum_{i} \outprod{f_{ki}}{e_{ki}} \biggr) 
\ket{\Delta x_k} . \label{eq:DukDxk}
\end{equation}
A similar expression for $\ket{\Delta v} = \Delta V \ket{0}$ can be derived from
Eq.\ (\ref{eq:Delta_V}):
\begin{subequations} \label{eq:DvDxk}
\begin{align}
\ket{\Delta v} & = \ket{v_{\Delta \Psi}} + \sum_{k=2}^{m} \sum_{i} \Delta
c_{ki} \ket{g_{ki}} \\ & = \ket{v_{\Delta \Psi}} + \sum_{k=2}^{m} \biggl(
\sum_{i} \outprod{g_{ki}}{e_{ki}} \biggr) \ket{\Delta x_k} ,
\end{align}
\end{subequations}
in which the creators of $\ket{g_{ki}}$ are defined as
\begin{equation}
g_{ki} = - \tilde{V}_k [ \{ e_{ki}, U_k^{-1} \} ] .
\end{equation}

\subsection{Subsystem geometry}

\label{sec:subsystem_geometry}

The preceding expressions for subsystem differences are useful primarily in the
context of \emph{geometric} structures that allow us to measure the
\emph{distance} between neighboring subsystem decompositions. Such a metric is
essential for both the definition of the time functional in Sec.\
\ref{sec:time_functional} and the variational formulation of dynamical stability
in Sec.\ \ref{sec:dynamical_stability}.

There are many ways to define such a distance, but the most suitable measure for
the present purposes is the Hilbert--Schmidt distance \cite{Bengtsson2006}. This
distance is based on the Hilbert--Schmidt inner product and norm
\begin{equation}
(A, B) = \frac12 \tr (A^{\dag} B) , \qquad \norm{A} = \sqrt{(A, A)} ,
\label{eq:HS_inprod}
\end{equation}
in which $A$ and $B$ are operators. The Hilbert--Schmidt distance
$D$ is then defined as
\begin{equation}
D (A, B) = \norm{A - B} .  \label{eq:HS_distance}
\end{equation}
The operators of interest are the subsystem projectors
\begin{equation}
\rho_{k} = \frac{\outprod{u_{k}}{u_{k}}}{\inprod{u_{k}}{u_{k}}} ,
\qquad
\rho_{k}' = \frac{\outprod{u_{k}'}{u_{k}'}}{\inprod{u_{k}'}{u_{k}'}} .
\end{equation}
The square of the Hilbert--Schmidt distance between the subsystem states
$\ket{u_k}$ and $\ket{u_k'}$ is thus given by
\begin{subequations}
\label{eq:D2k}
\begin{align}
D^2 (\rho_k, \rho_{k}') & = 1 - \tr (\rho_k \rho_{k}') \\ & =
\frac{\matelm{u_{k}'}{(1 - \rho_{k})}{u_{k}'}}{\inprod{u_{k}'}{u_{k}'}}  .
\end{align}
\end{subequations}
This satisfies $0 \le D^2 (\rho_k, \rho_{k}') \le 1$, which is the reason for
introducing the factor of $1/2$ in Eq.\ (\ref{eq:HS_inprod})
\cite{Bengtsson2006}.

Our primary interest is in the value of $D^2 (\rho_k, \rho_{k}')$ for small
subsystem differences $\ket{\Delta u_{k}} = \ket{u_{k}'} - \ket{u_{k}}$. Noting
that $(1 - \rho_{k}) \ket{u_{k}} = 0$, we can rewrite Eq.\ (\ref{eq:D2k}) as
\begin{equation}
D^2 (\rho_k, \rho_{k}') = 
\frac{\matelm{\Delta u_{k}}{(1 - 
\rho_{k})}{\Delta u_{k}}}{\inprod{u_{k}'}{u_{k}'}} ,
\end{equation}
which shows that $D^2 (\rho_k, \rho_{k}')$ is of order $\norm{\Delta u_{k}}^2
\equiv \inprod{\Delta u_{k}}{\Delta u_{k}}$. Indeed, since
$\inprod{u_{k}'}{u_{k}'} = \inprod{u_{k}}{u_{k}} + O (\norm{\Delta u_{k}})$, we
have
\begin{equation}
D^2 (\rho_k, \rho_{k}') = 
\frac{\matelm{\Delta u_{k}}{(1 - \rho_{k})}{\Delta u_{k}}}{\inprod{u_{k}}{u_{k}}}
+ O (\norm{\Delta u_{k}}^3) .
\end{equation}
The leading term in this expression is the familiar Fubini--Study metric
\cite{ProvostVallee1980, Wootters1981, Page1987, Berry1989, AnandanAharonov1990,
Anandan1991, Bengtsson2006}. The Hilbert--Schmidt distance is only one of
several large-scale measures of distance that lead to the Fubini--Study metric
in the limit of infinitesimal $\norm{\Delta u_{k}}$ \cite{ProvostVallee1980,
Bengtsson2006}, but it is generally the easiest of these to work with.

The next step is to extend this measure of distance to the subsystem
decomposition (\ref{eq:psi_pi}) as a whole. The simplest way to do this is to
construct a direct sum of the projectors $\rho_k$:
\begin{equation}
\rho \equiv \bigoplus_{k=1}^{m} \rho_{k} = 
\rho_1 \oplus \rho_2 \oplus \cdots \oplus \rho_m .
\end{equation}
The resulting operator $\rho$ is also a projector, since $\rho_k^2 = \rho_k$
implies $\rho^2 = \rho$. The Hilbert--Schmidt distance (\ref{eq:HS_distance})
between $\rho$ and $\rho'$ is then
\begin{subequations} \label{eq:HS_total}
\begin{align}
D^2 (\rho, \rho')  & = m - \tr (\rho \rho') \\ & =
\sum_{k=1}^{m} D^2 (\rho_k, \rho_{k}')  ,
\end{align}
\end{subequations}
which is clearly independent of the choice of permutation $\pi$ in Eq.\
(\ref{eq:psi_pi}). In the limit of infinitesimal $\norm{\Delta u_{k}}$, this
reduces to
\begin{equation}
D^2 (\rho, \rho') = \sum_{k=1}^{m} \frac{\matelm{\Delta u_{k}}{(1 - 
\rho_{k})}{\Delta u_{k}}}{\inprod{u_{k}}{u_{k}}} , \label{eq:D2_rho_FS}
\end{equation}
which is the Fubini--Study metric for the entire subsystem decomposition.

This result can be expressed more concisely by using a direct-sum representation
of vectors. For arbitrary subsystem kets $\ket{\varphi_k}$ and
$\ket{\chi_k}$, let their direct sum be denoted by the same symbol without the
subscript $k$:
\begin{equation}
\ket{\varphi} \equiv \bigoplus_{k=1}^{m} \ket{\varphi_k} , \qquad \ket{\chi}
\equiv \bigoplus_{k=1}^{m} \ket{\chi_k} . \label{eq:direct_sum_ket}
\end{equation}
It is convenient also to bury the normalization factor $\inprod{u_{k}}{u_{k}}$
inside the definition of the inner product:
\begin{equation}
\inprod{\varphi}{\chi} \equiv \sum_{k=1}^{m} 
\frac{\inprod{\varphi_{k}}{\chi_{k}}}{\inprod{u_{k}}{u_{k}}} .
\label{eq:direct_sum_inprod}
\end{equation}
This allows Eq.\ (\ref{eq:D2_rho_FS}) to be written simply as
\begin{equation}
\eta \equiv D^2 (\rho, \rho') = \matelm{\Delta u}{(1 - \rho)}{\Delta u} ,
 \label{eq:D2_rho_FS_simple}
\end{equation}
in which the letter $\eta$ is introduced as a concise symbol for this functional
of $\rho$ and $\ket{\Delta u}$.

\section{Time as a functional}

\label{sec:time_functional}


With this measure of distance in hand, we can now turn to the topic
of subsystem dynamics, beginning with the concept of \emph{time}.
This section starts by explaining why, in a system of interacting particles,
it is impossible for both $\ket{\psi}$ and the subsystems $\ket{u_k}$
to satisfy the Schr\"odinger equation.  Next, it is argued that in 
a closed system, the external time parameter $t$ of conventional quantum
mechanics is meaningless.  Instead, time should be defined internally
via the relations between changes in subsystems.  This is then used to 
construct a time \emph{functional}, which will be applied to calculations
of subsystem dynamics in Sec.\ \ref{sec:dynamical_stability}.

\subsection{Why interacting subsystems cannot satisfy the 
Schr\"odinger equation}

\label{sec:interacting_not_Schroedinger}

Let us start by examining whether it is possible for a closed system and its
subsystems to satisfy the Schr\"odinger equation. It is sufficient for this
purpose to study a two-subsystem decomposition $\ket{\psi} = \ket{u} \odot
\ket{v} = U \ket{v}$. The Schr\"odinger equation for the total system is
\begin{equation}
i \partial_{t} \ket{\psi} = H \ket{\psi} = HU \ket{v} ,  
\label{eq:Schr_psi}
\end{equation}
in which $H$ is the Hamiltonian.  But this is the same as
\begin{subequations}
\begin{align}
i \partial_{t} \ket{\psi} & = [H, U] \ket{v} + U H \ket{v} \\ & = [H, U] \ket{v}
+ \ket{u} \odot (H\ket{v}) . \label{eq:Schr_b}
\end{align}
\end{subequations}
If the operator $[H, U]$ is a creator, then $[H, U] \ket{v} = ([H, U] \ket{0})
\odot \ket{v}$, in which
\begin{equation}
[H, U] \ket{0} = HU \ket{0} - UH \ket{0} = (H - E_0) \ket{u} ,
\end{equation}
where $E_0$ is the energy of the vacuum. Assuming that $E_0 = 0$ (which is
necessary if $\ket{0}$ is to act as a time-independent multiplicative
identity), Eq.\ (\ref{eq:Schr_b}) can thus be written as
\begin{equation}
i \partial_{t} \ket{\psi} = (H \ket{u}) \odot \ket{v} + 
\ket{u} \odot (H \ket{v}) .
\end{equation}
A comparison with the differential identity
\begin{equation}
\partial_{t} \ket{\psi} = (\partial_{t} \ket{u}) \odot \ket{v} + 
\ket{u} \odot (\partial_{t} \ket{v})
\end{equation}
then shows that both $\ket{u}$ and $\ket{v}$ can satisfy the
Schr\"odinger equation.

But when is it true that $[H,U]$ is a creator? If $H$ conserves particle number,
it can be written as a polynomial (usually quadratic) function of the hopping
operators $a_{i}^{\dag} a_{j}$ (which reduce to number operators $N_i =
a_{i}^{\dag} a_{i}$ when $i = j$). Now the commutator of a
hopping operator and a creation operator is just another creation operator:
\begin{equation}
[ a_{i}^{\dag} a_{j} , a_{k}^{\dag} ] = \delta_{jk} a_{i}^{\dag} .
\label{eq:hop_commutator}
\end{equation}
Assuming that $H$ is \emph{linear} in the hopping operators---as would be the
case for a system of \emph{noninteracting} particles---this shows that the
commutator of $H$ with a creator always generates another creator. The
subsystems $\ket{u}$ and $\ket{v}$ can therefore satisfy the Schr\"odinger
equation in this case.

However, if pairs of particles interact, then $H$ also includes
the pair distribution operator \cite{Merzbacher1998}
\begin{equation}
P_{ij} = N_{i} N_{j} - \delta_{ij} N_{i} ,
\end{equation}
for which
\begin{equation}
[ P_{ij}, a_{k}^{\dag} ] = \delta_{ik} a_{k}^{\dag} N_{j} + 
\delta_{jk} a_{k}^{\dag} N_{i} .
\end{equation}
This contains both creation and annihilation operators. Hence, in a system of
interacting particles, $[H,U]$ is not a creator, and it is impossible in general
for all of $\ket{\psi}$, $\ket{u}$, and $\ket{v}$ to satisfy the Schr\"odinger
equation.


\subsection{Relational time in quantum mechanics}

An extensive literature on the topic of \emph{relational time} in quantum
mechanics also casts serious doubt on whether the time parameter $t$ in Eq.\
(\ref{eq:Schr_psi}) can have any meaning in a closed system (see, e.g., Refs.\
\cite{BarbourBertotti1982, PageWootters1983, Wootters1984, UnruhWald1989,
Pegg1991, Isham1992, Page1994, Barbour1994a, Barbour1994b,
GambiniPortoPullin2004a, GambiniPortoPullin2004b, GambiniPortoPullin2009,
Poulin2006, MilburnPoulin2006, AlbrechtIglesias2008, Arce2012, Moreva2014}). The
argument is very similar to that already given in Sec.\
\ref{sec:relational_properties}. Namely, within a closed system, one can observe
only changes in the relations between various subsystems; one does not have
access to any hypothetical absolute external time variable.

Page and Wootters \cite{PageWootters1983, Wootters1984, Page1994} have argued
that this gives rise to an effective energy superselection rule in which a
coherent superposition of different energy eigenstates is experimentally
indistinguishable from a statistical mixture. Page \cite{Page1994}, Poulin
\cite{Poulin2006}, and Milburn and Poulin \cite{MilburnPoulin2006} have extended
this approach by using group averaging of density operators to eliminate the
external time parameter $t$, thereby reducing a general unmixed state to a
statistical mixture of energy eigenstates.

A common strategy in this type of approach is to identify one subsystem
as a \emph{clock} and measure time via correlations between the clock
subsystem and other subsystems.  This gives rise to an effective decoherence
mechanism if the clock is of finite size \cite{GambiniPortoPullin2004a,
GambiniPortoPullin2004b, Poulin2006, MilburnPoulin2006, 
BartlettRudolphSpekkens2007}.

Barbour \cite{Barbour1994a} has argued that such approaches do not agree with
how time is defined operationally. In practice, we define time not by looking at
a single clock, but by using the time parameter $t$ to achieve the best fit to
all of the experimental information at our disposal. This \emph{ephemeris time}
concept was developed long ago by astronomers, but even today it is how time is
defined from a network of atomic clocks, all of which operate in different
environments and run at slightly different rates. From this perspective,
``ultimately the universe is the only clock'' \cite{Barbour1994a}.

Time is defined in the present paper by using the concept of ephemeris time in
the context of the geometric approach to quantum mechanics developed by Anandan
and Aharonov \cite{AnandanAharonov1990, Anandan1991}. These authors have
alluded to this concept themselves, even going so far as to say that ``The
parameter $t$ represents correlation between the Fubini--Study distances
determined by different clocks'' \cite{AnandanAharonov1990}.

However, the correlation between different clocks is found only in
these words, not in any of their equations. Their equations establish only a
relationship between external time and the Fubini--Study distance traveled by a
system evolving according to Schr\"odinger's equation \cite{AnandanAharonov1990,
Anandan1991}. During a ``measurement,'' however, the system can move a finite
distance in the projective Hilbert space during a time interval of zero
\cite{AnandanAharonov1988}. It is not clear how these disparate Hilbert-space
transport mechanisms are to be reconciled.  But this is of course just the 
old conundrum posed by von Neumann's axioms of time
evolution \cite{vonNeumann1955}.

This paper implements Anandan and Aharonov's idea mathematically by introducing
a \emph{time functional} that is optimized to achieve the best fit between
Schr\"odinger dynamics and the changes that occur in all subsystems. The actual
value of these changes is not determined by this functional; that task is left to
the principle of dynamical stability, to be discussed below in Sec.\
\ref{sec:dynamical_stability}.

\subsection{Definition of the time functional}

\label{sec:time_functional_defn}

The time functional can be defined using a simple
extension of the geometric concepts introduced previously in 
Sec.\ \ref{sec:subsystem_geometry}.  Consider two slightly different
subsystem decompositions, $\rho$ and $\rho'$.  If $\rho'$ differs from
$\rho$ only by a Schr\"odinger time evolution, the two decompositions
must be related by
\begin{equation}
\rho' \overset{?}{=} e^{-i \hat{H} \Delta \tau} \rho 
e^{i \hat{H} \Delta \tau} \label{eq:Schr_ideal}
\end{equation}
for some time interval $\Delta \tau$, in which
\begin{equation}
\hat{H} \equiv \bigoplus_{k=1}^{m} H \label{eq:H_hat}
\end{equation}
is the Hamiltonian in the direct-sum formalism. Of course, for arbitrary $\rho$
and $\rho'$, Eq.\ (\ref{eq:Schr_ideal}) will not be true, but we can try to get
as close as possible to such a description by minimizing the Hilbert--Schmidt
distance between the two sides of the equation. In other words, we can define a
function
\begin{subequations} \label{eq:lambda_defn}
\begin{align}
\lambda (\Delta \tau) & \equiv D^2 (e^{-i \hat{H} \Delta \tau} \rho 
e^{i \hat{H} \Delta \tau}, \rho') \\ & = 
D^2 (\rho, e^{i \hat{H} \Delta \tau} \rho' e^{-i \hat{H} \Delta \tau})
\end{align}
\end{subequations}
and seek the value of $\Delta \tau$ that minimizes this function.  This
special value, denoted $\Delta \tau = \Delta t$, provides the 
best fit between $\rho$ and $\rho'$ that can be expressed in
the language of Schr\"odinger dynamics.

Our only concern is the case of infinitesimal differences $\norm{\Delta u_k}$,
for which $\Delta t$ is also infinitesimal. We can therefore use the
Fubini--Study metric of Eqs.\ (\ref{eq:D2_rho_FS}) and
(\ref{eq:D2_rho_FS_simple}) to write
\begin{subequations}
\begin{align}
\lambda (\Delta \tau) & = \matelm{u'}{e^{-i \hat{H} \Delta \tau} (1 - \rho) 
e^{i \hat{H} \Delta \tau}}{u'} \\ & = \sum_{k=1}^{m}
\frac{\matelm{u_{k}'}{e^{-i H \Delta \tau} (1 - \rho_{k}) 
e^{i H \Delta \tau}}{u_{k}'}}{\inprod{u_{k}}{u_{k}}} .
\end{align}
\end{subequations}
Here the exponentials can be expanded in the usual way:
\begin{multline}
e^{-i \hat{H} \Delta \tau} (1 - \rho) 
e^{i \hat{H} \Delta \tau} = (1 - \rho) -i \Delta \tau [\hat{H}, 1-\rho] \\
- \tfrac12 \Delta \tau^2 [\hat{H}, [\hat{H}, 1 - \rho ]] + \cdots .
\end{multline}
If we treat $\Delta \tau$ and $\norm{\Delta u}$ as of the same
order and work to second order overall, the final result can 
be written as
\begin{equation}
\lambda (\Delta \tau) = \eta - 2 \Delta \tau \imag \inprod{\Delta u}{H} 
+ \Delta \tau^2 \inprod{H}{H} , \label{eq:lambda_quad1}
\end{equation}
in which $\lambda (0) = \eta$ was already defined in Eq.\
(\ref{eq:D2_rho_FS_simple}). The vector $\ket{H}$ in this expression is defined
as
\begin{subequations} \label{eq:H_ket_defn}
\begin{align}
\ket{H} & \equiv (1 - \rho) \hat{H} \ket{u} \\
& = \bigoplus_{k=1}^{m} (1 - \rho_k) H \ket{u_k} .
\end{align}
\end{subequations}
The minimum of the quadratic function (\ref{eq:lambda_quad1}) occurs at $\Delta
\tau = \Delta t$, in which
\begin{equation}
\Delta t = \frac{\imag \inprod{\Delta u}{H}}{\inprod{H}{H}} .
\label{eq:Delta_t}
\end{equation}
This is the desired expression giving the optimal time interval $\Delta t$ as a
functional of the subsystem change $\ket{\Delta u}$.

\subsection{Properties of the time functional}

\label{sec:properties_time_functional}

Note that the solution (\ref{eq:Delta_t}) can be used to rewrite 
Eq.\ (\ref{eq:lambda_quad1}) as
\begin{equation}
\lambda (\Delta \tau) = \eta + \Delta \tau (\Delta \tau - 2 \Delta t)
\inprod{H}{H} . \label{eq:lambda_quad2}
\end{equation}
When $\Delta \tau = \Delta t$, this function attains its
minimum value
\begin{subequations} \label{eq:lambda_min}
\begin{align}
\lambda (\Delta t) & = \eta - \Delta t^2  
\inprod{H}{H} \label{eq:lambda_min_a} \\
& = \eta - \frac{(\imag \inprod{\Delta u}{H})^{2}}{\inprod{H}{H}} .
\label{eq:lambda_min_b} 
\end{align}
\end{subequations}
Given the definition (\ref{eq:lambda_defn}) of $\lambda (\Delta \tau)$ as the
square of a distance, it seems obvious that this minimum value must satisfy
$\lambda (\Delta t) \ge 0$. However, it is not immediately clear from Eq.\
(\ref{eq:lambda_min}) that this is in fact the case.

To see explicitly that $\lambda (\Delta t)$ is indeed nonnegative, note that
\begin{equation}
(\imag \inprod{\Delta u}{H})^{2} \le \abs{\inprod{\Delta u}{H}}^{2} .
\label{eq:imag_inequality}
\end{equation}
This inequality in conjunction with Eq.\ (\ref{eq:lambda_min_b}) implies that
\begin{subequations}
\begin{align}
\lambda (\Delta t) & \ge \eta - 
\frac{\abs{\inprod{\Delta u}{H}}^{2}}{\inprod{H}{H}} \\ & =
\matelm{\Delta u}{(1 - \rho - \Pi_{H})}{\Delta u} ,
\end{align}
\end{subequations}
in which $\Pi_{H}$ is the projector
\begin{equation}
\Pi_{H} \equiv \frac{\outprod{H}{H}}{\inprod{H}{H}} .
\end{equation}
Because $\rho$ and $\Pi_{H}$ are orthogonal, the operator $(1 - \rho - \Pi_{H})$
is also a projector. This can be used to write
\begin{equation}
\matelm{\Delta u}{(1 - \rho - \Pi_{H})}{\Delta u} = \inprod{w}{w} \ge 0 ,
\end{equation}
in which
\begin{equation}
\ket{w} \equiv (1 - \rho - \Pi_{H}) \ket{\Delta u} .
\end{equation}
This proves that $\lambda (\Delta t) \ge 0$, and furthermore that a necessary
condition for $\lambda (\Delta t) = 0$ is $\ket{w} = 0$ or
\begin{equation}
(1 - \rho) \ket{\Delta u} = \Pi_{H} \ket{\Delta u} .
\end{equation}
However, this condition is not sufficient. Tracing back to the previous
inequality (\ref{eq:imag_inequality}), we see that $\real
\inprod{\Delta u}{H} = 0$ is also required.  Hence, in order to achieve
$\lambda (\Delta t) = 0$, it is necessary and sufficient that
\begin{equation}
(1 - \rho) \ket{\Delta u} = i C \ket{H} , \label{eq:minimum_condition}
\end{equation}
in which $C$ is a real constant. In other words, the component of $\ket{\Delta
u}$ that is orthogonal to $\ket{u}$ must be proportional to $\ket{H}$, with an
imaginary coefficient.

What is the significance of this? According to the definition
(\ref{eq:H_ket_defn}), the vector $\ket{H}$ is just the component of $\hat{H}
\ket{u}$ that is orthogonal to $\ket{u}$. Hence, the condition
(\ref{eq:minimum_condition}) says that in order to achieve $\lambda (\Delta t) =
0$, all subsystems must satisfy the Schr\"odinger equation, but only insofar as
the component of $\ket{\Delta u}$ orthogonal to $\ket{u}$ is concerned. [The
component of $\ket{\Delta u}$ that is parallel to $\ket{u}$ does not contribute
to the distance (\ref{eq:D2_rho_FS_simple}).]

Another way of expressing $\ket{H}$ is
\begin{equation}
\ket{H} = \bigoplus_{k=1}^{m} (H - \expect{H}_k) \ket{u_k} ,
\end{equation}
in which $\expect{H}_k$ is the mean value [cf.\ Eq.\
(\ref{eq:mean_value})]
\begin{equation}
\expect{H}_k \equiv \frac{\matelm{u_k}{H}{u_k}}{\inprod{u_k}{u_k}} .
\end{equation}
Hence, the inner product $\inprod{H}{H}$ can be written as
\begin{equation}
\inprod{H}{H} = \sum_{k=1}^{m} 
\frac{\matelm{u_k}{(H - \expect{H}_k)^{2}}{u_k}}{\inprod{u_k}{u_k}} .
\end{equation}
This provides a simple physical interpretation of $\inprod{H}{H}$: it is the
combined energy variance of all subsystems.  The corresponding standard
deviation is denoted $\Delta E \equiv \sqrt{\inprod{H}{H}}$.

The inequality $\lambda (\Delta t) \ge 0$ can thus be written in the alternative
form [cf.\ Eq.\ (\ref{eq:lambda_min_a})]
\begin{equation}
\Delta E^2 \Delta t^2 \le \eta .  \label{eq:time_energy}
\end{equation}
This looks vaguely like a time--energy uncertainty relation, except that the
inequality is pointing in the wrong direction---so actually it is nothing of the
kind. It simply says that, for a given squared distance $\eta = D^2 (\rho,
\rho')$, there is an upper bound on the optimal value of $\Delta \tau$ that can
be fitted to $\rho$ and $\rho'$ using Eq.\ (\ref{eq:Schr_ideal}). Furthermore,
because the optimal value $\Delta \tau = \Delta t$ is intimately related to
Schr\"odinger dynamics, the numerical value of $\Delta t$ depends on the energy
scale $\Delta E$ determined by the subsystem decomposition $\rho$.

To conclude this section, we may note that the time functional
(\ref{eq:Delta_t}) offers a very simple way of implementing the idea that
``ultimately the universe is the only clock.'' But of course, as mentioned
previously, the definition of such a clock tells us nothing about how the
subsystems evolve in time. Finding a way to define this time evolution is the
subject of the next section.

\section{Dynamical stability of subsystems}

\label{sec:dynamical_stability}


The most obvious criterion for defining subsystem dynamics is to maximize the
stability of the subsystem decomposition. In other words, we should choose the
dynamics such that the decomposition ``hops about the least'' \cite{[] [{, p.\
518.}] Fuchs2011}. This concept of \emph{dynamical stability} has a long
history. In the early days of quantum mechanics, Schr\"odinger used it in an
attempt to interpret particles as stable wave packets \cite{[] [{; English
translation in Ref.\ \cite{Schrodinger1982}.}] Schrodinger1926c, [] [{, pp.\
41--44.}] Schrodinger1982}. More recently, its importance for decoherence theory
has been repeatedly emphasized by Zeh \cite{Zeh1970, Zeh1971b, KublerZeh1973,
Zeh1973, Zeh1979, JoosZeh1985, Zeh2000, Zeh2003ch2, Zeh2006}, and the basic idea
has been developed extensively by Zurek under such names as the predictability
sieve \cite{Zurek1993a, ZurekHabibPaz1993}, einselection \cite{Zurek1982,
Zurek1998, Zurek2003}, the existential interpretation \cite{Zurek1993a,
Zurek1998, Zurek2003}, and quantum Darwinism \cite{Zurek2003, Zurek2014}.

In this section, the concept of dynamical stability is defined for the subsystem
decomposition (\ref{eq:psi_pi}) in terms of a dynamical stability functional
$\chi$. The time evolution of the subsystems is then determined by maximizing
$\chi$. For simplicity, the total system state $\ket{\psi}$ is initially assumed
to be independent of time. This analysis is then extended to the case of
time-dependent $\ket{\psi}$.

\subsection{Dynamical stability functional}

\label{sec:dynamical_stability_functional}

To simplify the description of dynamical stability, it is convenient to
introduce the dimensionless variable
\begin{equation}
\sigma \equiv \Delta E \Delta t = 
\frac{\imag \inprod{\Delta u}{H}}{\Delta E} . \label{eq:sigma_defn}
\end{equation}
The dynamical stability functional $\chi$ is then defined as
\begin{equation}
\chi \equiv \frac{\sigma^2}{\eta} = 
\frac{\Delta E^2 \Delta t^2}{D^2 (\rho, \rho')} \qquad (\eta \ne 0) .
\label{eq:chi_definition}
\end{equation}
The foundation for this definition is the inequality (\ref{eq:time_energy}),
which says that $0 \le \chi \le 1$. The \emph{principle of dynamical stability}
is implemented by holding $\rho$ fixed and varying $\rho'$ so as to maximize the
value of $\chi$. The decompositions $\rho$ and $\rho'$ could thus be regarded as
``initial'' and ``final,'' although this has the potential to be
misleading because it has nothing to do with the sign of $\Delta t$.

Maximizing $\chi$ with respect to variations in $\rho'$ simply requires that the
subsystems change as little as possible (as measured by the Fubini--Study
metric) in a given infinitesimal time interval $\Delta t$. According to the
results of Sec.\ \ref{sec:interacting_not_Schroedinger}, if $\ket{\psi}$ is
assumed to satisfy the Schr\"odinger equation, the upper limit $\chi = 1$ is
generally unattainable in a system of interacting particles.

Suppose now that $\rho'$ is varied by a small amount $\delta \rho$. (This should
perhaps be written as $\delta \rho'$, but the prime symbol can be omitted
because $\rho$ itself is not varied.) This will give rise to corresponding
variations $\delta \sigma$, $\delta \eta$, and $\delta \chi$, which are related
by
\begin{equation}
\delta \chi = \frac{(\sigma + \delta \sigma)^2}{\eta + \delta \eta} - 
\frac{\sigma^2}{\eta} = \frac{2 \sigma \delta \sigma - 
\chi \delta \eta + \delta \sigma^2}{\eta + \delta \eta} .
\end{equation}
To first order in small quantities, this reduces to
\begin{equation}
\delta \chi = \eta^{-1} (2 \sigma \delta \sigma - 
\chi \delta \eta) ,
\end{equation}
in which all variations are evaluated to first order in $\delta \rho$. The
stationary states of the dynamical stability functional are then given by 
$\delta \chi = 0$ or
\begin{equation}
\frac{\delta \eta}{\eta} = 2 \frac{\delta \sigma}{\sigma} 
\qquad (\sigma \ne 0) , \label{eq:eta_sigma_stationary}
\end{equation}
in which $\sigma \ne 0$ can always be assumed because we have
no interest in the minima of $\chi$.

\subsection{Time-independent \texorpdfstring{$\Psi$}{Psi}}

\label{sec:time_independent_psi}

Let us now apply the principle of dynamical stability to the special case in
which the total system state $\Psi$ is assumed to be independent of time, so
that $\Delta \Psi \equiv \Psi' - \Psi = 0$. This implies that $V_{\Delta \Psi} =
0$ in the general expression (\ref{eq:Delta_V}) for $\Delta V \equiv \Delta
U_1$, which simplifies the analysis considerably.

The quantities to be varied are the ``final'' subsystem exponents $\ket{x_k'} =
X_k' \ket{0}$ for the quasiclassical subsystems ($k \ne 1$). As before, it is
convenient to use a direct-sum representation for the differences $\ket{\Delta
x_k} = \ket{x_k'} - \ket{x_k}$:
\begin{equation}
\ket{\Delta x} = \bigoplus_{k=2}^{m} \ket{\Delta x_k} . \label{eq:Delta_x_sum}
\end{equation}
In contrast to the definition of $\ket{\Delta u}$ [see Eq.\
(\ref{eq:direct_sum_ket})], the value $k = 1$ is \emph{not} included in the
definition of $\ket{\Delta x}$. There are two reasons for this. First, as noted
in Sec.\ \ref{sec:subsystem_differences}, only the subsystems with $k \ne 1$ are
treated as independently variable; the subsystem $k = 1$ is entirely determined
by $\Psi$ and the other subsystems. Second, it is not generally even possible to
define $X_1 = \ln U_1$, since $\Psi$ need not be quasiclassical.

The dimensionless time interval (\ref{eq:sigma_defn}) can now be expressed in
terms of the \emph{independent} variables $\ket{\Delta x}$ as
\begin{equation}
\sigma = \frac{\imag \inprod{\Delta u}{H}}{\Delta E} \equiv 
\imag \inprod{\Delta x}{\sigma} , \label{eq:sigma_def}
\end{equation}
in which the components of the vector $\ket{\sigma}$ are given
by [cf.\ Eqs.\ (\ref{eq:DukDxk}), (\ref{eq:DvDxk}), (\ref{eq:H_ket_defn})]
\begin{multline}
\inprod{e_{ki}}{\sigma} = \frac{1}{\Delta E} \biggl[ 
\frac{\matelm{f_{ki}}{(1 - \rho_k)H}{u_k}}{\inprod{u_k}{u_k}} \\ +
\frac{\matelm{g_{ki}}{(1 - \rho_v)H}{v}}{\inprod{v}{v}}
\biggr] . \label{eq:sigma_ket}
\end{multline}
In this expression, $\rho_v \equiv \rho_1$ and $\ket{v} \equiv \ket{u_1}$.
The squared distance $\eta$ can be written likewise as
\begin{equation}
\eta = \matelm{\Delta x}{\hat{\eta}}{\Delta x} ,
\label{eq:eta_operator}
\end{equation}
in which the matrix elements of the operator $\hat{\eta}$ are
given by [cf.\ Eq.\ (\ref{eq:D2_rho_FS_simple})]
\begin{multline}
\matelm{e_{ki}}{\hat{\eta}}{e_{k'i'}} = \delta_{kk'} 
\frac{\matelm{f_{ki}}{(1 - \rho_k)}{f_{ki'}}}{\inprod{u_k}{u_k}} \\ +
\frac{\matelm{g_{ki}}{(1 - \rho_v)}{g_{k'i'}}}{\inprod{v}{v}} .
\label{eq:eta_hat}
\end{multline}
The operator $\hat{\eta}$ is clearly positive ($\hat{\eta} \ge 0$), and it can
be made positive definite ($\hat{\eta} > 0$) if we agree to exclude the vacuum
state \footnote{The vacuum state can be excluded from the subsystem exponents
$\ket{x_k}$ because its inclusion has no effect other than to change the
normalization of $\ket{u_k}$. If this choice is made, one can readily verify
using Eqs.\ (\ref{eq:fki}) and (\ref{eq:switch_basis}) that $\hat{\eta} > 0$.}
from the orthonormal basis $\{ \ket{e_{ki}} \}$ used to define the subsystem
exponents in Eq.\ (\ref{eq:xk_cki}).

If we now vary $\ket{x'}$ by $\ket{\delta x}$ (holding $\ket{x}$ fixed),
$\ket{\Delta x}$ also varies by $\ket{\delta x}$. The resulting first-order
variations in $\sigma$ and $\eta$ are
\begin{subequations}
\begin{align}
\delta \sigma & = \imag \inprod{\delta x}{\sigma} = 
\frac{\inprod{\delta x}{\sigma} - \inprod{\sigma}{\delta x}}{2i} , \\
\delta \eta & = \matelm{\delta x}{\hat{\eta}}{\Delta x} + 
\matelm{\Delta x}{\hat{\eta}}{\delta x} .
\end{align}
\end{subequations}
Substituting these expressions into the stationary-state condition
(\ref{eq:eta_sigma_stationary}) gives
\begin{equation}
\matelm{\delta x}{\hat{\eta}}{\Delta x} + \matelm{\Delta x}{\hat{\eta}}{\delta
x} = -i C (\inprod{\delta x}{\sigma} - \inprod{\sigma}{\delta x}) ,
\end{equation}
in which $C \equiv \eta / \sigma$ is real.  Because the variation
$\ket{\delta x}$ is arbitrary, we can partition this equation
in the usual way \cite[pp.\ 764--765]{Messiah1962} to obtain
\begin{equation}
\matelm{\delta x}{\hat{\eta}}{\Delta x} = -i 
C \inprod{\delta x}{\sigma} , \label{eq:eta_sigma_linear}
\end{equation}
together with its complex conjugate. Removing the arbitrary vector $\bra{\delta
x}$ gives the linear algebraic equation
\begin{equation}
\hat{\eta} \ket{\Delta x} = -i C \ket{\sigma} ,
\end{equation}
in which $\hat{\eta}$ is positive definite and therefore invertible.
All stationary states of the dynamical stability functional $\chi$ 
with $\chi > 0$ are thus given explicitly by
\begin{equation}
\ket{\Delta x} = -i C \hat{\eta}^{-1} \ket{\sigma} .
\label{eq:Dx_soln}
\end{equation}

Upon substituting this result back into the definitions of $\sigma$, $\eta$, and
$\chi$, we find
\begin{subequations}
\begin{align}
\sigma & = C \matelm{\sigma}{\hat{\eta}^{-1}}{\sigma} , \\
\eta & = C^2 \matelm{\sigma}{\hat{\eta}^{-1}}{\sigma} , \\
\chi & = \matelm{\sigma}{\hat{\eta}^{-1}}{\sigma} . \label{eq:chi_solution}
\end{align}
\end{subequations}
The only degree of freedom in the solution (\ref{eq:Dx_soln}) is the value of
the real constant $C = \eta / \sigma = \sigma / \chi$. Since $\chi$ is
independent of $C$, one can find the value of $C$ from a given time interval
$\Delta t$ simply by calculating $C = \Delta E \Delta t / \chi$.
The sign of $\Delta t$ can be positive or negative, but its magnitude
should always be chosen small enough that $\eta \ll 1$ (or else
the approximations used in deriving the basic equations are
no longer valid).

Thus, for a given sign and magnitude of $\Delta t$, there is only \emph{one}
stationary state of the dynamical stability functional with $\chi > 0$. This
strongly suggests that this stationary state is the unique global maximum of
$\chi$. A proof of this conjecture is given in Appendix \ref{app:chi_maximum}.

The solution (\ref{eq:Dx_soln}) can be viewed as a differential equation
for $\ket{x}$, since
\begin{equation}
\frac{\partial \ket{x}}{\partial t} = \lim_{\Delta t \to 0}
\frac{\ket{\Delta x}}{\Delta t} = -i \Delta E \frac{\hat{\eta}^{-1} \ket{\sigma}}{\matelm{\sigma}{\hat{\eta}^{-1}}{\sigma}} , \label{eq:dxdt}
\end{equation}
in which the limit $\Delta t \to 0$ is somewhat redundant because it has been
assumed throughout the derivation. This equation can be integrated to obtain
$\ket{x}$ as a function of $t$; in practice, this is done by using Eq.\
(\ref{eq:Dx_soln}) repeatedly for small but finite intervals $\Delta t$.
Numerical calculations on simple models (see Sec.\ \ref{sec:model_calculations})
show that the change in $\ket{x}$ over a fixed time interval $T$ does indeed
converge (quadratically) to a definite value in the limit $\Delta t \to 0$.

The time evolution generated by Eq.\ (\ref{eq:dxdt}) is deterministic. That is,
the final subsystem decomposition is uniquely determined by the initial one, and
if the subsystems are propagated forward and backward over a finite interval
(by changing the sign of $\Delta t$ at the far end), the returning solution
always converges to its initial value. Hence, even though the differential
equation (\ref{eq:dxdt}) is nonlinear, it does not exhibit any of the lack of
determinism so characteristic of standard quantum mechanics.

\subsection{Time-dependent \texorpdfstring{$\Psi$}{Psi}}

\label{sec:time_dependent_psi}

The next step is to lift the restriction $\Delta \Psi = 0$ that was imposed in
Sec.\ \ref{sec:time_independent_psi}. Because the total system is closed,
$\ket{\Delta \psi}$ is assumed to follow the Schr\"odinger equation (to
first order in the time functional $\Delta t$):
\begin{equation}
\ket{\Delta \psi} = -i \Delta t H \ket{\psi} = -i \Delta t H \Psi \ket{0} .
\label{eq:Delta_psi_Schr}
\end{equation}
This does not imply that $\Delta \Psi = -i \Delta t H \Psi$, because $H
\Psi$ is not a creator.  Rather, we have
\begin{equation}
\Delta \Psi = -i \Delta t (H \cdot \Psi) , \label{eq:Delta_Psi_Schr}
\end{equation}
in which $H \cdot \Psi$ denotes the creator defined by $(H \cdot \Psi) \ket{0}
\equiv H \ket{\psi}$. Upon substituting this result into Eq.\ (\ref{eq:DvDxk}),
we obtain
\begin{equation}
\ket{\Delta v} = -i \Delta t \ket{v_{H \cdot \Psi}} + \sum_{k=2}^{m} 
\biggl( \sum_{i} \outprod{g_{ki}}{e_{ki}} \biggr) \ket{\Delta x_k} .
\label{eq:Delta_v_mod}
\end{equation}
The dimensionless interval $\sigma$ then takes the form
\begin{align}
\sigma & = \Delta E \Delta t = 
\frac{\imag \inprod{\Delta u}{H}}{\Delta E} \nonumber \\ & = 
\imag \inprod{\Delta x}{\sigma_0} + \frac{\Delta t}{\Delta E} \real \biggl[
\frac{\matelm{v_{H \cdot \Psi}}{(1 - \rho_v)H}{v}}{\inprod{v}{v}} \biggr] ,
\end{align}
in which $\ket{\sigma_0}$ relabels the vector introduced previously in Eq.\
(\ref{eq:sigma_ket}).  Noting that $\Delta t$ now appears on both
sides of the equation, we can combine these terms to obtain
\begin{equation}
\omega \Delta E \Delta t = \imag \inprod{\Delta x}{\sigma_0} ,
\end{equation}
in which $\omega$ is the real constant
\begin{equation}
\omega \equiv 1 - \frac{1}{\Delta E^2} \real \biggl[
\frac{\matelm{v_{H \cdot \Psi}}{(1 - \rho_v)H}{v}}{\inprod{v}{v}} \biggr] .
\label{eq:omega_E}
\end{equation}
At this point it is convenient to redefine the vector $\ket{\sigma}$ so as to
obtain the same outward appearance as Eq.\ (\ref{eq:sigma_def}):
\begin{equation}
\sigma = \Delta E \Delta t = \frac{\imag \inprod{\Delta x}{\sigma_0}}{\omega}
\equiv \imag \inprod{\Delta x}{\sigma} , \label{eq:sigma_redef}
\end{equation}
in which $\ket{\sigma}$ is just a renormalized version of Eq.\ 
(\ref{eq:sigma_ket}):
\begin{multline}
\inprod{e_{ki}}{\sigma} = \frac{1}{\omega \Delta E} \biggl[ 
\frac{\matelm{f_{ki}}{(1 - \rho_k)H}{u_k}}{\inprod{u_k}{u_k}} \\ +
\frac{\matelm{g_{ki}}{(1 - \rho_v)H}{v}}{\inprod{v}{v}}
\biggr] . \label{eq:sigma_ket_redef}
\end{multline}
Hence, the time evolution of $\Psi$ affects the variable $\sigma$ only 
through the renormalization factor $\omega$.

However, its effect on $\eta$ is more profound. Because $\ket{\Delta v}$ is
now linear in $\Delta t$, $\eta$ in Eq.\
(\ref{eq:D2_rho_FS_simple}) becomes a quadratic function of $\Delta t$. When
expressed in terms of $\sigma$, this quadratic dependence
takes the form
\begin{equation}
\eta (\Delta t) = \eta_0 + 2 \beta \sigma + \kappa \sigma^2 ,
\label{eq:eta_Delta_t}
\end{equation}
in which 
$\eta (\Delta t) = D^2 (\rho, \rho') = \matelm{\Delta u}{(1 - \rho)}{\Delta u}$
is the function defined in Eq.\ (\ref{eq:D2_rho_FS_simple}) and $\eta_0 \equiv
\eta(0) = \matelm{\Delta x}{\hat{\eta}}{\Delta x}$ relabels the quantity defined
earlier in Eq.\ (\ref{eq:eta_operator}). Although $\eta$ now depends on $\Delta
t$, one should note carefully that $\eta (\Delta t) \ne \lambda (\Delta t)$ [see
Eq.\ (\ref{eq:lambda_min})].

The new functional $\beta$ in Eq.\ (\ref{eq:eta_Delta_t}) is defined by
\begin{equation}
\beta = \imag \inprod{\Delta x}{\beta} ,
\end{equation}
in which $\ket{\beta}$ is the vector
\begin{equation}
\inprod{e_{ki}}{\beta} = \frac{1}{\Delta E}
\frac{\matelm{g_{ki}}{(1 - \rho_v)}{v_{H \cdot \Psi}}}{\inprod{v}{v}} .
\label{eq:beta_ket_E}
\end{equation}
The dimensionless constant $\kappa$ in Eq.\ (\ref{eq:eta_Delta_t}) is 
defined by
\begin{equation}
\kappa = \frac{1}{\Delta E^2} \frac{\matelm{v_{H \cdot \Psi}}{(1 - \rho_v)}{v_{H
\cdot \Psi}}}{\inprod{v}{v}} . \label{eq:kappa_E}
\end{equation}
With these results, we can now construct the dynamical stability functional
$\chi = \sigma^2 / \eta$ just as before, in which $\eta \equiv \eta (\Delta t)$.

\subsection{Real matrix representation}

\label{sec:real_matrix}

There are several ways to solve the variation problem for $\chi$. The method
used here has the advantage of requiring very little modification when the
definition of distance is changed below in Sec.\ \ref{sec:superselection}.

When $\eta$ in Eq.\ (\ref{eq:eta_Delta_t})
is expressed as a function of the expansion coefficients
$\Delta c_{ki} = \inprod{e_{ki}}{\Delta x}$ introduced in Eq.\ (\ref{eq:Dxk}),
the result can be written as
\begin{equation}
\eta = \mu + \nu ,
\end{equation}
in which $\mu$ has the same form as $\eta_0 = \matelm{\Delta
x}{\hat{\eta}}{\Delta x}$:
\begin{equation}
\mu = \sum_{ki} \sum_{k' i'} \Delta c_{ki}^{*} \Delta c_{k'i'}
\mu_{ki,k'i'} . \label{mu_def}
\end{equation}
However, $\nu$ is qualitatively different:
\begin{equation}
\nu = \real \left( \sum_{ki} \sum_{k' i'} \Delta c_{ki}^{*} \Delta c_{k'i'}^{*}
\nu_{ki,k'i'} \right) . \label{nu_def}
\end{equation}
Here the matrix elements $\mu_{ki,k'i'}$ and $\nu_{ki,k'i'}$ are given by
\begin{multline}
\mu_{ki,k'i'} = \delta_{kk'} 
\frac{\matelm{f_{ki}}{(1 - \rho_k)}{f_{ki'}}}{\inprod{u_k}{u_k}} +
\frac{\matelm{g_{ki}}{(1 - \rho_v)}{g_{k'i'}}}{\inprod{v}{v}} \\ +
\frac{1}{2} (\beta_{ki} \sigma_{k'i'}^{*} + \sigma_{ki} \beta_{k'i'}^{*}
+ \kappa \sigma_{ki} \sigma_{k'i'}^{*}) \label{eq:mu_matrix}
\end{multline}
and
\begin{equation}
\nu_{ki,k'i'} = - \frac{1}{2} (\beta_{ki} \sigma_{k'i'} + \sigma_{ki} \beta_{k'i'}
+ \kappa \sigma_{ki} \sigma_{k'i'}) , \label{eq:nu_matrix}
\end{equation}
in which $\beta_{ki} = \inprod{e_{ki}}{\beta}$ and $\sigma_{ki} =
\inprod{e_{ki}}{\sigma}$. Because $\nu$, unlike $\mu$ and $\eta_0$, is not a
sesquilinear form, the stationary-state equation (\ref{eq:eta_sigma_stationary})
no longer reduces to a linear algebraic equation for the complex coefficients
$\Delta c_{ki}$.

However, one can put Eq.\ (\ref{eq:eta_sigma_stationary}) into the 
form of a linear algebraic equation simply by separating
\begin{equation}
\Delta c_{ki} = \Delta c_{ki}' + i \Delta c_{ki}'' ,
\end{equation}
and working with the real variables $\Delta c_{ki}'$
and $\Delta c_{ki}''$.  If the real and imaginary parts
of $\mu_{ki,k'i'}$ and $\nu_{ki,k'i'}$ are likewise separated, 
$\eta$ can be expressed in block matrix notation as
\begin{equation}
\eta = 
\begin{pmatrix}
\Delta c' & \Delta c''
\end{pmatrix}
\begin{pmatrix}
\mu' + \nu' & -\mu'' + \nu'' \\
\mu'' + \nu'' & \mu' - \nu'
\end{pmatrix}
\begin{pmatrix}
\Delta c' \\
\Delta c''
\end{pmatrix} ,
\end{equation}
in which all matrix elements are real.  

It is convenient to write this equation in a quasi-Dirac notation:
\begin{equation}
\eta = \pmatelm{\Delta x}{\tilde{\eta}}{\Delta x} ,
\end{equation}
in which the rounded ket vector is represented by the
real column matrix
\begin{equation}
\pket{\Delta x} =
\begin{pmatrix}
\Delta c' \\
\Delta c''
\end{pmatrix} ,
\end{equation}
and the operator $\tilde{\eta}$ is represented by the
real symmetric matrix
\begin{equation}
\tilde{\eta} =
\begin{pmatrix}
\mu' + \nu' & -\mu'' + \nu'' \\
\mu'' + \nu'' & \mu' - \nu'
\end{pmatrix} . \label{eq:eta_tilde}
\end{equation}
This matrix is symmetric because $\mu'$, $\nu'$, and $\nu''$ are 
symmetric, whereas $\mu''$ is antisymmetric.

A similar representation can be introduced for the dimensionless
time interval
\begin{equation}
\sigma = \imag \inprod{\Delta x}{\sigma} = 
\imag \left( \sum_{ki} \Delta c_{ki}^{*} \sigma_{ki} \right) ,
\end{equation}
if we separate $\sigma_{ki} = \sigma_{ki}' + i \sigma_{ki}''$ just
as for $\Delta c_{ki}$.  This is written in quasi-Dirac notation as
\begin{equation}
\sigma = \pinprod{\Delta x}{\sigma} = \pinprod{\sigma}{\Delta x} ,
\end{equation}
in which the matrix representation for $\pket{\sigma}$ is
\begin{equation}
\pket{\sigma} =
\begin{pmatrix}
\sigma'' \\
-\sigma'
\end{pmatrix} . \label{eq:pket_sigma}
\end{equation}

\subsection{Dynamically stable subsystem changes}

\label{sec:solve_dynamical_stability}

The dynamical stability functional $\chi$ now has a form very similar
to that found in Sec.\ \ref{sec:time_independent_psi}:
\begin{equation}
\chi = \frac{\sigma^2}{\eta} = 
\frac{\pinprod{\Delta x}{\sigma}^2}{\pmatelm{\Delta x}{\tilde{\eta}}{\Delta x}} .
\end{equation}
When $\pket{\Delta x}$ is varied by $\pket{\delta x}$, the stationary states are
determined by $\delta \chi = 0$ or [cf.\ Eq.\ (\ref{eq:eta_sigma_linear})]
\begin{equation}
\pmatelm{\delta x}{\tilde{\eta}}{\Delta x} = C \pinprod{\delta x}{\sigma} ,
\label{eq:eta_sigma_linear_real}
\end{equation}
in which $C \equiv \sigma / \chi = \eta / \sigma$.  Because $\pket{\delta x}$
can range over the whole vector space, this is equivalent to
\begin{equation}
\tilde{\eta} \pket{\Delta x} = C \pket{\sigma} , 
\label{eq:eta_sigma_linear_real_ket}
\end{equation}
which can be solved as before to obtain the dynamically stable subsystem change
\begin{equation}
\pket{\Delta x} = C \tilde{\eta}^{-1} \pket{\sigma} .
\label{eq:Delta_x_real}
\end{equation}
Note that the factor of $-i$ in Eq.\ (\ref{eq:Dx_soln}) is absent here
because it is embedded into the definition (\ref{eq:pket_sigma}) of
$\pket{\sigma}$.

The qualitative properties of this solution are again very similar to the
solution (\ref{eq:Dx_soln}) found in Sec.\ \ref{sec:time_independent_psi}. In
particular, the time evolution generated by Eq.\ (\ref{eq:Delta_x_real}) remains
deterministic.

If $\Psi$ happens to be an energy eigenstate with energy $E$, we have $H \cdot
\Psi = E \Psi$ and thus
\begin{equation}
\ket{v_{H \cdot \Psi}} = E \ket{v_{\Psi}} = E \ket{v} .
\end{equation}
In this case
\begin{equation}
(1 - \rho_v) \ket{v_{H \cdot \Psi}} = E (1 - \rho_v) \ket{v} = 0 ,
\end{equation}
which implies that $\omega = 1$, $\ket{\beta} = 0$, and $\kappa = 0$.
Consequently $\sigma$ and $\eta$ are exactly the same as when 
$\Delta \Psi = 0$, and there is no difference between the 
present results and those of Sec.\ \ref{sec:time_independent_psi}.
This is reassuring because it is precisely what we would expect 
when time evolution does not change the ray that $\ket{\psi}$
belongs to.

\subsection{Model calculations and special cases}

\label{sec:model_calculations}

As a tool for developing insight, it is helpful to run some numerical
calculations on simple models and see how well the general principles of the
theory hold up in practice. The model used here was the extended Hubbard model
\cite{[] [{, pp.\ 22 and 403.}] Mahan2000} for small one-dimensional lattices of
interacting fermions. Tests were run on both spinless fermions (with
nearest-neighbor interactions) and spin $1/2$ fermions (with on-site and
nearest-neighbor interactions). For fermions, the algebra of the $\psi$ product
can easily be implemented using bitwise operations in the binary representation
of Eq.\ (\ref{eq:binary_notation}).

As noted already at the end of Sec.\ \ref{sec:time_independent_psi}, convergence
tests of evolution over finite time intervals show that the subsystem dynamics
is indeed deterministic, with the solutions converging quadratically in $\Delta
t$. This remains true for the case of time-dependent $\Psi$.

An interesting test case is obtained by setting all terms derived from $\ket{v}$
equal to zero in Eqs.\ (\ref{eq:eta_hat}), (\ref{eq:omega_E}),
(\ref{eq:sigma_ket_redef}), (\ref{eq:beta_ket_E}), (\ref{eq:kappa_E}),
(\ref{eq:mu_matrix}), and (\ref{eq:nu_matrix}). This eliminates all constraints
on $\Delta \Psi$, thereby converting the constrained variation problem to an
unconstrained variation. With no constraints on the quasiclassical subsystems,
one would expect the solutions of Eq.\ (\ref{eq:Delta_x_real}) to have the
absolute maximum value of $\chi = 1$, corresponding to the limiting case of
Schr\"odinger dynamics for all subsystems [cf.\ Eq.\
(\ref{eq:minimum_condition})]. This is precisely what happens.

A similar result is obtained if one keeps all terms derived from $\ket{v}$ but
sets the particle interaction potential to zero. This again yields $\chi = 1$
and Schr\"odinger dynamics for all subsystems. As noted by Wiseman
\cite{Wiseman2004}, such a limit is physically uninteresting because it
turns each particle into an isolated universe having no connection with anything
else. However, it is a crucial test for the logical coherence of the theory, in
that it establishes the consistency of assuming Schr\"odinger dynamics for the
state $\ket{\psi}$ of a closed system.

\subsection{The number of subsystems is dynamically essential}

\label{sec:arbitrary_number}

The number of subsystems $m$ has so far been treated as an arbitrary
parameter. But what is the significance of this number? Does it play an active
part in determining the subsystem dynamics, or is its role more passive?

This question addresses the distinction between \emph{trivial} and
\emph{nontrivial} subsystems. A trivial subsystem is one that remains in a pure
vacuum state ($\ket{u_k} = \ket{0}$) as time evolves. A trivial subsystem
has no observable properties (see Sec.\ \ref{sec:observables}), so it
makes no difference whether it is included in the subsystem decomposition. The
value of $\chi$ is also unchanged by the addition of trivial subsystems. If
trivial subsystems are allowed by the principle of dynamical stability, then the
number $m$ plays no essential role in the dynamics, because one can add trivial
subsystems to any given subsystem decomposition without changing any observable
property.

However, vacuum subsystems are \emph{not} dynamically stable in systems of
interacting particles. All quasiclassical subsystems, including vacuum
subsystems, are coupled to each other by the terms derived from $\ket{v}$ in the
matrix (\ref{eq:eta_hat}). A vacuum subsystem satisfies the time-dependent
Schr\"odinger equation, and we know already from Secs.\
\ref{sec:interacting_not_Schroedinger} and
\ref{sec:dynamical_stability_functional} that such a time dependence is not
dynamically stable in a system of interacting particles. Because $\chi < 1$ (see
Sec.\ \ref{sec:dynamical_stability_functional}), there is always room to
increase $\chi$ by allowing an initial vacuum subsystem to evolve into a
nonvacuum final subsystem. Hence, in a system of interacting particles, the
number $m$ plays an essential role in the subsystem dynamics, because there are
no trivial subsystems.

On the other hand, for noninteracting particles, dynamically stable subsystem
decompositions always have $\chi = 1$. The most general such decomposition can
be obtained by choosing an independent solution of the Schr\"odinger equation
for each subsystem. Because the vacuum state satisfies the Schr\"odinger
equation, trivial subsystems are dynamically stable. Hence, for noninteracting
particles, the number $m$ need not be the same as the number of
nontrivial subsystems. (However, this case is physically uninteresting, as noted
in Sec.\ \ref{sec:model_calculations}.)

In conclusion, for the physically interesting case of interacting particles, the
number of subsystems $m$ is an essential determining factor for the subsystem
dynamics. The value of $m$ is arbitrary, but some number must be chosen
in order to apply the principle of dynamical stability.

\section{Reference frames and superselection rules}

\label{sec:superselection}

Thus far, we have seen no sign of any deviation from strict determinism in the
subsystem dynamics. This result seems to be in tension with the lack of
determinism exhibited by ordinary quantum mechanics. However, up to this point
it has also been assumed that there are in principle no restrictions on
observable quantities. It is therefore of interest to consider the effect of
restrictions arising from the lack of an external reference frame, which were
discussed briefly in Sec.\ \ref{sec:relational_properties}.  In standard quantum
mechanics, it is well known that the lack of a reference frame generally gives rise to a superselection rule \cite{BartlettRudolphSpekkens2007} together with 
associated classical variables.

\subsection{Lack of phase reference}

\label{sec:phase_reference}

Rather than discussing reference frames in general, this paper focuses on a
particular example relevant to nonrelativistic fermions, namely the number
superselection rule arising from the lack of a phase reference
\cite{BartlettRudolphSpekkens2007}. Lack of a phase reference simply means that
the phase transformation
\begin{equation}
\ket{\psi} \to e^{i N \phi} \ket{\psi} \label{eq:phase_psi}
\end{equation}
has no observable consequences, where $N$ is the operator for the total number
of particles and $\phi$ is any real number. If particle number is conserved
($[H, N] = 0$), then this symmetry is maintained over time, since
\begin{equation}
e^{-i H t} \ket{\psi} = e^{-i N \phi} e^{-i H t} e^{i N \phi} \ket{\psi} .
\end{equation}
Of course, in the present theory, observables are associated with the
subsystems rather than $\ket{\psi}$ (Sec.\ \ref{sec:observables}).
To see the effect on the subsystems, note that the transformation
(\ref{eq:phase_psi}) is equivalent to
\begin{equation}
\Psi \to e^{i N \phi} \Psi e^{-i N \phi} , \label{eq:phase_Psi}
\end{equation}
because $N \ket{0} = 0$.  But this is equivalent to applying
the same phase transformation to every subsystem:
\begin{equation}
U_k \to e^{i N \phi} U_k e^{-i N \phi} \qquad 
(k = 1, 2, \ldots, m) . \label{eq:phase_Uk}
\end{equation}

A crucial difference between this phase shift and the Schr\"odinger dynamics
problem studied in Sec.\ \ref{sec:interacting_not_Schroedinger} is that both
sides of the mapping (\ref{eq:phase_Uk}) are creators [see Eq.\
(\ref{eq:hop_commutator})]. Hence, a phase shift applied to the total state
$\Psi$ propagates directly to all of the subsystems. This is analogous to
Lubkin's description of superselection rules in standard quantum mechanics
\cite{Lubkin1970}.

\subsection{Equivalence classes of subsystem decompositions}

In the direct-sum formalism, applying the phase shift $\ket{u_k} \to e^{i N
\phi} \ket{u_k}$ to all subsystems $k$ is the same as applying the phase shift
$\ket{u} \to e^{i \hat{N} \phi} \ket{u}$ to the direct sum of subsystems, in
which [cf.\ Eq.\ (\ref{eq:H_hat})]
\begin{equation}
\hat{N} \equiv \bigoplus_{k=1}^{m} N .
\end{equation}
If this phase shift has no observable consequences, the relevant
mathematical object is not the individual subsystem decomposition
$\ket{u}$ but rather the \emph{equivalence class}
\begin{equation}
[u] \equiv \{ \exp (i \hat{N} \phi) \ket{u} : 0 \le \phi < 2\pi \} .
\end{equation}
The corresponding equivalence class for a subsystem projector $\rho$ is
\begin{equation}
[\rho] \equiv \{ \exp (i \hat{N} \phi) \rho 
\exp (-i \hat{N} \phi) : 0 \le \phi < 2\pi \} . \label{eq:rho_phase_orbit}
\end{equation}
Such equivalence classes are also referred to as \emph{phase orbits}
\footnote{The name ``orbit'' is commonly used in this context; see, e.g., Refs.\
\cite{Bengtsson2006} and \cite{Isham1999}.}. In a system without a phase
reference, all of the preceding theory must be reformulated in terms of orbits
rather than individual subsystem decompositions.

\subsection{Distance between phase orbits}

\label{sec:phase_orbit_distance}

The first step is to define a suitable measure of distance between phase
orbits. The distance between $[\rho]$ and $[\rho']$ can be defined simply as the
minimum distance \footnote{It is interesting to note that the Fubini--Study
metric can also be derived from such a minimum principle
\cite{ProvostVallee1980}.} between any two elements of these orbits:
\begin{equation}
D^2 ([\rho], [\rho']) \equiv \min_{\theta, \phi}
D^2 (e^{i \hat{N} \theta} \rho e^{-i \hat{N} \theta},
e^{i \hat{N} \phi} \rho' e^{-i \hat{N} \phi})  .
\end{equation}
One of these phase shifts is redundant, so we can write this definition
more simply as
\begin{equation}
D^2 ([\rho], [\rho']) = \min_{\phi} \lambda (\phi) ,
\label{eq:D2_lambda}
\end{equation}
in which
\begin{equation}
\lambda (\phi) =  D^2 (\rho, e^{i \hat{N} \phi} \rho' e^{-i \hat{N} \phi}) .
\label{eq:lambda_phi}
\end{equation}
Here it is worthwhile to pause and note the similarity between $\lambda (\phi)$
and the function $\lambda (\Delta \tau)$ introduced previously in Eq.\
(\ref{eq:lambda_defn}).  This similarity means that much of the following
derivation will be almost identical to that given in
Sec.\ \ref{sec:time_functional_defn}.  Consequently, only a brief
outline of the results is presented.

For small changes $\norm{\Delta u}$, Eq.\ (\ref{eq:lambda_phi}) reduces to
\begin{equation}
\lambda (\phi) = 
\matelm{u'}{e^{-i \hat{N} \phi} (1 - \rho) e^{i \hat{N} \phi}}{u'} .
\end{equation}
After expanding the right-hand side to second order in small quantities,
we obtain the quadratic function
\begin{equation}
\lambda (\phi) = \eta_0 - 2 \phi \imag \inprod{\Delta u}{N} 
+ \phi^2 \inprod{N}{N} , \label{eq:lambda_phi_quad}
\end{equation}
in which $\eta_0 = \lambda (0)$ and $\ket{N} \equiv (1 - \rho) \hat{N} \ket{u}$.
The function (\ref{eq:lambda_phi_quad}) has a minimum at $\phi = \varphi$, in
which
\begin{equation}
\varphi = \frac{\inprod{\Delta u}{N}}{\inprod{N}{N}} .
\end{equation}
The value of $\lambda (\phi)$ at the minimum is
\begin{equation}
\lambda (\varphi) = \eta_0 - 
\frac{(\imag \inprod{\Delta u}{N})^{2}}{\inprod{N}{N}} .
\label{eq:lambda_phi_min}
\end{equation}
But this minimum value is just the desired distance (\ref{eq:D2_lambda}) between
the two orbits:
\begin{equation}
D^2 ([\rho], [\rho']) = \matelm{\Delta u}{(1 - \rho)}{\Delta u} - 
\frac{(\imag \inprod{\Delta u}{N})^{2}}{\inprod{N}{N}} . 
\label{eq:D2_phase_orbit}
\end{equation}

As before, the symbol $\eta$ is used to refer to the square of the
basic measure of distance:
\begin{equation}
\eta \equiv D^2 ([\rho], [\rho']) = \lambda (\varphi) .
\end{equation}
It is convenient to write this more concisely as
\begin{equation}
\eta = \eta_0 - \xi^2 ,
\end{equation}
in which $\xi$ is the functional
\begin{equation}
\xi \equiv \frac{\imag \inprod{\Delta u}{N}}{\Delta N} = 
\imag \inprod{\Delta x}{\xi} , \quad \Delta N \equiv \sqrt{\inprod{N}{N}} ,
\label{eq:xi_def}
\end{equation}
and $\ket{\xi}$ is the vector
\begin{multline}
\inprod{e_{ki}}{\xi} = \frac{1}{\Delta N} \biggl[ 
\frac{\matelm{f_{ki}}{(1 - \rho_k)N}{u_k}}{\inprod{u_k}{u_k}} \\ +
\frac{\matelm{g_{ki}}{(1 - \rho_v)N}{v}}{\inprod{v}{v}}
\biggr] . \label{eq:xi_ket}
\end{multline}
Again, it is worth noting the close similarity between these quantities and
those defined in Secs.\ \ref{sec:time_functional} and
\ref{sec:dynamical_stability}.

Sometimes it is necessary to calculate $D^2 ([\rho], [\rho'])$ in situations
where $\norm{\Delta u}$ is not small. (See, for example, the last paragraph in
Sec.\ \ref{sec:dynamical_stability_phase_orbits}.) This case is considered in
Appendix \ref{app:phase_orbit_distance}.

\subsection{Time functional for phase orbits}

A time functional suitable for phase orbits can now be derived by minimizing
the function [cf.\ Eqs.\ (\ref{eq:lambda_defn}), (\ref{eq:lambda_phi})]
\begin{equation}
\lambda (\phi, \Delta \tau) \equiv D^2 (\rho, e^{i \hat{N} \phi} 
e^{i \hat{H} \Delta \tau} \rho' e^{-i \hat{H} \Delta \tau} e^{-i \hat{N} \phi})
\end{equation}
with respect to both $\phi$ and $\Delta \tau$. Given that $[H, N] = 0$, this
reduces in the case of small $\norm{\Delta u}$ to
\begin{equation}
\lambda (\phi, \Delta \tau) = 
\matelm{u'}{e^{-i \hat{A}} (1 - \rho) e^{i \hat{A}}}{u'} ,
\label{eq:lambda_A}
\end{equation}
in which the operator $\hat{A}$ is defined by
\begin{equation}
\hat{A} (\phi, \Delta \tau) \equiv \hat{H} \Delta \tau + \hat{N} \phi .
\end{equation}
When $\norm{\Delta u}$ is small, $\phi$ and $\Delta \tau$ can also be treated as
small quantities of the same order, and Eq.\ (\ref{eq:lambda_A}) can
be expanded as usual to obtain the quadratic approximation
\begin{equation}
\lambda (\phi, \Delta \tau) = \eta_0 - 2 \imag \inprod{\Delta u}{A} 
+ \inprod{A}{A} ,
\end{equation}
in which $\eta_0 = \lambda (0, 0)$ and $\ket{A} = (1 - \rho) \hat{A} \ket{u}$.

The minimum of this function occurs at $(\phi, \Delta \tau) = (\varphi, \Delta
t)$, in which $\varphi$ and $\Delta t$ satisfy the system of equations
\begin{equation}
\begin{pmatrix}
\inprod{N}{N} & \inprod{N}{H} \\
\inprod{H}{N} & \inprod{H}{H}
\end{pmatrix}
\begin{pmatrix}
\varphi \\
\Delta t
\end{pmatrix}
=
\begin{pmatrix}
\imag \inprod{\Delta u}{N} \\
\imag \inprod{\Delta u}{H}
\end{pmatrix} .
\end{equation}
The matrix on the left is real and symmetric, because $[H, N] = 0$ implies
$\inprod{N}{H} = \inprod{H}{N}$.  Upon inverting this matrix, we
find the desired time functional
\begin{equation}
\Delta t = \frac{\inprod{N}{N} \imag \inprod{\Delta u}{H} -
\inprod{H}{N} \imag \inprod{\Delta u}{N}}{\inprod{N}{N} \inprod{H}{H} -
\inprod{N}{H} \inprod{H}{N}} . \label{eq:Delta_t_messy}
\end{equation}
This solution is well defined as long as $\inprod{N}{N} > 0$, $\inprod{H}{H} >
0$, and (by the Schwarz inequality) $\ket{H} \ne c \ket{N}$, where $c$ is any
constant. If $\ket{H} = c \ket{N}$, this simply means that Schr\"odinger
dynamics cannot move the subsystems out of the initial phase orbit, so $\Delta
t$ is ill defined (at least to first order in small quantities).

Equation (\ref{eq:Delta_t_messy}) looks considerably more complicated than
the previous functional (\ref{eq:Delta_t}), but it can be simplified by
introducing the operator
\begin{equation}
K \equiv H - \frac{\inprod{H}{N}}{\inprod{N}{N}} N .
\end{equation}
Here $K$ is just the component of $H$ that is orthogonal to $N$, in the sense
that $\inprod{K}{N} = 0$. In terms of $K$, the time functional
(\ref{eq:Delta_t_messy}) is simply
\begin{equation}
\Delta t = \frac{\imag \inprod{\Delta u}{K}}{\inprod{K}{K}} .
\label{eq:Delta_t_K}
\end{equation}
The denominator satisfies $\inprod{K}{K} \le \inprod{H}{H}$, equality occurring
if and only if $\inprod{H}{N} = 0$. In geometric terms, $\Delta K \equiv
\sqrt{\inprod{K}{K}}$ is the length of $\ket{K}$, which is the component of
$\ket{H}$ orthogonal to $\ket{N}$. Physically, $\Delta K$ is a renormalized
energy uncertainty, just what remains of $\Delta E$ after the unphysical part of
$H$ is removed (``unphysical'' because it has no observable consequences for
this particular $\ket{u}$).

The phase angle $\varphi$ at the minimum of $\lambda (\phi, \Delta \tau)$
can be written likewise as
\begin{equation}
\varphi = \frac{\imag \inprod{\Delta u}{L}}{\inprod{L}{L}} , \qquad
L \equiv N - \frac{\inprod{N}{H}}{\inprod{H}{H}} H .
\end{equation}
The value of $\lambda (\phi, \Delta \tau)$ at the minimum is
\begin{multline}
\lambda (\varphi, \Delta t) = \matelm{\Delta u}{(1 - \rho)}{\Delta u} \\ - 
\frac{(\imag \inprod{\Delta u}{N})^{2}}{\inprod{N}{N}} -
\frac{(\imag \inprod{\Delta u}{K})^{2}}{\inprod{K}{K}} ,
\end{multline}
which is similar to Eqs.\ (\ref{eq:lambda_min}) and (\ref{eq:lambda_phi_min}).
This can also be written as
\begin{equation}
\lambda (\varphi, \Delta t) = \lambda (\varphi, 0) - \Delta K^2 \Delta t^2 ,
\end{equation}
in which $\lambda (\varphi, 0)$ is the same as Eq.\ (\ref{eq:lambda_phi_min}):
\begin{equation}
\lambda (\varphi, 0) = \matelm{\Delta u}{(1 - \rho)}{\Delta u} -
\frac{(\imag \inprod{\Delta u}{N})^{2}}{\inprod{N}{N}} . 
\label{eq:lambda_phi_0}
\end{equation}

If we follow the argument used in Sec.\ \ref{sec:properties_time_functional}, it
is easy to prove that $\lambda (\varphi, \Delta t) \ge 0$, in which a necessary
condition for equality is
\begin{equation}
(1 - \rho) \ket{\Delta u} = (\Pi_{K} + \Pi_{N})  \ket{\Delta u} ,
\end{equation}
where $\Pi_{K}$ and $\Pi_{N}$ are the projectors for $\ket{K}$ and $\ket{N}$.
A necessary and sufficient condition for $\lambda (\varphi, \Delta t) = 0$ is
\begin{equation}
(1 - \rho) \ket{\Delta u} = i C_{K} \ket{K} + i C_{N} \ket{N} ,
\end{equation}
in which the numbers $C_{K}$ and $C_{N}$ are real.

\subsection{Dynamical stability of phase orbits}

\label{sec:dynamical_stability_phase_orbits}

With the modified time functional (\ref{eq:Delta_t_K}) in hand, we can now apply
the principle of dynamical stability in much the same way as before (see Sec.\
\ref{sec:dynamical_stability}). Many parts of the previous analysis can be
carried over to the case of phase orbits simply by replacing $H \to K$ and
$\Delta E \to \Delta K$. Thus, for example, the dimensionless time interval
$\sigma$ is redefined as [cf.\ Eqs.\ (\ref{eq:sigma_def}),
(\ref{eq:sigma_redef})]
\begin{equation}
\sigma \equiv \Delta K \Delta t = 
\frac{\imag \inprod{\Delta u}{K}}{\Delta K} = 
\imag \inprod{\Delta x}{\sigma} ,
\end{equation}
in which the components of $\ket{\sigma}$ are [cf.\ Eq.\ 
(\ref{eq:sigma_ket_redef})]
\begin{multline}
\inprod{e_{ki}}{\sigma} = \frac{1}{\omega \Delta K} \biggl[ 
\frac{\matelm{f_{ki}}{(1 - \rho_k)K}{u_k}}{\inprod{u_k}{u_k}} \\ +
\frac{\matelm{g_{ki}}{(1 - \rho_v)K}{v}}{\inprod{v}{v}}
\biggr] ,
\end{multline}
and the renormalization factor $\omega$ is [cf.\ Eq.\ (\ref{eq:omega_E})]
\begin{equation}
\omega = 1 - \frac{1}{\Delta K^2} \real \biggl[
\frac{\matelm{v_{H \cdot \Psi}}{(1 - \rho_v)K}{v}}{\inprod{v}{v}} \biggr] .
\end{equation}
Likewise, $\ket{\beta}$ and $\kappa$ in Eqs.\ (\ref{eq:beta_ket_E}) and
(\ref{eq:kappa_E}) are redefined as
\begin{equation}
\inprod{e_{ki}}{\beta} = \frac{1}{\Delta K}
\frac{\matelm{g_{ki}}{(1 - \rho_v)}{v_{H \cdot \Psi}}}{\inprod{v}{v}} ,
\end{equation}
\begin{equation}
\kappa = \frac{1}{\Delta K^2} \frac{\matelm{v_{H \cdot \Psi}}{(1 - \rho_v)}{v_{H
\cdot \Psi}}}{\inprod{v}{v}} .
\end{equation}
The only truly new parameter to arise is
\begin{equation}
\theta \equiv \frac{1}{\Delta N \Delta K} \real \biggl[
\frac{\matelm{v_{H \cdot \Psi}}{(1 - \rho_v) N}{v}}{\inprod{v}{v}} \biggr] .
\end{equation}

Given these definitions, the squared distance between neighboring phase
orbits is [cf.\ Eq.\ (\ref{eq:eta_Delta_t})]
\begin{equation}
\eta = \eta_0 - \xi^2 + 2 \sigma (\beta - \theta \xi) + 
\sigma^2 (\kappa - \theta^2) ,
\end{equation}
in which $\eta_0 = \matelm{\Delta x}{\hat{\eta}}{\Delta x}$ and $\xi$ is defined
in Eq.\ (\ref{eq:xi_def}). Here it should be noted that $\Psi$ is treated as
time dependent (see Sec.\ \ref{sec:time_dependent_psi})
and $\eta$ refers to the quantity
\begin{equation}
\eta = \eta(\Delta t) \equiv \lambda (\varphi, 0) \ne \lambda (\varphi, \Delta t) ,
\end{equation}
in which $\lambda (\varphi, 0)$ is given in Eq.\ (\ref{eq:lambda_phi_0}).

It is convenient now to follow the approach of Sec.\ \ref{sec:real_matrix} and
write $\eta = \mu + \nu$, in which $\mu$ and $\nu$ are defined in Eqs.\
(\ref{mu_def}) and (\ref{nu_def}). The only difference is that the matrix
elements $\mu_{ki,k'i'}$ and $\nu_{ki,k'i'}$ in Eqs.\ (\ref{eq:mu_matrix}) and
(\ref{eq:nu_matrix}) are modified to become
\begin{multline}
\mu_{ki,k'i'} = \delta_{kk'} 
\frac{\matelm{f_{ki}}{(1 - \rho_k)}{f_{ki'}}}{\inprod{u_k}{u_k}} +
\frac{\matelm{g_{ki}}{(1 - \rho_v)}{g_{k'i'}}}{\inprod{v}{v}} \\ +
\frac{1}{2} [ (\beta_{ki} \sigma_{k'i'}^{*} + \sigma_{ki} \beta_{k'i'}^{*})
+ (\kappa - \theta^2) \sigma_{ki} \sigma_{k'i'}^{*} \\ - \xi_{ki} \xi_{k'i'}^{*}
- \theta (\xi_{ki} \sigma_{k'i'}^{*} + \sigma_{ki} \xi_{k'i'}^{*}) ] ,
\end{multline}
\begin{multline}
\nu_{ki,k'i'} = - \frac{1}{2} [ (\beta_{ki} \sigma_{k'i'} + \sigma_{ki} \beta_{k'i'})
+ (\kappa - \theta^2) \sigma_{ki} \sigma_{k'i'} \\ - \xi_{ki} \xi_{k'i'}
- \theta (\xi_{ki} \sigma_{k'i'} + \sigma_{ki} \xi_{k'i'}) ] .
\end{multline}
Aside from this change of definition, all of the subsequent analysis in Secs.\
\ref{sec:real_matrix} and \ref{sec:solve_dynamical_stability} follows through in
the same way as before. Because the solution (\ref{eq:Delta_x_real}) has the
same mathematical structure as before, it gives rise to the same qualitative
behavior too. That is, the dynamics of phase orbits is also deterministic.

%
%

Of course, this does not mean that the subsystem dynamics is totally unchanged.
The main difference arises because $\Delta K$ is generally smaller than $\Delta
E$. The subsystems therefore tend to evolve in time more slowly, because all
physically unobservable changes (lying entirely within a given orbit) have been
filtered out.

Test calculations show that the phase-orbit dynamics obtained by integrating
Eq.\ (\ref{eq:Delta_x_real}) over a finite time interval has the correct
limiting behavior (i.e., Schr\"odinger subsystem dynamics) for the special cases
discussed in Sec.\ \ref{sec:model_calculations}. To demonstrate this, one must
use the phase-orbit distance formulas derived in Appendix
\ref{app:phase_orbit_distance} for the case of large $\norm{\Delta u}$.

\section{Subsystem permutations}

\label{sec:subsystem_permutations}


Thus far we have considered only one of the $m!$ possible permutations of the
subsystems in Eq.\ (\ref{eq:psi_pi}). The next step is to consider the set of
all permutations.

\subsection{Influence of permutations on dynamics}

\label{sec:permutations_influence_dynamics}

According to the definition of an observable in Sec.\ \ref{sec:observables}, a
permutation of the subsystems merely rearranges the labels $k$ on the numbers
$\expect{A}_k$ in Eq.\ (\ref{eq:mean_value}). But the subsystem labels $k$ are
not themselves observable, so a permutation cannot affect the value of any
observable quantity at any given time.

On the other hand, permutation of the subsystems does affect the value of the
product $\Psi_{\pi}$ in Eq.\ (\ref{eq:psi_pi}). This has no \emph{direct} effect
on any observable quantity, but $\Delta \Psi_{\pi} = \Psi_{\pi}' - \Psi_{\pi}$
determines the value of the subsystem change $\Delta V$ [see Eq.\
(\ref{eq:Delta_V})] generated by given changes $\Delta U_k$ in the
quasiclassical subsystems ($k \ne 1$).

This means that subsystem permutations do alter the subsystem \emph{dynamics}.
Starting from a given initial subsystem decomposition $\rho$ or phase orbit
$[\rho]$, a single time step (\ref{eq:Delta_x_real}) will in general carry each
permutation into a \emph{different} final state. In other words, the permutation
symmetry of observables is \emph{broken} by the dynamics. Subsystem permutations
are thus qualitatively different from the phase transformations considered in
Sec.\ \ref{sec:phase_reference}.

But this suggests that it might be possible, at least in principle, to use the
effect of permutations on dynamics to obtain information about subsystem
permutations from the time evolution of observables. Given unlimited information
about the observables of all subsystems (which is an unrealistic assumption, as
will be discussed in Sec.\ \ref{sec:information_theory}), one might even be able
to deduce which individual permutation was consistent with a given set of
experimental data.

\subsection{Subsystem ordering in orthodox quantum mechanics}

\label{sec:orthodox_subsystem}

It is important to pause here and note that this effect of subsystem
permutations on dynamics is not limited to the context of the present dynamical
stability formalism. It is a general consequence of the noncommutative algebra of
fermion creation operators that holds even in orthodox quantum mechanics,
although this aspect of the theory has received little prior attention.

Consider, for example, the products $\Psi_a = U_1 U_2$ and $\Psi_b = U_2 U_1$ of
two creators $U_1$ and $U_2$. Since $\Psi_a$ and $\Psi_b$ generally belong to
different rays in projective Hilbert space, they will of course evolve in
different ways under the Schr\"odinger equation. That is really all there is to
it.

An obvious objection to this conclusion is that $\Psi_a$ and $\Psi_b$ do
\emph{not} belong to different rays if $U_1$ and $U_2$ have a definite number of
fermions. In that case, $\Psi_a$ differs from $\Psi_b$ by at most a physically
meaningless overall sign [see Eq.\ (\ref{eq:signed_product})]. More generally,
this is also true if every subsystem $U_k$ is even or odd [see Eq.\
(\ref{eq:creator_commute_anti})]---or, in other words, if each $U_k$ has a
definite univalence, where the univalence $W$ is the number of fermions modulo
2:
\begin{equation}
W \equiv N_{\mathrm{f}} \pmod 2 . \label{eq:univalence}
\end{equation}
Therefore, if the subsystems in orthodox quantum mechanics are required to
satisfy a fermion-number or univalence superselection rule
\cite{WickWightmanWigner1952, HegerfeldtKrausWigner1968, WickWightmanWigner1970,
Wightman1995}, the Schr\"odinger dynamics of $\Psi$ does not depend on
the order in which the subsystems are multiplied.

An answer to this objection can be found in the growing consensus
\cite{AharonovSusskind1967a, AharonovSusskind1967b, AharonovRohrlich2005,
Mirman1969, Mirman1970, Mirman1979, Lubkin1970, Zeh1970, Zurek1982,
GiuliniKieferZeh1995, Giulini2003a, Giulini2009a, WeinbergVol1,
DowlingBartlettRudolphSpekkens2006, BartlettRudolphSpekkens2007, Earman2008}
that most (if not all) superselection rules are pragmatic expressions of
practical limitations on experimental capabilities rather than fundamental laws
of nature. This suggests that, at the level of fundamental theory, the
superposition principle should be taken seriously, not lightly brushed aside
\cite{Zeh2003ch2}.

The practical limitation that gives rise to a particle-number superselection
rule in orthodox quantum mechanics is just the lack of a phase reference
discussed in Sec.\ \ref{sec:phase_reference} \cite{BartlettRudolphSpekkens2007}.
Since this limitation was already taken into account in the analysis of Sec.\
\ref{sec:superselection} and the discussion of Sec.\
\ref{sec:permutations_influence_dynamics}, it does not alter the conclusion that
subsystem permutations can have an observable effect on subsystem dynamics.

\subsection{Significance of a univalence superselection rule}

\label{sec:univalence_ssr}

Thus, if one wishes to eliminate the dependence of dynamics on subsystem
permutations, one must introduce the univalence superselection rule as an
independent axiom; it cannot be derived from the phase symmetry
(\ref{eq:phase_Uk}).  This rule would require all quasiclassical
subsystems to be even, but subsystem $U_1$ could be even or odd. One can easily
verify that the univalence of all subsystems is conserved by the time evolution
(\ref{eq:Delta_x_real}) if $[H, W] = 0$.

Introducing such a rule makes it easier to combine two subsystems into one.
Suppose, for example, we have two subsystems localized in adjacent regions of
coordinate space. In such a case it seems natural to talk about using the $\psi$
product to merge these subsystems. However, this cannot be done if they are
separated in the product (\ref{eq:psi_pi}) by a subsystem that commutes with
neither of them, even if that subsystem is localized far away in coordinate
space.  The univalence superselection rule therefore provides a natural 
framework for discussing composition and decomposition of subsystems.

However, at present the question of whether to introduce such an axiom is
simply left open. The basic structure of the theory that follows does not depend
on this choice.

\section{A bare-bones theory of information}

\label{sec:information_theory}

At this stage the basic theory of subsystem dynamics is more or less complete.
The next step is to construct a theory of information (or, in other words, a
theory of measurement) that connects the subsystem dynamics to the experiences
of observers. This paper develops such a theory only at a very rudimentary
level, focusing mostly on qualitative questions such as ``Whose
information?'' and ``Information about what?''~\cite{Bell1990} Detailed
investigations of this theory of information are left for future work.

\subsection{Bayesian inference in the present moment}

\label{sec:Bayesian_inference}

The present theory of information is essentially just a theory of Bayesian
inference for individual observers treated as subsystems of a closed system. The
use of Bayesian inference means that this theory has much in common with QBism
\cite{CavesFuchsSchack2002, CavesFuchsSchack2007, Fuchs2003, Fuchs2010,
Mermin2012a, *Mermin2012b, FuchsSchack2013, FuchsMerminSchack2014, Mermin2014b,
Mermin2014a, *Mermin2014c, Mermin2014d, Mermin2016}. However, it differs from
QBism in its assignment of subsystem vectors $\ket{u_k}$ to all observers whose
experiences are being described.

Here the implementation of Bayesian inference is controlled by two fundamental
principles: (1)~Each observer experiences directly only the beables associated
with one subsystem $\ket{u_k}$. All else must be inferred. In particular, the
existence of other subsystems is inferred from the influence of those subsystems
on the dynamics of $\ket{u_k}$. (2)~Each observer experiences directly only the
beables associated with a single moment of time (a \emph{moment} being defined
as an infinitesimal interval of time). All else must be inferred. In
particular, the dynamics of all subsystems in the past and future of the present
moment is purely an inference.

The meaning, significance, and historical context of these statements are
elaborated in this subsection. Their mathematical implications are discussed in
Sec.\ \ref{sec:backbone}.

Let us start with the case in which inferences are drawn from the experiences of
only one observer. In this case, principle (1) says that the existence of other
subsystems is inferred from their influence on the dynamics of the observer's
subsystem $\ket{u_k}$. Interactions between particles are crucial in this
regard. If the particles do not interact, the time evolution of $\ket{u_k}$ is
independent of the other subsystems, so nothing at all can be inferred about the
properties of other subsystems.

The idea that an observer can only ``measure'' the state of his own
subsystem was proposed long ago by London and Bauer \cite{LondonBauer1939a,
*LondonBauer1939b, WheelerZurek1983}. They described this capacity of an
observer as a ``faculty of introspection.'' London and Bauer's concept of
measurement is therefore very different from that of von Neumann
\cite{vonNeumann1955}, in which the consciousness of the observer somehow
directly perceives the state of the outside world, even though the observer is
expressly excluded from this state.

In the present theory, the ``faculty of introspection'' does not lead to any
collapse of the total state vector $\ket{\psi}$. Instead, the observer merely
takes note of the beables for subsystem $\ket{u_k}$ (or some subset thereof)
during the present moment of time. This information is then used by the
observer to perform a Bayesian updating of the probabilities that he
ascribes to the various possible subsystem decompositions of $\ket{\psi}$. Here
Bayesian updating is just the usual process of replacing prior probabilities
with posterior probabilities, in which the words ``prior,'' ``posterior,'' and
``updating'' refer to a direction of logical inference, not to a direction in
time. The total state $\ket{\psi}$ is not assumed to be known, since all
subsystems are treated as initially unknown.

The idea that all observations are fundamentally \emph{self}-observations may
seem strange from the perspective of orthodox measurement theory
\cite{vonNeumann1955}, which requires an observer to always be \emph{outside}
the observed system \cite{Wheeler1977}. Indeed the orthodox description is
ostensibly the most natural one, since we intuitively regard our sense of vision
as a direct perception of the world around us. However, our susceptibility to
optical illusions shows clearly that the three-dimensional world we see is
actually an inference based on very incomplete two-dimensional
information provided by the retinas \cite{Pinker1997}.

The fact that observers are treated as subsystems does not mean that the problem
of consciousness must be solved before this theory can be used. The theory
merely limits what can be \emph{known} by any observer to the properties of a
single subsystem. However, the consequences of imposing such a limit can be
evaluated by studying simple non-biological subsystems. According to Wheeler
\cite{Wheeler1981}, it is ``not consciousness but the distinction between the
probe and the probed [that is] central to the elemental quantum act of
observation.'' A similar remark was made by Heisenberg \cite{[{}] [{, p.\ 58.}]
Heisenberg1930}: ``The observing system need not always be a human being; it may
also be an inanimate apparatus, such as a photographic plate.''

Principle (2) bridges the gap between the ``block universe'' concept of time
\cite{Price1996, Price2013, Zeh2007} used in most physical theories and the
subjective flow of time that each of us experiences. According to Carnap
\cite{Carnap1963}, Bergson's criticism of the block universe picture
\cite{Bergson1910, Bergson1999, Ridley2014} was deeply troubling to
Einstein:
\begin{quote}
Once Einstein said that the problem of the Now worried him seriously. He
explained that the experience of the Now means something special for man,
something essentially different from the past and the future, but that this
important difference does not and cannot occur within physics. That this
experience cannot be grasped by science seemed to him a matter of painful but
inevitable resignation. \ldots
We both agreed that this was not a question of a defect for which science
could be blamed, as Bergson thought.
\end{quote}
The idea that the present moment \emph{does} play a crucial role in physics
was emphasized by Wheeler in connection with his ``delayed-choice'' experiments
\cite{Wheeler1978, Wheeler1980b, Wheeler1983b}:
\begin{quote}
The ``past'' is theory. The past has no existence except as it is recorded in
the present.
\end{quote}
Wheeler's central message is that everything we say about the past is
necessarily an \emph{inference} based on records in the present. This is of
course a platitude for historians and paleontologists, but it carries important
lessons for experimental physicists as well \cite{Jacques2007}.

Everett \cite{Everett1957, Everett1973, DeWittGraham1973}, Bell \cite{Bell1971,
Bell1976, Bell1981}, Barbour \cite{Barbour1994d, Barbour1994b, Barbour2000}, and
Page \cite{Page1994} have suggested ways of taking this into account by
restructuring quantum theory around the records contained in the present value
of $\ket{\psi}$. Mermin \cite{Mermin1998b, Mermin2014b, Mermin2014a,
*Mermin2014c}, Hartle \cite{Hartle2005}, and Smolin \cite{Smolin2013} have all
recently called attention to the physical significance of the present. But Zeh has
remarked that ``physics does not even offer any conceptual means for deriving
the concept of a present that would objectively separate the past from the
future'' \cite{Zeh2007}.

Here this concept is not \emph{derived}; rather, it is built into the theory of
information as a fundamental axiom about the \emph{subjective} experiences of
observers. The subjective nature of the Now was stressed by Zeh \cite{Zeh2007},
and its importance for the foundations of both classical and quantum physics has
been emphasized especially by Mermin \cite{Mermin2014b, Mermin2014a,
*Mermin2014c}.

It should be noted that the infinitesimal time interval referred to in principle
(2) is different from the finite time interval (of ``perhaps a few tenths of a
second'' \cite{Mermin1998b}) that is associated with our intuitive
perception of the present. As discussed by Barbour \cite{Barbour1994d,
Barbour1994b} and Hartle \cite{Hartle2005}, the \emph{perceived} duration of the
present moment is probably related to the way in which information about the
immediate past is stored and processed in the brain.

\subsection{Mathematical backbone for inferences}

\label{sec:backbone}

Let us now examine some mathematical implications of the fundamental principles
discussed in Sec.\ \ref{sec:Bayesian_inference}. The first step is to consider
\emph{ideal} observers, each of whom is fully aware of every detail of the state
of her subsystem. This is of course unrealistic, as an observer's experience
will in general tell her only the values of some subset of the beables
associated with her subsystem. The consequences of this further restriction are
discussed below in Sec.\ \ref{sec:rest_of_skeleton}.

To begin, let us review the way in which the principle of dynamical stability
was used in Sec.\ \ref{sec:dynamical_stability}. The problem solved there was a
\emph{time evolution problem}. The initial subsystems $\ket{u_k}$ were all
assumed to be known; the unknown quantities were the final subsystems
$\ket{u_k'}$. The initial and final subsystems were assumed to differ only
infinitesimally. The known subsystems were held fixed and the unknown subsystems
were varied so as to maximize the dynamical stability functional $\chi$, subject
to the constraint that the total states $\ket{\psi}$ and $\ket{\psi'}$ satisfy
the Schr\"odinger equation (\ref{eq:Delta_psi_Schr}).

The problem of concern now is an \emph{environmental stabilization problem}, in
which environmental subsystems are used to stabilize the overall subsystem
decomposition. The difference lies in the classification of known and unknown
subsystem states. Let us write the total number of subsystems as
\begin{equation}
m = s + r ,
\end{equation}
in which $s$ is the number of subsystems that are classified as observers;
the remaining $r$ subsystems are regarded as parts of the environment. For
the environmental stabilization problem, all observer states $\ket{u_k}$ and
$\ket{u_k'}$ are treated as known quantities (defined by the experiences of the
observers), whereas the environmental states $\ket{u_k}$ and $\ket{u_k'}$ are
treated as unknown. As before, the initial and final subsystems are assumed to
differ only infinitesimally. As before, the known subsystems are held fixed and
the unknown subsystems are varied so as to maximize $\chi$, subject to the
constraint that $\ket{\psi}$ and $\ket{\psi'}$ satisfy the Schr\"odinger
equation.

The solutions of the environmental stabilization problem are pairs $(\ket{u},
\ket{u'})$ of subsystem decompositions. Each such pair has its own
(infinitesimal) time interval $\Delta t$, which is the duration of the present
moment associated with that pair. This is always an \emph{inferred} quantity
because $\Delta t$ is defined only for the subsystem decomposition as a whole,
not for individual subsystems. Once the environmental stabilization problem has
been solved, we can use time evolution to extrapolate any given pair $(\ket{u},
\ket{u'})$ into the past and future of $\Delta t$.

From their definition as variation problems, the time evolution and
environmental stabilization problems each lead to a set of equations in which
the number of equations is equal to the number of unknown variables. However,
there is a big difference in the difficulty of these sets of equations. The time
evolution problem leads to a set of linear equations (see Sec.\
\ref{sec:solve_dynamical_stability}), whereas the environmental stabilization
problem leads to a set of nonlinear equations.

The latter set of equations is not written out explicitly here, but its
qualitative properties can easily be seen by returning to the model system
introduced in Sec.\ \ref{sec:model_calculations} (i.e., a system of interacting
fermions). In this model, each state $\ket{u_k}$ or $\ket{u_k'}$ can be regarded
as a function of $(2^d - 1)$ independent complex variables or $2(2^d - 1)$
independent real variables, where $d = \dim \mathcal{H}_{\mathrm{f}}$. The
dynamical stability functional $\chi$ is a rational function of these variables.
The environmental stabilization problem therefore requires one to find the
common zeros of a set of rational functions. This is a difficult but well
defined problem in algebraic geometry \cite{Cox2007}.

However, since the author has no training in this field, the surest route to
progress is to publicize the problem and invite experts in algebraic geometry to
work on it. For this reason, the quest for explicit solutions shall not be
pursued any further here. The existence of solutions is simply taken for
granted. Indeed, since the environmental stabilization problem is nonlinear, it
may have many solutions [in contrast to the time evolution problem, for which
the set of linear equations (\ref{eq:eta_sigma_linear_real_ket}) has a unique
solution (\ref{eq:Delta_x_real})]. If no univalence superselection rule is
imposed (see Sec.\ \ref{sec:univalence_ssr}), there will in general be a
different set of solutions for each possible permutation of the subsystems, thus
further increasing the total number of solutions.

Assuming that the environmental stabilization problem can be solved, how do we
use its solutions to perform Bayesian inference? Consider the set of pairs
$(\ket{u}, \ket{u'})$ of subsystem decompositions that satisfy the environmental
stabilization problem for all possible choices of observer states. In this set,
the observer states are \emph{not} required to agree with the experiences of the
observers. However, to fix the scale of the subsystem changes, the Fubini--Study
distance between the sets of observer states in $\ket{u}$ and $\ket{u'}$ is
required to be the same as that given by the experiences of the observers.

The first step in the Bayesian inference problem is to assign a prior
probability to each pair $(\ket{u}, \ket{u'})$ in this set. The choice of prior
probability is subjective \cite{Appleby2005a, Appleby2005b, Jeffrey2004},
although with some work it may be possible to reduce the degree of subjectivity
to a choice of symmetry principle \cite{Jaynes1989, Jaynes2003}. Posterior
probabilities are then obtained simply by setting to zero the probabilities of
all pairs $(\ket{u}, \ket{u'})$ that have the wrong observer states (i.e.,
states inconsistent with the experiences of the observers) and renormalizing.

The outcome of this inference problem is reminiscent of Wheeler's 
definition of reality
\cite{Wheeler1981}:
\begin{quote}
The vision of what we call ``reality'' is in large measure of a pale and
theoretic hue. It is framed by a few iron posts of true observation---themselves
also resting on theory for their meaning---but most of the walls and towers in
the vision are of papier-m{\^ a}ch{\' e}, plastered in between those posts by an
immense labor of imagination and theory.
\end{quote}
Here the iron posts are the observer subsystems in the present moment;
everything else is inferred. In general, an observer's experiences in the
present moment have a definiteness that is lacking in her memory of her
inferences about that moment \footnote{The observer's inferences about the
present moment were made in the inferred past, when the present was regarded as
part of the future. Sentences such as this one illustrate how difficult it is to
talk consistently about the past and future as inferences. The difficulty is
that our language takes the reality of the past and future for granted. For most
of this paper, this problem is dealt with in the simplest possible way, by not
striving for absolute consistency.}. This plays a role similar to the
``reduction of the wave packet'' in orthodox quantum mechanics
\cite{CohTan1977}. 

It must be emphasized that this is only an analogy. No actual ``jump'' is
supposed to take place; there is only a contrast between expectations and
experiences. The analogy with the orthodox description of wave-packet reduction
\cite{CohTan1977} is actually rather distant. A closer analogy can be found in
the dynamical reduction models of Ghirardi and others \cite{BassiGhirardi2003}.
In these models the reduction process is continuous, and the beables are defined
in terms of expectation values (rather than eigenvalues) of operators, just as
in Sec.\ \ref{sec:observables}. In the present paper, however, the beables are
not limited to mass density in coordinate space, and the effective ``reduction''
takes place only at the subsystem level, leaving the Schr\"odinger dynamics
of the total system untouched.

\subsection{How many subsystems?}

\label{sec:how_many_subsystems}

Are there any general criteria for choosing the subsystem numbers $s$ and $r$?
This question is easiest to answer for the number of environmental subsystems
$r$. As argued below, the sharpness of Bayesian inferences can be maximized by
choosing the smallest possible value: namely, $r = 1$ \footnote{The value $r=0$
must be excluded because the resulting state $\ket{\psi}$ would not satisfy the
Schr{\" o}dinger equation.}. The choice of the number of observers $s$ requires
a longer answer, as it occupies the borderland between subjectivity and
objectivity.

The value $r = 1$ is favored by Occam's razor, which can be regarded as a
corollary of Bayesian probability theory \cite{Nemenman2015}. The basic argument
is that, for any given value of $s$, the number of solutions to the
environmental stabilization problem can be expected to increase rapidly as a
function of $r$. This statement is plausible because both the number of
equations and the degree of the polynomials involved are increasing functions of
$r$. It is, however, not possible to prove this claim without actually
solving the environmental stabilization problem. The sharpest Bayesian inferences
are thus expected to be given by $r = 1$.

The most obvious value to choose for the number of observers is also $s =
1$. This choice seems to be consistent with each person's intuitive view of the
world. If we assume that Darwinian natural selection has instilled in us a rough
facsimile of the above Bayesian inference process as the basis for our
perception of the world around us (which is, admittedly, a big assumption), then
it would be difficult to explain how natural selection could settle on any value
other than $s = 1$.

This instinctive value---the one used for the automatic 
subconscious inferences that underlie our perception of the world---cannot be
changed. We cannot separate these subconscious inferences from our raw sense
impressions any more than we could measure the bare (unrenormalized) charge of
an electron in quantum electrodynamics. However, the value of primary concern
here is not this instinctive value but the value of $s$ used for conscious
mathematical calculations in quantum mechanics.

The value most commonly chosen in this context is also $s = 1$. A division
of the world into observer and observed is the foundation for
the quantum theory of measurement used by Heisenberg \cite{[{}] [{, p.\ 58.}]
Heisenberg1930}, von Neumann \cite{vonNeumann1955}, and many others. As Zeh has
noted \cite{Zeh2003ch2, Zeh1999}, it is never strictly necessary to introduce
any other subsystems.

However, this choice leaves one open to a charge of \emph{solipsism}, because
each of us has experiences that include encounters with other human beings.
Choosing $s = 1$ confines each observer to a hermitage, within which the
experiences of others are treated only as inferences, never as primary data.
Different observers will therefore always base their worldviews on mutually
exclusive subsystem decompositions. This leads inevitably to Wheeler's question
\cite{Wheeler1979, Wheeler1983b}:
\begin{quote}
What keeps these images of something ``out there''  from degenerating
into separate and private universes: one observer, one universe; another
observer, another universe?
\end{quote}
Wheeler's answers are cryptic but instructive:
\begin{quote}
That is prevented by the very solidity of those iron posts, the elementary acts
of observership-participancy.

That is the importance of Bohr's point that no observation is an observation
unless we can communicate the results of that observation to others in plain
language. \cite{Bohr1958}
\end{quote}
Translated from Wheelerian poetry into the language of the present theory, the
first answer says that although different observers always have different
experiences, these experiences---the iron posts---are not affected by the choice
of $s$. A subsystem decomposition is just a tool observers use to draw
inferences from their experiences. The only thing affected by the choice of
$s$ is the set of inferences. Choosing $s > 1$ facilitates intersubjective
agreement by allowing multiple observers to have equal status within the theory.

The second answer says that if we are to believe that the image of an outside
world depicted by these inferences is anything more than a mirage or a
hallucination, the essential features of this image must be independent of the
value of $s$ used to perform the inferences. Indeed, we can take the set of
those features that are robust under increases of $s$ as the
\emph{definition} of what is objective. Here the word ``objective'' is used in
roughly the sense defined by Primas \cite[p.\ 352]{Primas1983}:
\begin{quote}
That is, objectivity can never mean anything else but \emph{conditional
intersubjective agreement}, conditioned by a jointly accepted context. The
criterion of objectivity is that the perceptions can be shared.
\end{quote}
The idea that ``objective'' properties are necessarily contextual dates back at
least to Bohr's thoughts on complementarity \cite{Bohr1928, Bohr1958}. Here the
``jointly accepted context'' is the entire structure of the present theory of
information. The ``perceptions [that] can be shared'' are the common features of
the worldview that emerges in the limit of many observers. Communication between
observers therefore plays a central role in establishing which elements of the
theory are meaningful \cite{Wheeler1979, Wheeler1983b, Wheeler1986b,
Wheeler1988}.

To flesh out these answers, it is necessary to explain how the theory works when
$s > 1$. The most egalitarian approach would be to assign all probabilities
as a team. A minimum requirement for forming such a team is to get all teammates
to agree to assign unit probability to the hypothesis that ``you are a being
very much like myself, with your own private experience'' \cite{Mermin2014d}.
Prior probabilities are assigned by team consensus and can be regarded as an
expression of gambling commitments by the team as a whole. Inferences drawn from
such calculations can then be used as the basis for group decisions.

Of course, this egalitarian approach is somewhat in tension with the
subconscious inferences that define each observer's intuitive worldview. Members
of such a team might still be wise to make \emph{personal} decisions by
assigning less than unit probability to the hypotheses that the other teammates
always tell the truth and never make mistakes. If the reported experiences of
these teammates are weighted accordingly, the resulting monarchical structure is
no longer fully egalitarian. However, it is not solipsist either, as long as
these weights are nonzero.

This type of monarchy is the closest point of approach between QBism and the
present theory. QBism requires quantum mechanics to be a ``single-user theory''
\cite{Fuchs2010, FuchsSchack2013, FuchsMerminSchack2014, Mermin2014d} but also
emphasizes that communications with others can (and should) form part
of the basis for single-user quantum state assignments. The two theories are not
directly comparable because the words ``user'' (in QBism) and ``observer'' (in
this paper) have different mathematical implications. In particular, the
single-user theory of the QBist is not equivalent to choosing $s = 1$ in the
present theory, because according to the principles of Sec.\
\ref{sec:Bayesian_inference} (which are not part of QBism), the choice $s =
1$ does not allow the experiences of more than one observer to be included in
any way.

It should also be noted that the QBist arguments for requiring
quantum mechanics to be a single-user theory do not rule out the possibility
of choosing $s > 1$ in the present theory. The basic argument is that ``my
internal personal experience is inaccessible to you except insofar as I attempt
to represent it to you verbally, and vice-versa''~\cite{Mermin2014d}. But this
is irrelevant here, because neither classical nor quantum Bayesian theory has any
direct contact with internal personal experience; it can only deal with what can
be distilled out of that experience and represented \emph{mathematically}
\footnote{Private experiences may influence the choice of prior probabilities in
a way that cannot be described mathematically. However, it is assumed here that
a precondition for forming a team is consensus on the method of defining prior
probabilities.}. But what can be described mathematically can also be
communicated to another person ``in plain language'' \cite{Bohr1958}. So there
is no reason why these mathematical representations of personal experiences
cannot be pooled and used as the basis for group inferences, in the manner
described above.

\subsection{Strong dynamical stability}

\label{sec:strong_dynamical_stability}

As shown in Sec.\ \ref{sec:how_many_subsystems}, the sharpness of Bayesian
inferences can be optimized with respect to the number of environmental
subsystems by choosing $r = 1$. However, given that the environmental
stabilization problem of Sec.\ \ref{sec:backbone} has not yet been solved, it is
not possible at this stage to say whether the resulting inferences would be
sharp enough to be comparable to the predictions of orthodox quantum mechanics.
The situation could be similar to that pointed out by Zurek \cite{Zurek1993a} in
connection with the consistent histories formulation of quantum mechanics
\cite{Griffiths2002, Omnes1999, GellMannHartle2014, Hohenberg2010}, where the
consistency conditions alone are not sufficiently selective to single out
emergent classical behavior.

Given this possibility, it is of interest to consider whether there are any
means available for further sharpening of inferences. The method discussed here
is based on a strengthening of the principle of dynamical stability. The basic
idea is similar to the concept of the ``predictability sieve'' introduced
by Zurek \cite{Zurek1993a, ZurekHabibPaz1993}.

The environmental stabilization problem of Sec.\ \ref{sec:backbone} involves
holding the observer states fixed and maximizing $\chi$ with respect to
variations in the environmental states. The observer states can be chosen
arbitrarily, subject only to the (implicit) requirement of leading to a
mathematically well defined variation problem.

But is this complete freedom of choice necessary? Might it be
possible to narrow down the choices by maximizing $\chi$ with respect to
\emph{all} subsystem states, including those of the observers? The choice of
observer states would then no longer be completely arbitrary, but if this
variation problem has a sufficiently large number of solutions, it may still be
possible to account for the actual experiences of observers. 

For obvious reasons, the variational principle thus defined is called the
\emph{strong} principle of dynamical stability. Subsystem decompositions derived
from this principle are maximally stable, in the sense that they are constrained
solely by the requirement that $\ket{\psi}$ and $\ket{\psi'}$ satisfy the
Schr\"odinger equation. Given that our experience of the world does have a
certain stability, it is at least plausible that this experience is congruent
with strong dynamical stability.

The selective power of strong dynamical stability is most noticeable in a
context where the observer's perception of her state is assumed not to be
infinitely precise. Given that an observer's experience is only capable of
selecting a certain continuous range of states, strong dynamical stability may
be able to narrow the possibilities down to a much smaller (perhaps finite)
subset. This would lead to sharper Bayesian inferences.

\subsection{The rest of the skeleton: Complementarity and ``phenomenon''}

\label{sec:rest_of_skeleton}

Let us now consider the effect of removing the restriction to ideal observers
imposed in Sec.\ \ref{sec:backbone}. What an observer actually perceives is not
the state of her subsystem but some subset of the beables for that subsystem.
The perceived values of these beables may be consistent with more than one state
$\ket{u_k}$, so the state must be \emph{inferred} from the beables. This imposes
another layer of inference, thus making the inferences drawn by a real observer
less sharp than those drawn by an ideal observer.

The reason why all beables are not perceived is probably Darwinian. Survival
requires an efficient mechanism for drawing inferences about the outside world,
and it is most efficient to focus attention on those beables with the greatest
signal-to-noise ratio and ignore the rest. This signal-to-noise ratio can be
greatly enhanced by sense organs, so the set of perceived beables may
vary between organisms with a different evolutionary history. However, the
degree of variation is also constrained by efficiency considerations, since slowly
changing beables are easier to keep track of than rapidly changing ones. All
beables change more slowly under the strong principle of dynamical stability
(Sec.\ \ref{sec:strong_dynamical_stability}), but some still change more slowly
than others. This limits the possible sets of beables that it is worth
developing sense organs for.

The question of which beables are most stable---thus readily seized upon by
Darwinian natural selection---is essentially just the ``preferred basis
problem'' of decoherence theory \cite{Schlosshauer2007}. 
Given that particle interactions are typically functions of number operators in
coordinate space, the expectation values of such operators (see Sec.\
\ref{sec:observables}) will play the most prominent role in a strongly
interacting system. This yields a description similar to the mass-density
beables of dynamical reduction theories \cite{GhirardiPearleRimini1990,
GhirardiGrassiBenatti1995, BassiGhirardi2003}. However, in general one must
consider the dynamics generated by the total Hamiltonian, not just the
interaction Hamiltonian. This typically shifts the arena from coordinate space
to phase space \cite{Schlosshauer2007}.

Bohr's principle of complementarity \cite{Bohr1928, Bohr1958} is an immediate
corollary of these Darwinian restrictions on the subset of perceived beables.
Given that all inferences are performed in the present moment, it is simply 
impossible for an observer to infer definite values for all beables of her
environment from a limited set of her own beables. But the set of sharply
inferred environmental beables may change with her experiences in the present
moment---depending, for example, on which piece of measurement apparatus she is
looking at.

The fact that all subsystem beables are defined in the present theory, but not
all of them play an active role in the worldview of observers, is similar to
ideas used previously in modal interpretations of quantum mechanics. As noted by
Vink \cite{Vink1993} and Bub \cite{Bub1995a, Bub1995b}, in modal
interpretations it is possible to assign definite values to all possible
observables without violating the Bell--Kochen--Specker theorem \cite{Bell1966,
KochenSpecker1967, Mermin1993} for the \emph{measured} values of observables.
The axiomatic foundations of the theory can thus be simplified by allowing the
process of measurement (or, in the present case, Bayesian inference from a
Darwinian subset) to take up some of the burden that would otherwise be
shouldered by a restriction of the allowed beables.

As discussed in Sec.\ \ref{sec:how_many_subsystems}, our conscious experiences
are largely based on subconscious inferences about the structure of our
environment. We seem to perceive a ``real'' three-dimensional butterfly rather
than a pair of two-dimensional butterfly-shaped patterns on our retinas. In
order to match our conscious experiences to the mathematical structure of the
present theory of information, we have to infer the latter description from the
former. This additional layer of (inverse) inference makes the theory even more
complicated. However, such a description is arguably more reasonable than
assuming that we directly perceive the three-dimensional butterfly, which is the
approach used in orthodox quantum mechanics.

This type of automatic subconscious inference (i.e., an inference about the
environment based on a Darwinian-restricted set of observer beables) is an
example of what Bohr and Wheeler call a ``phenomenon'' \cite{Bohr1949,
Wheeler1978, Wheeler1979, Wheeler1980b}. A phenomenon is
subjective in the sense that it depends on the sensory apparatus of the
observer. For example, although the inferred ``blackening of a grain of silver
bromide'' constitutes a phenomenon for us \cite{Wheeler1980b}, it
presumably would not be regarded as such by a Kentucky cave shrimp, which lives
its entire life underground and has no eyes.

\subsection{Comparison with other quantum theories}

\label{sec:compare_other_quantum}

The essence of the theory of information outlined here is its reliance upon
inferences from the properties of a limited number of subsystems in the present
moment. In some ways this is similar to the Everett interpretation
\cite{Everett1957, Everett1973}, which also uses the properties of one subsystem
to make inferences about others and also relies on memories and records in the
present to make inferences about the past. However, Everett defines subsystems
using the traditional tensor product of vector spaces, and he defines the
present to be an instant rather than a moment (i.e., a point on the real line
rather than an interval). 

Because Everett's subsystems are entangled, he introduces the relative-state
concept as a new axiom \cite{vonNeumann1955} that allows him to infer the state
of the outside world from the instantaneous state of the observer. However, this
inference assumes knowledge of the total state $\ket{\psi}$. It is not clear how
such an instantaneous inference scheme can get off the ground if the observer
starts in a state of ignorance of $\ket{\psi}$. That is, it is not clear why an
observer using such a scheme would have any justification for believing that
knowledge of his own present state says anything about the state of his
environment.

By contrast, the subsystems in the present theory are defined to be unentangled
quantum objects, and inferences are based on an infinitesimal time interval
rather than a single point in time. The information about dynamics contained in
this interval then allows an observer to infer something about the state of the
environment even when the observer has no prior information about the
environment. Of course, it is not clear whether this scheme can get off the
ground either; at this stage, it is only a conjecture that the solution of the
environmental stabilization problem (in either the original form of Sec.\
\ref{sec:backbone} or the strong form of Sec.\
\ref{sec:strong_dynamical_stability}) will provide inferences sharp enough to
say anything significant about the environment. However, it is reasonable to
assume that inferences from a moment in time would be considerably sharper than
instantaneous inferences.

Bell has said that the really novel element in Everett's theory is ``a
repudiation of the concept of the `past','' although he acknowledges that this
interpretation might not be accepted by Everett \cite{Bell1981}. Bell did not
endorse this interpretation, which Kucha{\v r} has described as ``solipsism of
an instant'' \footnote{See Kucha{\v r}'s discussion with Page on p.\ 296 of
Ref.\ \cite{Page1994}.}.  For the reasons described above, this label
might indeed be warranted for a truly instantaneous inference scheme.

However, the momentary inference scheme adopted here does not require a
wholesale repudiation of the concept of the past. Rather, it places limits on
what can be said about the past. It remains to be seen whether these limits are
in quantitative agreement with the experiences of observers.

This approach also places limits on what can be said about the future. As noted
near the end of Sec.\ \ref{sec:backbone}, the contrast between an observer's
inferences about the future and her experiences in successive present moments
has an effect roughly comparable to the quantum jumps of orthodox quantum
theory.

The definition of subsystems used by Everett is the same as that used in nearly
all other formulations of quantum mechanics. The vector spaces entering into the
tensor product have no dynamics and can be chosen arbitrarily at different
times. Because the resulting subsystem states are not objects (see Sec.\
\ref{sec:intro_define_subsystems}), it is meaningless to compare their
observable properties directly with those of the present theory. The subsystems
in the two theories simply refer to different things. Given this qualitative
difference, perhaps the only meaningful test of the present theory would be a
direct comparison between theory and experiment.

\subsection{Is anything missing?}

When one is contemplating the possible outcomes of such a comparison, two other
qualitative differences between this theory and standard quantum mechanics stand
out. One is that the quasiclassical subsystems used here have indefinite
particle numbers (see Sec.\ \ref{sec:invertible_important}). Therefore, they
cannot directly replicate standard textbook problems for distinguishable
subsystems with definite particle numbers. Instead, these subsystems seem to
enforce a compliance with Bohr's emphasis on the ``whole experimental
arrangement'' \cite{Bohr1949}. That is, they would only be able to describe a
hydrogen atom as a part of a larger subsystem, not as a subsystem by itself.

Another difference is the way in which observable (or beable) quantities were
defined in Sec.\ \ref{sec:observables}. This definition is based on what would
conventionally be described as expectation values rather than eigenvalues. The
physical picture is closest to the ``density of stuff'' interpretation
\cite{Bell1990} used in the dynamical reduction models of Ghirardi and others
\cite{GhirardiPearleRimini1990, GhirardiGrassiBenatti1995, BassiGhirardi2003},
but with the continuous ``reduction'' occurring in the subsystem states
$\ket{u_k}$ rather than the total system state $\ket{\psi}$ (see Sec.\
\ref{sec:backbone}).

Taken together, these differences raise doubts as to whether the present theory
would be able to replicate the innumerable successful predictions of orthodox
quantum mechanics. A particular concern is whether there is any element of this
theory comparable to the discrete eigenvalue spectra predicted for the results
of ideal measurements in orthodox quantum mechanics.

However, the predictions of dynamical reduction models are well known to be
experimentally indistinguishable (given the current state of experimental
capabilities) from those of orthodox quantum mechanics \cite{BassiGhirardi2003}.
This shows it is unnecessary to define beables as eigenvalues. Since the
definition of beables used in Sec.\ \ref{sec:observables} is just a
generalization of the one used in dynamical reduction models, it is not
unreasonable to expect it to yield comparable results. Testing this conjecture
is a key challenge for future research.

\section{Conclusions}

\label{sec:conclusions}

This paper arose from the observation that ordinary many-particle quantum
mechanics has a hitherto unnoticed mathematical structure that can be
interpreted as an unentangled subsystem decomposition. This structure relies on
the superposition of different numbers of particles, but it also permits a full
description of the equivalence relation that leads to a particle-number
superselection rule in orthodox quantum mechanics. The goal of the paper was to
take this structure seriously and see what it leads to. Can the elimination of
entanglement between subsystems help to resolve some of the conceptual
difficulties at the heart of quantum mechanics?

To build on this foundation, one must link the subsystem states to the
experiences of observers. The first step is the definition of time as a
functional of subsystem changes. This functional can then be embedded into a
dynamical stability functional that describes the subsystem dynamics. The
resulting subsystem decompositions change by the smallest amount consistent with
Schr\"odinger dynamics for the total system. This change is deterministic.

The observable or ``beable'' properties of subsystems are defined as expectation
values of the conventional operators in many-particle quantum mechanics. These
beables include, but are not limited to, the mass density functions used in
dynamical reduction theories.  An observer could in principle experience
any beable associated with her subsystem.  However, for Darwinian reasons
it is assumed that only those beables with the greatest signal-to-noise ratio
are experienced.

An observer's experiences are also limited to the present moment of time. These
experiences are the ``iron posts'' upon which our concept of reality is based.
From them, an observer can infer the existence of an outside world together with
a past and future of the present moment. These inferences are extremely useful
tools that are indispensable for us to ``orient ourselves in the labyrinth of
sense impressions,'' but they always remain an ``arbitrary creation of the human
(or animal) mind'' \cite{Einstein1936, Einstein1950, Einstein1954}. The
resulting image of the world is thus unavoidably subjective. Objectivity emerges
only in the limit of many observers.

Much work remains to be done in order to fill in the details of
the theory of information outlined here.  The most important
task is to solve the variation problem in which an environmental
subsystem is used to stabilize the dynamics of the observer 
subsystems in the present moment.  Only then will it be possible
to see whether the inferences derived in this way are in sufficiently 
close agreement with experience to be useful.

\begin{acknowledgments}
I wish to thank Mike Burt for many stimulating conversations over the years, as
well as Yik Man Chiang and Avery Ching for guidance on the theory of
several complex variables. I also thank Gerald Bastard, Shengwang Du,
Brian Ridley, Ping Sheng, Henry Tye, and Yi Wang for helpful comments on an
earlier version of the manuscript.
\end{acknowledgments}

\appendix

\section{Associativity of the \texorpdfstring{$\psi$}{psi} product}

\label{app:associative}

The $\psi$ product (\ref{eq:psi_product}) is associative if
\begin{equation}
(\ket{u^{(k)}} \odot \ket{v^{(l)}}) \odot \ket{w^{(m)}} = \ket{u^{(k)}} \odot
(\ket{v^{(l)}} \odot \ket{w^{(m)}}) \label{eq:associative1}
\end{equation}
for all $\ket{u^{(k)}} \in S (\mathcal{H}^{k})$, $\ket{v^{(l)}} \in S
(\mathcal{H}^{l})$, and $\ket{w^{(m)}} \in S (\mathcal{H}^{m})$. The explicit
form of the associativity condition given by the definition
(\ref{eq:psi_product}) is
\begin{multline}
c(k, l) c(k+l, m) S [ S (\ket{u^{(k)}} \otimes \ket{v^{(l)}}) \otimes
\ket{w^{(m)}} ] \\ = c(k, l+m) c(l, m) S [ \ket{u^{(k)}} \otimes S (\ket{v^{(l)}}
\otimes \ket{w^{(m)}}) ] .  \label{eq:associative2}
\end{multline}
This appendix examines the implications of this constraint for the function $c(k, l)$.

Now it is well known that \cite{Szekeres2004,
Abraham1988, KostrikinManin1989, LoomisSternberg1990}
\begin{equation}
S ( \ket{u} \otimes \ket{v} ) = S ( \ket{u} \otimes S \ket{v} ) = S ( S \ket{u}
\otimes \ket{v} )
\end{equation}
for all $\ket{u} \in \mathcal{H}^{p}$ and $\ket{v} \in \mathcal{H}^{q}$.
Equation (\ref{eq:associative2}) consequently
reduce to the condition
\begin{equation}
c(k, l) c(k+l, m) = c(k, l+m) c(l, m) ,  \label{eq:associative3}
\end{equation}
which must be satisfied in order for the $\psi$ product to be associative.

To solve this equation, it is convenient to follow the procedure suggested on
p.\ 401 of Ref.\ \cite{Abraham1988}. Define a function $f (l)$ recursively by
the relation
\begin{equation}
f (l + 1) = c(1, l) f (l) . \label{eq:psi_def}
\end{equation}
In the special case $k = 1$, Eq.\ (\ref{eq:associative3}) gives
\begin{equation}
\frac{c(l+1, m)}{c(l, m)} = \frac{c(1, l+m)}{c(1, l)} = \frac{f (l+m+1) f
(l)}{f (l+m) f (l+1)} , \label{eq:c_ratio_psi}
\end{equation}
in which $c(1, l) = f (l+1) / f (l)$ was used in the last step. But this
is just a special case of a more general relation
\begin{equation}
\frac{c(l+k, m)}{c(l, m)} = \frac{f (l+m+k) f
(l)}{f (l+m) f (l+k)} , \label{eq:induction}
\end{equation}
which can be proved by mathematical induction.
In the identity
\begin{equation}
\frac{c(l+k+1, m)}{c(l, m)} = \frac{c(l+k+1, m)}{c(l+k, m)} \frac{c(l+k, m)}{c(l, m)},
\end{equation}
one can replace the first term using Eq.\ (\ref{eq:c_ratio_psi}) (with $l
\rightarrow l + k$) and the second term using Eq.\ (\ref{eq:induction}).
This gives
\begin{align}
\frac{c(l+k+1, m)}{c(l, m)} & = \frac{f (l+k+m+1) f (l+k)}{f (l+k+m)
f (l+k+1)} \nonumber \\ & \quad \times \frac{f (l+m+k) f (l)}{f (l+m) f
(l+k)} \nonumber \\ & = \frac{f (l+m+k+1) f (l)}{f (l+m) f (l+k+1)} ,
\end{align}
which shows that Eq.\ (\ref{eq:induction}) holds for $k+1$ whenever it holds
for $k$. The initial condition (\ref{eq:c_ratio_psi}) then establishes that
(\ref{eq:induction}) holds for all $k \ge 1$. (Of course, it is trivially valid
when $k = 0$ too.)

Substituting $k = q - l$ in Eq.\ (\ref{eq:induction}) then gives
\begin{equation}
\frac{c(q, m)}{c(l, m)} = \frac{f (q+m) f
(l)}{f (l+m) f (q)} , 
\end{equation}
which becomes
\begin{equation}
\frac{c(q, m)}{c(1, m)} = \frac{f (q+m) f
(1)}{f (m+1) f (q)}
\end{equation}
when $l = 1$. Replacing $c(1, m) = f (m+1) / f (m)$ then yields the
desired solution
\begin{equation}
c(q, m) = \frac{f (q+m) f (1)}{f (q) f (m)} , \label{eq:cqm}
\end{equation}
showing that $c(q, m) = c(m, q)$.

When $m = 0$, this reduces to
\begin{equation}
c(q, 0) = c(0, q) = \frac{f (1)}{f (0)} ,
\end{equation}
which is independent of $q$. The value of $c(q, 0)$ can be fixed by
requiring the vacuum state $\ket{0}$ to serve as a multiplicative identity for
the $\psi$ product [cf.\ Eq.\ (\ref{eq:mult_ident})]:
\begin{equation}
\ket{0} \odot \ket{\Psi} = \ket{\Psi} \odot \ket{0} = \ket{\Psi} \qquad
\forall \ket{\Psi} \in \mathcal{F}_{s} (\mathcal{H})  .
\end{equation}
Imposing this condition when $\ket{\Psi} = \ket{u^{(q)}}$ gives $c(q, 0) = 1$ or
$f (0) = f (1)$. The result $\abs{f (0)} = \abs{f (1)}$ can also be derived from
cluster decomposition, as shown in Appendix
\ref{app:cluster_decomposition}.

The recursive definition (\ref{eq:psi_def}) of $f (l)$ leaves one value of
$f (l)$ that can be chosen arbitrarily. It is convenient to choose $f (1)
= 1$, thus reducing Eq.\ (\ref{eq:cqm}) to Eq.\ (\ref{eq:cf}), which was
to be proved.

\section{Cluster decomposition theorem}

\label{app:cluster_decomposition}

This appendix contains a proof of Theorem \ref{thm:cluster}, which is about the
cluster decomposition property $\inprod{st}{uv} = \inprod{s}{u} \inprod{t}{v}$
of the inner product in $\mathcal{F}_{s} (\mathcal{H})$. This inner product is
derived from the inner product in $\mathcal{F} (\mathcal{H})$ and the definition
of the $\psi$ product in Eqs.\ (\ref{eq:psi_product}) and (\ref{eq:cf}).

To define the inner product in $\mathcal{F} (\mathcal{H})$, let
\begin{equation}
\pket{\alpha_1 \alpha_2 \cdots \alpha_n} = \ket{\alpha_1} \otimes \ket{\alpha_2}
\otimes \cdots \otimes \ket{\alpha_n}
\end{equation}
denote a general tensor-product state in $\mathcal{H}^{n}$, where
$\ket{\alpha_{k}} \in \mathcal{H}$. The set $\{ \ket{\alpha_{k}} \}$ is not
assumed to be linearly independent or normalized. The rounded bracket on the
ket $\pket{\alpha_1 \alpha_2 \cdots \alpha_n}$ distinguishes this
unsymmetrized tensor product from the symmetrized product $\ket{\alpha_1 \alpha_2
\cdots \alpha_n}$ defined in Eqs.\ (\ref{eq:ket_general}) and
(\ref{eq:ket_relation}).

The inner product of two such tensor products is defined in the usual way
as \cite{CohTan1977}
\begin{equation}
\newinprod{\alpha_1 \cdots \alpha_n}{\beta_1 \cdots \beta_n} =
\inprod{\alpha_1}{\beta_1} \cdots
\inprod{\alpha_n}{\beta_n} ,
\end{equation}
where $\inprod{\alpha_k}{\beta_k}$ is the inner product in $\mathcal{H}$. Now
let $\{ \ket{e_i} \}$ be an orthonormal basis in $\mathcal{H}$. The
corresponding tensor products
\begin{equation}
\pket{e_{i_1} \cdots e_{i_n}} = \ket{e_{i_1}}
\otimes \cdots \otimes \ket{e_{i_n}} \label{eq:ebasis}
\end{equation}
therefore form an orthonormal basis in $\mathcal{H}^{n}$, since
\begin{equation}
\newinprod{e_{i_1} \cdots e_{i_n}}{e_{j_1} \cdots e_{j_n}} = \delta_{i_1 j_1}
\cdots \delta_{i_n j_n} .
\end{equation}

These elementary results can now be used to evaluate the inner product
$\inprod{s t}{u v}$ of the vectors $\ket{s t} = \ket{s} \odot \ket{t}$ and
$\ket{u v} = \ket{u} \odot \ket{v}$ in Theorem \ref{thm:cluster}. The kets
$\ket{s} \in \mathcal{F}_{s} (\mathcal{H}_1)$ and $\ket{t} \in \mathcal{F}_{s}
(\mathcal{H}_2)$ are expanded as
\begin{equation}
\ket{s} = \sum_{k} \ket{s^{(k)}} , \qquad \ket{t} = \sum_{l} \ket{t^{(l)}} ,
\end{equation}
in which $\ket{s^{(k)}} \in S (\mathcal{H}^{k})$ and $\ket{t^{(l)}} \in S
(\mathcal{H}^{l})$. Hence
\begin{equation}
\ket{s t} = \sum_{k} \sum_{l} \ket{s^{(k)}} \odot \ket{t^{(l)}} = \sum_{k}
\sum_{l} \ket{s^{(k)} t^{(l)}} ,
\end{equation}
in which $\ket{s^{(k)} t^{(l)}} \in S (\mathcal{H}^{k+l})$. With a similar
expansion for $\ket{u v}$, we can write
\begin{equation}
\inprod{s t}{u v} = \sum_{kl} \sum_{mn} \inprod{s^{(k)} t^{(l)}}{u^{(m)}
v^{(n)}} . \label{eq:stuvklmn}
\end{equation}
Here $\inprod{s^{(k)} t^{(l)}}{u^{(m)} v^{(n)}}$ is zero
unless $k+l = m+n$, since states with different numbers of particles are
orthogonal.

But due to the orthogonality of the subspaces $\mathcal{H}_1$ and
$\mathcal{H}_2$, a stronger restriction is possible: $\inprod{s^{(k)}
t^{(l)}}{u^{(m)} v^{(n)}}$ is zero unless $k = m$ and $l = n$. Thus
\begin{equation}
\inprod{s t}{u v} = \sum_{kl} \inprod{s^{(k)} t^{(l)}}{u^{(k)}
v^{(l)}} . \label{eq:stuvkl}
\end{equation}
The definition (\ref{eq:psi_product}) of the $\psi$ product can now be used
to write
\begin{equation}
\ket{u^{(k)} v^{(l)}} = c(k, l) S (\ket{u^{(k)}} \otimes \ket{v^{(l)}}) .
\end{equation}
Here associativity requires $c(k, l)$ to have the form derived in Appendix
\ref{app:associative}:
\begin{equation}
c(k, l) = \frac{f(k+l)}{f(k) f(l)} , \qquad f(1) = 1 , \label{eq:ckl}
\end{equation}
but the function $f(k)$ has not yet been determined.

The inner product (\ref{eq:stuvkl}) is therefore
\begin{equation}
\inprod{s t}{u v} = \sum_{kl} \abs{c(k, l)}^2 (\bra{s^{(k)}} \otimes
\bra{t^{(l)}}) S^{\dagger} S (\ket{u^{(k)}} \otimes \ket{v^{(l)}}) .
\end{equation}
Since $S$ is an orthogonal projector, $S^{\dagger} S = S^2 = S$, thus
\begin{equation}
\inprod{s t}{u v} = \sum_{kl} \abs{c(k, l)}^2 (\bra{s^{(k)}} \otimes
\bra{t^{(l)}}) S (\ket{u^{(k)}} \otimes \ket{v^{(l)}}) . \label{eq:stuv_expand}
\end{equation}
Here $(\ket{u^{(k)}} \otimes \ket{v^{(l)}}) \in \mathcal{H}^{k+l}$, so the
definition of $S$ gives
\begin{equation}
S (\ket{u^{(k)}} \otimes \ket{v^{(l)}}) = \frac{1}{(k+l)!} \sum_{\sigma}
\varepsilon (\sigma) \sigma (\ket{u^{(k)}} \otimes \ket{v^{(l)}}) . \label{eq:Skl}
\end{equation}

Now if the product $(\ket{u^{(k)}} \otimes \ket{v^{(l)}})$ is expanded in the
unsymmetrized basis of Eq.\ (\ref{eq:ebasis}) (with $n = k + l$), the first
$k$ vectors $\ket{e_i}$ will be from $\mathcal{H}_1$ and the last $l$ vectors
will be from $\mathcal{H}_2$. The same is true for the product $(\ket{s^{(k)}}
\otimes \ket{t^{(l)}})$. The only permutations $\sigma$ in equation
(\ref{eq:Skl}) that contribute nonvanishing terms to equation
(\ref{eq:stuv_expand}) are therefore those that do not interchange any of the
first $k$ vectors with the last $l$ vectors. These permutations are of the form
$\sigma = \sigma_1 \sigma_2$, where $\sigma_1$ is any permutation of the first
$k$ vectors and $\sigma_2$ is any permutation of the last $l$ vectors.

The orthogonality of the basis vectors (\ref{eq:ebasis}) therefore reduces Eq.\ (\ref{eq:stuv_expand}) to
\begin{multline}
\inprod{s t}{u v} = \sum_{kl} \frac{\abs{c(k, l)}^2}{(k + l)!} \sum_{\sigma_1}
\varepsilon (\sigma_1) \matelm{s^{(k)}}{\sigma_1}{u^{(k)}} \\ \times
\sum_{\sigma_2} \varepsilon (\sigma_2) \matelm{t^{(l)}}{\sigma_2}{v^{(l)}} ,
\label{eq:stuv_expand2}
\end{multline}
in which $\varepsilon (\sigma_1 \sigma_2) = \varepsilon (\sigma_1) \varepsilon
(\sigma_2)$ was used. Now $\ket{u^{(k)}} \in S(\mathcal{H}^{k})$, so $\sigma_1
\ket{u^{(k)}} = \varepsilon (\sigma_1) \ket{u^{(k)}}$, and likewise $\sigma_2
\ket{v^{(l)}} = \varepsilon (\sigma_2) \ket{v^{(l)}}$. But $[\varepsilon
(\sigma)]^2 = 1$ for any $\sigma$, hence
\begin{align}
\inprod{s t}{u v} & = \sum_{kl} \frac{\abs{c(k, l)}^2}{(k + l)!} \sum_{\sigma_1}
\inprod{s^{(k)}}{u^{(k)}} \sum_{\sigma_2} \inprod{t^{(l)}}{v^{(l)}} \\ & =
\sum_{kl} \babs{\frac{f(k+l)}{f(k) f(l)}}^2 \frac{k! l!}{(k + l)!}
\inprod{s^{(k)}}{u^{(k)}} \inprod{t^{(l)}}{v^{(l)}} . \label{eq:stuv_expand3}
\end{align}

At this point, choosing $\abs{f(k)} = \sqrt{k!}$ eliminates all of the numerical
factors, yielding
\begin{equation}
\inprod{s t}{u v} = \sum_{k} \inprod{s^{(k)}}{u^{(k)}} \sum_{l}
\inprod{t^{(l)}}{v^{(l)}} .
\end{equation}
Each factor can be rewritten as
\begin{equation}
\sum_{k} \inprod{s^{(k)}}{u^{(k)}} = \sum_{k} \sum_{m} \inprod{s^{(k)}}{u^{(m)}}
= \inprod{s}{u} ,
\end{equation}
since $\inprod{s^{(k)}}{u^{(m)}} = 0$ when $k \ne m$. Therefore $\inprod{s t}{u
v} = \inprod{s}{u} \inprod{t}{v}$, thus proving the ``if'' part of the cluster
decomposition theorem \ref{thm:cluster}.

Conversely, suppose it is given that $\inprod{s t}{u v} = \inprod{s}{u}
\inprod{t}{v}$ for all $\ket{s}, \ket{u} \in \mathcal{F}_{s} (\mathcal{H}_1)$
and all $\ket{t}, \ket{v} \in \mathcal{F}_{s} (\mathcal{H}_2)$, where
$\mathcal{H}_1$ and $\mathcal{H}_2$ are orthogonal subspaces of $\mathcal{H}$.
What does this tell us about $f(k)$?

Under the given conditions, we are free to choose $\ket{u} = \ket{u^{(k)}}$ and
$\ket{v} = \ket{v^{(l)}}$ for any values of $k$ and $l$. Equation
(\ref{eq:stuv_expand3}) then reduces to
\begin{equation}
\inprod{s t}{u v} = \babs{\frac{f(k+l)}{f(k) f(l)}}^2 \frac{k! l!}{(k + l)!}
\inprod{s^{(k)}}{u^{(k)}} \inprod{t^{(l)}}{v^{(l)}} ,
\end{equation}
with no summation on $k$ and $l$.  Likewise
\begin{equation}
\inprod{s}{u} \inprod{t}{v} = \inprod{s^{(k)}}{u^{(k)}} \inprod{t^{(l)}}{v^{(l)}} .
\end{equation}
Hence, $\inprod{s t}{u v} = \inprod{s}{u}
\inprod{t}{v}$ in all such cases only if
\begin{equation}
\babs{\frac{f(k+l)}{f(k) f(l)}}^2 =
\frac{(k + l)!}{k! l!} \qquad \forall k, l \ge 0 .
\end{equation}
Setting $l = 1$ and using $f(1) = 1$ then gives
\begin{equation}
\abs{f(k+1)}^2 = (k + 1) \abs{f(k)}^2 .
\end{equation}
This defines $\abs{f(k)}^2$ recursively as
\begin{equation}
\abs{f(k)}^2 = k! \abs{f(1)}^2 = k! \qquad (k \ge 1) ,
\end{equation}
while  $k = 0$ gives $\abs{f(1)}^2 = \abs{f(0)}^2$. Hence,
$\abs{f(k)} = \sqrt{k!}$ for all $k$, thus completing the proof of Theorem
\ref{thm:cluster}.

\section{Algebraic closure conditions for bosons}

\label{app:rigged}

This appendix describes the construction of the boson vector space
$\mathcal{F}_{\psi} (\mathcal{H}_{\rmb})$ mentioned at the end of Sec.\
\ref{sec:creation}. The objective is to identify a subspace of $\mathcal{F}_{s}
(\mathcal{H}_{\rmb})$ that is a good match for the algebra of the boson $\psi$
product. This algebra is easiest to describe using the Segal--Bargmann
representation of Fock space \cite{Bargmann1961, Schweber1962, Folland1989,
Hall2000}, which is often used in the definition of coherent boson states
\cite{Glauber1963c, KlauderSkagerstam1985, NegeleOrland1998}.

Let us start by establishing a concise notation. Standard basis kets in Fock
space are written as $\ket{n}$, where $n = (n_1, n_2, \ldots, n_b) \in
\mathbb{N}^b$ is a vector of nonnegative integers, $b$ is the dimension of the
single-boson Hilbert space $\mathcal{H}_{\rmb}$, and $n_i$ is the number of
bosons in the single-particle state $i$. In multi-index notation
\cite{ReedSimonVol1, Bargmann1961}, powers of a complex vector $z =
(z_1, \ldots, z_b) \in \mathbb{C}^b$ are written as $z^n = z_1^{n_1} \cdots
z_b^{n_b}$, and likewise for powers of the vector $a^{\dagger} = (a_1^{\dagger},
\ldots, a_b^{\dagger})$ of boson creation operators. This allows the normalized
basis ket $\ket{n}$ to be written simply as $\ket{n} = (n !)^{-1/2}
(a^{\dagger})^n \ket{0}$ [see Eqs.\ (\ref{eq:normalization}) and
(\ref{eq:create_vacuum})], where $n! = n_1 ! \cdots n_b !$.

A general vector in $\mathcal{F}_{s} (\mathcal{H}_{\rmb})$ has
the form $\ket{f} = \sum_{n} c_n \ket{n}$, where $c_n = \inprod{n}{f}$. The Fock
space $\mathcal{F}_{s} (\mathcal{H}_{\rmb})$ is also required to be a Hilbert
space, the members of which must satisfy $\norm{f} < \infty$, in which
$\norm{f}^2 = \inprod{f}{f} = \sum_{n} \abs{c_n}^2$. For reasons to be explained
below, the Hilbert-Fock space defined in this way is also written as
$\mathcal{H}_1$.

The Segal--Bargmann representation of $\ket{f}$ is defined by the
expression [cf.\ Eq.\ (\ref{eq:uU0})]
\begin{equation}
\ket{f} = F (a^{\dagger}) \ket{0}  ,
\end{equation}
in which the function $F(z)$ is defined by the power series
\begin{equation}
F (z) = \sum_{n} \frac{c_n}{\sqrt{n !}} z^n .
\end{equation}
If $\norm{f} < \infty$, $F(z) \in \mathbb{C}$ is an entire holomorphic function
of $z = x + i y$, where $x, y \in \mathbb{R}^b$. This representation of ket
vectors by entire functions is a powerful advantage of the Segal--Bargmann
theory.

In the Segal--Bargmann representation, the inner product of two vectors
$\ket{f}$ and $\ket{g}$ is given by the integral
\begin{equation}
\inprod{f}{g} = \int F^*(z) G(z) \rho (z) \, \dzarea , \label{eq:Bargmann_ip}
\end{equation}
in which $\rho (z) = \pi^{-b} \exp (- \abs{z}^2)$, $\abs{z}^2 = \abs{z_1}^2 +
\cdots + \abs{z_b}^2$, and $\dzarea = \rmd x_1 \, \rmd y_1 \cdots \rmd x_b \,
\rmd y_b$. Functions of the form $F(z) = \exp (\frac12 \gamma z^2 + \alpha \cdot
z)$ are of special interest, where $\gamma \in \mathbb{C}$, $\alpha \in
\mathbb{C}^b$, $\alpha \cdot z = \alpha_1 z_1 + \cdots + \alpha_b z_b$, and $z^2
= z \cdot z$. This function is normalizable (i.e., $\norm{f} < \infty$) if and
only if $\abs{\gamma}^2 < 1$ \cite{Bargmann1961}. A general bound on all
normalizable states is given by the Schwarz inequality \cite{Bargmann1961}:
\begin{equation}
\abs{F(z)} \le \exp (\tfrac12 \abs{z}^2) \norm{f} \qquad
\forall z \in \mathbb{C}^b . \label{eq:Schwarz_Fz}
\end{equation}
This implies that the Hilbert-Fock space $\mathcal{H}_1 = \mathcal{F}_{s}
(\mathcal{H}_{\rmb})$ is a poor match for the algebra of the $\psi$ product,
since $\ket{f} \odot \ket{g}$ is represented by the product $F(z) G(z)$ [cf.\
Eq.\ (\ref{eq:uvUV})]. For example, the product of $F(z) = \exp (\frac12 \gamma
z^2 + \alpha \cdot z)$ and $G(z) = \exp (\frac12 \delta z^2 + \beta \cdot z)$ is
normalizable if and only if $\abs{\gamma + \delta}^2 < 1$, but this condition is
violated by many pairs of states with $\abs{\gamma}^2 < 1$ and $\abs{\delta}^2 <
1$.

To find a suitable vector space for the algebra of the $\psi$ product, it is
helpful to consider the family of vectors $\ket{f_k}$ defined by
\begin{equation}
F_k (z) = F (k z) , \qquad \inprod{n}{f_k} = k^{\abs{n}} \inprod{n}{f} ,
\label{eq:Fk_def}
\end{equation}
in which $k$ is a positive integer (i.e., $k \in \mathbb{N}_{+}$) and $\abs{n}
\equiv n_1 + \cdots + n_b$. One can easily see that $\inprod{f_k}{g_k} =
\inprod{f}{g}_k$, in which $\inprod{f}{g}_k$ denotes the inner product
\begin{equation}
\inprod{f}{g}_k = \int F^*(z) G(z) 
\rho_k (z) \, \dzarea , 
\end{equation}
where $\rho_k (z) = (\pi k^2)^{-b} \exp (- \abs{z}^2 / k^2 )$. This gives rise
to a countable family of norms $\norm{f}_k = (\inprod{f}{f}_k)^{1/2}$; note that
$\norm{f}_1$ is the same as the norm $\norm{f}$ defined by the inner product
(\ref{eq:Bargmann_ip}). The set of vectors with $\norm{f}_k < \infty$ forms a
Hilbert space, which is denoted $\mathcal{H}_k$. The statement $\ket{f} \in
\mathcal{H}_k$ is the same as $\ket{f_k} \in \mathcal{H}_1$.

According to Eqs.\ (\ref{eq:Schwarz_Fz}) and (\ref{eq:Fk_def}), all vectors in
$\mathcal{H}_k$ must satisfy
\begin{equation}
\abs{F(z)} \le \exp (\abs{z}^2 / 2 k^2) \norm{f}  \qquad
\forall z \in \mathbb{C}^b . \label{eq:Schwarz_Fz_k}
\end{equation}
Conversely, to show that $\ket{f} \in \mathcal{H}_k$, it is sufficient to find
numbers $0 \le A < \infty$ and $0 \le \lambda < 1$ such that \cite{Bargmann1961}
\begin{equation}
\abs{F (z)} \le A \exp (\lambda \abs{z}^2 / 2 k^2) \qquad 
\forall z \in \mathbb{C}^b . \label{eq:F_in_Hk}
\end{equation}
From these results it is easy to see that
\begin{equation}
\mathcal{H}_{k+1} \subset \mathcal{H}_k ,
\end{equation}
since Eq.\ (\ref{eq:Schwarz_Fz_k}) with $k \to k + 1$ yields an inequality of
the type (\ref{eq:F_in_Hk}), with $A = \norm{f}$ and $\lambda = k^2 / (k+1)^2$.

Let us now define a vector space $\mathcal{F}_{\psi} = \mathcal{F}_{\psi}
(\mathcal{H}_{\rmb})$ as the intersection of the Hilbert spaces $\mathcal{H}_k$
for all $k \in \mathbb{N}_{+}$. $\mathcal{F}_{\psi}$ is defined by the countable
family of norms $\{ \norm{f}_k \}$, but it cannot be defined by any single norm.
This space is therefore a Fr\'echet space \cite{ReedSimonVol1, Horvath2012},
not a Hilbert space. Such vector spaces are familiar from the rigged Hilbert
space formalism of quantum mechanics
\cite{Antoine1998, Antoine2009b1, Bohm1993, Bohm1978,
Bogolubov1975, Ballentine2015},
which can be used to provide a rigorous justification for the Dirac bra-ket
formalism.

The subscript on $\mathcal{F}_{\psi}$ is intended to suggest that this space is
a suitable arena for the algebra of the $\psi$ product. To show this, we need to
prove that $\ket{h} = \ket{f} \odot \ket{g}$ belongs to $\mathcal{F}_{\psi}$
whenever $\ket{f}$ and $\ket{g}$ do. In other words, we must show that $\ket{h}
\in \mathcal{H}_k$ for all $k \in \mathbb{N}_{+}$ whenever $\ket{f} \in
\mathcal{H}_q$ and $\ket{g} \in \mathcal{H}_q$ for all $q \in \mathbb{N}_{+}$.
But this is easily done, since Eq.\ (\ref{eq:Schwarz_Fz_k}) gives inequalities
$\abs{F(z)} \le \exp (\abs{z}^2 / 2 q^2) \norm{f}$ and $\abs{G(z)} \le \exp
(\abs{z}^2 / 2 q^2) \norm{g}$; the product $H(z) = F(z) G(z)$ thus satisfies
$\abs{H (z)} \le A \exp (\lambda \abs{z}^2 / 2 k^2)$, where $A = \norm{f}
\norm{g}$ and $\lambda = 2 k^2 / q^2$. According to Eq.\ (\ref{eq:F_in_Hk}),
this implies that $\ket{h} \in \mathcal{H}_k$ as long as we are free to choose
$\lambda < 1$, i.e., $q > \sqrt{2} k$. But this can be done for any $k \in
\mathbb{N}_{+}$, by the definition of $\mathcal{F}_{\psi}$.

It should be clear from the above derivation that the algebra of the $\psi$
product cannot be accommodated within any vector space defined by a finite
number of norms. Hence, the move from Hilbert space to Fr\'echet space is
necessary for boson systems.

What type of vectors belong to $\mathcal{F}_{\psi}$? It was noted above that
$F(z) = \exp (\frac12 \gamma z^2 + \alpha \cdot z)$ is in $\mathcal{H}_1$ if and
only if $\abs{\gamma}^2 < 1$. However, Eqs.\ (\ref{eq:Schwarz_Fz_k}) and
(\ref{eq:F_in_Hk}) show that it belongs to $\mathcal{F}_{\psi}$ if and only if
$\gamma = 0$. The only exponential functions in $\mathcal{F}_{\psi}$ are
therefore those of the form $F(z) = \exp (\alpha \cdot z)$, for arbitrary
$\alpha \in \mathbb{C}^b$. But these are just the coherent states
\begin{equation}
\ket{\alpha} = \exp (\alpha \cdot a^{\dagger}) \ket{0} , \label{eq:coherent}
\end{equation}
which can be defined as eigenvectors of the boson annihilation operators $a_i$
(i.e., $a_i \ket{\alpha} = \alpha_i \ket{\alpha}$) \cite{Glauber1963c}. Note
that the operator $\exp (\alpha \cdot a^{\dagger})$ in Eq.\ (\ref{eq:coherent})
is easy to invert; its inverse is $\exp (-\alpha \cdot a^{\dagger})$.

Bargmann called the functions $F (z) = \exp (\alpha \cdot z)$ ``principal
vectors'' and showed that they are complete (although not orthogonal), in the
sense that finite linear combinations of them are dense in $\mathcal{H}_1$
\cite{Bargmann1961}. This completeness is usually expressed as the integral
\cite{Klauder1960, Glauber1963c}
\begin{equation}
\frac{1}{\pi^b} \int \outprod{\alpha}{\alpha} 
\exp (- \abs{\alpha}^2)  \, \rmd^{2b} \alpha = 1 .
\end{equation}
The monomials $F(z) = (n!)^{-1/2} z^n$ also form a complete orthonormal basis
\cite{Bargmann1961}, corresponding to the original basis $\ket{n} = (n !)^{-1/2}
(a^{\dagger})^n \ket{0}$ in Fock space.

Finally, note from Eq.\ (\ref{eq:Fk_def}) that if $\ket{f} \in
\mathcal{F}_{\psi}$, then as $\abs{n} \to \infty$, $\inprod{n}{f}$ must decrease
faster than $\exp (- \kappa \abs{n})$ for any positive value of $\kappa$. This
rate of decrease is even faster than that of the sequences of rapid descent
encountered in connection with Schwartz spaces $\mathcal{S}$ \cite{Gelfand1968,
ReedSimonVol1, Simon1971, Bogolubov1975, Bohm1978}.

\section{Different types of particles}

\label{app:different}

Consider a system containing two types of particles, labeled $A$
and $B$. If the corresponding single-particle Hilbert spaces are
$\mathcal{H}_{A}$ and $\mathcal{H}_{B}$, the vector space of the whole 
system can be defined as the tensor-product space
\begin{equation}
\mathcal{G} = \mathcal{F}_{s} (\mathcal{H}_{A}) \otimes \mathcal{F}_{s}
(\mathcal{H}_{B}) . \label{eq:combined_space}
\end{equation}
That is, a general vector $\ket{u} \in \mathcal{G}$ is a linear combination of
tensor products $\ket{u_{A}} \otimes \ket{u_{B}}$, where
$\ket{u_{A}} \in \mathcal{F}_{s} (\mathcal{H}_{A})$ and $\ket{u_{B}} \in
\mathcal{F}_{s} (\mathcal{H}_{B})$.

One can define a $\psi$ product in $\mathcal{G}$ by
letting the $\psi$ product of Sec.\ \ref{sec:psi_product} act in
parallel on the subspaces $\mathcal{F}_{s} (\mathcal{H}_{A})$ and
$\mathcal{F}_{s} (\mathcal{H}_{B})$. That is, the $\psi$ product of two
simple tensor products $\ket{u} = \ket{u_{A}} \otimes \ket{u_{B}}$ and $\ket{v}
= \ket{v_{A}} \otimes \ket{v_{B}}$ is defined to be
\begin{multline}
(\ket{u_{A}} \otimes \ket{u_{B}}) \odot (\ket{v_{A}} \otimes \ket{v_{B}}) \\ =
(\ket{u_{A}} \odot \ket{v_{A}}) \otimes (\ket{u_{B}} \odot \ket{v_{B}}) .
\end{multline}
This is then extended to arbitrary vectors $\ket{u}, \ket{v} \in \mathcal{G}$ by
multilinearity. The algebra thus defined is associative, which follows directly
from the associativity of the $\psi$ product in $\mathcal{F}_{s}
(\mathcal{H}_{A})$ and $\mathcal{F}_{s} (\mathcal{H}_{B})$.

From this definition, it is a simple exercise to show that the cluster
decomposition property of Theorem \ref{thm:cluster} is valid in $\mathcal{G}$ if
it holds in both $\mathcal{F}_{s} (\mathcal{H}_{A})$ and $\mathcal{F}_{s}
(\mathcal{H}_{B})$. The equivalence between the algebra of the $\psi$ product
and the algebra of creation operators discussed in Sec.\ \ref{sec:creation}
likewise remains valid in systems with more than one type of particle.

\section{Invertibility theorem}

\label{app:invertibility_theorem}

The first step in the proof of Theorem \ref{thm:invertible} is to show that a
boson-fermion creator $U : \mathcal{E} \to \mathcal{E}$ is invertible if and
only if its associated boson creator $U_0 : \mathcal{F}_{\psi}
(\mathcal{H}_{\rmb}) \to \mathcal{F}_{\psi} (\mathcal{H}_{\rmb})$ is invertible.
The basic reason for this can be seen intuitively from the matrix of creators
(\ref{eq:U_bf}). The determinant of this triangular matrix is $\det U =
(U_0)^{2^d}$, in which $d = \dim \mathcal{H}_{\rmf}$. If we assume that the
standard theorems of matrix algebra can be extended to this matrix of commuting
operators, then $U^{-1}$ exists if and only if $U_0^{-1}$ exists.

A more explicit argument is as follows. Because $U_0$ is a boson creator, we can
use the multi-index notation of Appendix \ref{app:rigged} to write $U_0 = F
(a^{\dagger})$, in which $F(z)$ is an entire function of $z \in \mathbb{C}^b$
and $b = \dim \mathcal{H}_{\rmb}$. Invertibility of $U_0$ thus requires that $1
/ F(z)$ is also an entire function, or that $F(z) \ne 0$ for finite $z$. But if
$F(z) = 0$ at $z = \alpha^{*}$, the coherent state $\ket{\alpha}$ in Eq.\
(\ref{eq:coherent}) is orthogonal to every vector in the image of $U_0$, as
shown below. Hence, $U_0$ cannot be surjective (or onto) in this case. This also
implies that $U$ is not surjective, because $\ket{\alpha} \otimes
\ket{0}_{\rmf}$ is orthogonal to the image of $U$. Invertibility of $U_0$ is
therefore necessary for invertibility of $U$. Its sufficiency follows
immediately from Eqs.\ (\ref{eq:Z}) and (\ref{eq:U_inverse_series}).

To clarify the orthogonality relation mentioned above, let us start by writing
$U_0^{\dagger} = \tilde{F} (a)$, in which $a = (a_1, \ldots, a_b)$ is a vector
of boson annihilation operators and $\tilde{F} (z^{*}) \equiv [F(z)]^{*}$ is an
entire function of $z^{*}$. Given that $F(z) = 0$ at $z = \alpha^{*}$, we have
$\tilde{F} (\alpha) = 0$ and thus $\ket{U_0^{\dagger} \alpha} \equiv
U_0^{\dagger} \ket{\alpha} = \tilde{F} (a) \ket{\alpha} = \tilde{F} (\alpha)
\ket{\alpha} = 0$, because $\ket{\alpha}$ is an eigenket of $a$. For any
$\ket{x} \in \mathcal{F}_{\psi} (\mathcal{H}_{\rmb})$ we then have
$\inprod{U_0^{\dagger} \alpha}{x} = \matelm{\alpha}{U_0}{x} = 0$. But
$\matelm{\alpha}{U_0}{x} = 0$ for all $\ket{x} \in \mathcal{F}_{\psi}
(\mathcal{H}_{\rmb})$ is precisely the statement that $\ket{\alpha}$ is
orthogonal to every vector in the image of $U_0$.

The second step in the proof of Theorem \ref{thm:invertible} is to show that
$U_0$ is invertible if and only if $\ket{u_0}$ is a coherent state. The
starting point is the condition $F(z) \ne 0$ established above. Now it is well
known in the theory of a single complex variable $z \in \mathbb{C}$ that every
entire function $F(z)$ with no zeros can be written as $F(z) = \exp [G (z)]$,
where $G (z)$ is another entire function \cite{[] [{, Part II, p.\ 3.}]
Knopp1996, [] [{, pp.\ 123--124, 376.}] Lang1999, [] [{, pp.\ 27, 206.}]
Forster1981}. This is equivalent to the existence of a global logarithm of such
a function $F(z)$, which depends essentially on whether the domain of $F$ is
simply connected. The single-variable proof given in Ref.\ \cite{Forster1981}
can also be extended to the case of entire functions of several complex
variables $z \in \mathbb{C}^b$ \cite{[] [{, p.\ 248.}] Kaup1983}. In order for
$U_0$ to be invertible, it is therefore necessary that $F (z) = \exp [G (z)]$
for some entire function $G(z)$.

However, according to the results of Appendix \ref{app:rigged}, if $\ket{u_0}
\in \mathcal{F}_{\psi} (\mathcal{H}_{\rmb})$, then $G(z)$ can only be a linear
function of $z$. That is, $F (z)$ must be proportional to $\exp(\alpha \cdot z)$
for some $\alpha \in \mathbb{C}^b$, and $\ket{u_0}$ must be proportional to one
of the coherent states $\ket{\alpha}$ defined in Eq.\ (\ref{eq:coherent}). The
necessity of $\ket{u_0}$ being a coherent state is therefore established.

To demonstrate its sufficiency, we only need to note that the operator $\exp
(\alpha \cdot a^{\dagger})$ appearing in Eq.\ (\ref{eq:coherent}) is invertible,
its inverse being given by $\exp (-\alpha \cdot a^{\dagger})$. Thus, $U_0$ is
invertible whenever $\ket{u_0}$ is a coherent state. This concludes the proof of
Theorem \ref{thm:invertible}.

\section{Creator identities}

\label{app:creator_identities}

A useful identity for the symmetrized product of three creators $A$, 
$B$, and $C$ is 
\begin{equation}
\{ A, \{ B, C \} \} = \{ \{ A, B \} , C \} .  \label{eq:ABC_sym}
\end{equation}
This can be derived simply by writing out the definition of 
the symmetrized products, which leads to the general
operator identity
\begin{equation}
\{ A, \{ B, C \} \} - \{ \{ A, B \} , C \} = \frac14 [[ A, C ] , B ] .
\label{eq:ABC_op_iden}
\end{equation}
Given that $A$, $B$, and $C$ are creators, the right-hand
side vanishes due to Eq.\ (\ref{eq:creator_commutator}),
yielding the identity in Eq.\ (\ref{eq:ABC_sym}).

A useful corollary of this identity is the equivalence
\begin{equation}
B = \{ U, A \} \quad \Leftrightarrow \quad A = \{ U^{-1}, B \} ,
\label{eq:switch_basis}
\end{equation}
in which $A$ and $B$ are creators and $U$ is an invertible creator.
For example, the leftward implication can be derived from
\begin{equation}
\{  U, A \} = \{ U, \{ U^{-1}, B \} \} = \{ \{ U, U^{-1} \} , B \} = B ,
\end{equation}
since $\{ U, U^{-1} \} = 1$.

\section{Proof that \texorpdfstring{$\chi$}{chi} has a global maximum}

\label{app:chi_maximum}

In Sec.\ \ref{sec:time_independent_psi} it was shown that, for a given value of
$\Delta t$, the dynamical stability functional has only one stationary state
with $\chi > 0$. To prove that this is indeed the global maximum of $\chi$, we
can follow the approach used in Eq.\ (\ref{eq:imag_inequality}) to obtain
the inequality
\begin{equation}
\chi = 
\frac{(\imag \inprod{\Delta x}{\sigma})^2}{\matelm{\Delta x}{\hat{\eta}}{\Delta x}}
\le \frac{\matelm{\Delta x}{\Sigma}{\Delta x}}{\matelm{\Delta x}{\hat{\eta}}{\Delta x}}
\equiv \gamma , \label{eq:chi_gamma}
\end{equation}
in which $\Sigma \equiv \outprod{\sigma}{\sigma}$.
Varying the functional $\gamma$ leads to the generalized eigenvalue
equation
\begin{equation}
\Sigma \ket{\Delta x} = \gamma \hat{\eta} \ket{\Delta x} ,
\end{equation}
which is well defined because $\hat{\eta} > 0$. However, because $\Sigma$ is a
projector of rank one, it has only one eigenvector with eigenvalue $\gamma >
0$.  This is just
\begin{equation}
\ket{\Delta x} = -i C \hat{\eta}^{-1} \ket{\sigma} , 
\label{eq:gamma_eigenvector}
\end{equation}
where $C$ is an arbitrary \emph{complex} number. Since the functional $\gamma$
is bounded from above by its maximum eigenvalue, this eigenvalue is the global
maximum of $\gamma$.

Looking back now at Eq.\ (\ref{eq:chi_gamma}), we see that choosing $C$ to be
\emph{real} turns the inequality $\chi \le \gamma$ into an equality, and also
makes the eigenvector (\ref{eq:gamma_eigenvector}) identical to the stationary
state (\ref{eq:Dx_soln}) of $\chi$.  Hence, the global maximum of $\gamma$
is also the global maximum of $\chi$, and the conjecture is proved.

\section{Distance between phase orbits}

\label{app:phase_orbit_distance}

The calculation of $D^2 ([\rho], [\rho'])$ in Sec.\
\ref{sec:phase_orbit_distance} was based on the assumption that $\norm{\Delta
u}$ is small. If this is not true, we must return to Eqs.\ (\ref{eq:D2_lambda})
and (\ref{eq:lambda_phi}) and calculate the function
\begin{align}
\lambda (\phi) & =  m - \tr (\rho e^{i \hat{N} \phi} \rho' e^{-i \hat{N} \phi}) \\
& = m - \sum_{k=1}^{m} \tr (\rho_k e^{i N \phi} \rho_k' e^{-i N \phi})
\end{align}
without any approximations.  This can be done by using the resolution of the
identity $\sum_{n} \Pi_{n} = 1$, in which $\Pi_{n}$ is the projector for the 
$n$-particle subspace [cf.\ Eq.\ (\ref{eq:symmetrizer})].  The result is
\begin{equation}
\lambda (\phi) = m - \sum_{n,n'} M_{n n'} e^{i (n - n') \phi} ,
\label{eq:lambda_phi_M}
\end{equation}
in which
\begin{align}
M_{n n'} & \equiv \sum_{k=1}^{m} \tr (\rho_k \Pi_{n} \rho_k' \Pi_{n'})
\label{eq:M_rho} \\ & = \sum_{k=1}^{m} \frac{\matelm{u_k}{\Pi_{n}}{u_k'}
\matelm{u_k'}{\Pi_{n'}}{u_k}}{\inprod{u_k}{u_k} \inprod{u_k'}{u_k'}} .
\end{align}
This matrix is hermitian, as can be seen from Eq.\ (\ref{eq:M_rho}). The
function (\ref{eq:lambda_phi_M}) can therefore be written as $\lambda (\phi) =
m - G_0 + 2 g (\phi)$, in which $G_l \equiv \sum_{n} M_{n+l, n}$ and
\begin{equation}
g (\phi) = - \sum_{l>0} \real (G_l) \cos (l \phi) + 
\sum_{l>0} \imag (G_l) \sin (l \phi) .
\end{equation}
Hence, in Eq.\ (\ref{eq:D2_lambda}), minimizing $\lambda (\phi)$ is the same as
minimizing $g (\phi)$. This is easy to do in fermion systems with small $d =
\dim \mathcal{H}_{\rmf}$, because $l \le d$. The minimum of $g (\phi)$ can then
be found quickly using a simple grid search and Newton's method.



%

\end{document}